\documentclass[12pt,prd]{article}
\pdfoutput=1
\usepackage{jheppub}
\usepackage{rotating}
\usepackage{array}
\usepackage{amsmath}
\usepackage[normalem]{ulem}
\usepackage{slashed}
\usepackage{booktabs}
\usepackage[pdftex,table]{xcolor}
\usepackage{units}
\usepackage{xfrac}
\usepackage{mathtools}
\usepackage{empheq}
\usepackage[]{units}
\usepackage{multirow}
\usepackage{amssymb}
\usepackage{url}
\usepackage{comment}

\usepackage{tikz}

\def\beq{\begin{equation}}
\def\eeq{\end{equation}}
\newcommand{\bea}{\begin{eqnarray}\begin{aligned}}
\newcommand{\eea}{\end{aligned}\end{eqnarray}}
\def\bitem{\begin{itemize}}
\def\eitem{\end{itemize}}

\definecolor{darkpurple}{rgb}{0.5, 0.2, 0.8}
\definecolor{darkblue}{rgb}{0.0, 0.0, 0.8}
\definecolor{darkgreen}{rgb}{0.0, 0.4, 0.0}
\definecolor{darkred}{rgb}{0.5, 0.0, 0.0}
\usepackage{hyperref}
\hypersetup{
    linktocpage,
     colorlinks,
     citecolor=darkgreen,
     linkcolor= darkgreen,
     urlcolor=darkgreen
}

\usepackage[title]{appendix}
\usepackage{soul}
\usepackage{float}

\def \GeV{{\mathrm{GeV}}}
\def \TeV{{\mathrm{TeV}}}

\abstract{Anomaly detection techniques are growing in importance at the Large Hadron Collider (LHC), motivated by the increasing need to search for new physics in a model-agnostic way. In this work, we provide a detailed comparative study between a well-studied unsupervised method called the autoencoder (AE) and a weakly-supervised approach based on the Classification Without Labels (CWoLa) technique. We examine the ability of the two methods to identify a new physics signal at different cross sections in a fully hadronic resonance search. By construction, the AE classification performance is independent of the amount of injected signal. In contrast, the CWoLa performance improves with increasing signal abundance. When integrating these approaches with a complete background estimate, we find that the two methods have complementary sensitivity. In particular, CWoLa is effective at finding diverse and moderately rare signals while the AE can provide sensitivity to very rare signals, but only with certain topologies. We therefore demonstrate that both techniques are complementary and can be used together for anomaly detection at the LHC.

}

\begin{document}
\title{Comparing Weak- and Unsupervised Methods for Resonant Anomaly Detection}

\author[1]{Jack H. Collins,} 
\author[2,3]{Pablo Mart\'in-Ramiro,} 
\author[3,4]{Benjamin Nachman,}
\author[5]{and David Shih}
\affiliation[1]{\normalsize SLAC National Accelerator Laboratory, Stanford University, Stanford, CA 94309, USA}
\affiliation[2]{\normalsize Instituto de F\'isica Te\'orica, IFT-UAM/CSIC, Universidad Aut\'onoma de Madrid, 28049 Madrid, Spain}
\affiliation[3]{\normalsize Physics Division, Lawrence Berkeley National Laboratory, Berkeley, CA 94720, USA}
\affiliation[4]{\normalsize Berkeley Institute for Data Science, University of California, Berkeley, CA 94720, USA}
\affiliation[5]{\normalsize NHETC, Department of Physics and Astronomy, Rutgers University, Piscataway, NJ 08854, USA}

\emailAdd{jcollins@slac.stanford.edu}
\emailAdd{pmartin.ramiro@predoc.uam.es}
\emailAdd{bpnachman@lbl.gov}
\emailAdd{shih@physics.rutgers.edu}

\preprint{
\begin{flushright}
SLAC--PUB--17558 \\
IFT-UAM/CSIC-21-24
\end{flushright}
}

\maketitle

\section{Introduction}
\label{sec:intro}

The LHC has the potential to address many of the most fundamental questions in physics. Despite all the searches for physics beyond the Standard Model (BSM) conducted by ATLAS \cite{atlasexoticstwiki, atlassusytwiki} and CMS \cite{cmsexoticstwiki, cmssusytwiki, cmsb2gtwiki}, no significant evidence of new physics has been found so far. These searches are designed to target specific new physics signals that would be produced by particular, well-motivated theoretical models. However, it is not feasible to perform a dedicated analysis for every possible topology and therefore some potential signals may be missed. This motivates the introduction of new methods that are less reliant on model assumptions and that are sensitive to a broad spectrum of new physics signatures.

A variety of machine-learning assisted anomaly detection techniques have been proposed that span the spectrum from completely supervised to completely unsupervised\footnote{Citation block taken from the Living Review~\cite{2102.02770}. Background model dependent, non-machine learning models have also been studied experimentally - see Ref.\cite{sleuth,Abbott:2000fb,Abbott:2000gx,Abbott:2001ke,Aaron:2008aa,Aktas:2004pz,Cranmer:2005zn,Aaltonen:2007dg,Aaltonen:2007ab,Aaltonen:2008vt,CMS-PAS-EXO-14-016,CMS-PAS-EXO-10-021,CMS:2020ohc,Aaboud:2018ufy,ATLAS-CONF-2014-006,ATLAS-CONF-2012-107}.}~\cite{DAgnolo:2018cun,Collins:2018epr,Collins:2019jip,DAgnolo:2019vbw,Farina:2018fyg,Heimel:2018mkt,Roy:2019jae,Cerri:2018anq,Blance:2019ibf,Hajer:2018kqm,DeSimone:2018efk,Mullin:2019mmh,1809.02977,Dillon:2019cqt,Andreassen:2020nkr,Nachman:2020lpy,Aguilar-Saavedra:2017rzt,Romao:2019dvs,Romao:2020ojy,knapp2020adversarially,collaboration2020dijet,1797846,1800445,Amram:2020ykb,Cheng:2020dal,Khosa:2020qrz,Thaprasop:2020mzp,Alexander:2020mbx,aguilarsaavedra2020mass,1815227,pol2020anomaly,Mikuni:2020qds,vanBeekveld:2020txa,Park:2020pak,Faroughy:2020gas,Stein:2020rou,Kasieczka:2021xcg,Chakravarti:2021svb,Batson:2021agz,Blance:2021gcs,Bortolato:2021zic} (see Refs.~\cite{Nachman:2020ccu,Kasieczka:2021xcg} for an overview). Two promising approaches are CWoLa Hunting~\cite{Collins:2018epr, Collins:2019jip} and deep autoencoders (AE)~\cite{Farina:2018fyg, Heimel:2018mkt, Cerri:2018anq, Roy:2019jae, Blance:2019ibf}:
\begin{itemize}
\item CWoLa Hunting is a weakly-supervised anomaly detection technique that uses the idea of Classification Without Labels (CWoLa)~\cite{Metodiev:2017vrx} and trains a classifier to distinguish two statistical mixed samples (typically a signal region and a sideband region when used to search for new physics~\cite{Collins:2018epr, Collins:2019jip}) with different amounts of (potential) signal. The output of this classifier can then be used to select signal-like events. This method has already been tested in a real search by the ATLAS collaboration \cite{collaboration2020dijet}.

\item Autoencoders are the basis for a fully-unsupervised anomaly detection technique that has been widely explored and used in many real-world scenarios. A deep autoencoder is a neural network that learns to compress data into a small latent representation and then reconstruct the original input from the compressed version. The AE can be trained directly on a background-rich sample to learn the features of background events and reconstruct them well. By contrast, it will struggle to reconstruct anomalous (e.g. signal) events. The reconstruction loss, defined by some chosen distance measure between the original and reconstructed event, can then be used as a classification score that selects anomalous events.

\end{itemize}
To date, there has not been a direct and detailed comparison between these two methods.\footnote{Recently,  the  authors  of  the  Tag  N'  Train  method~\cite{Amram:2020ykb}  also  made  comparisons  between  these approaches with the aim of combining them.  Our study has the orthogonal goal of directly comparing the two approaches in detail as distinct methods to understand their complementarity.} The goal of this paper will be to provide such a comparison, describe the strengths and weaknesses of the two approaches, and highlight their areas of complementarity.

We will focus on the new physics scenario where a signal is localized in one known dimension of phase space (in this case, the dijet invariant mass) on top of a smooth background. While CWoLa Hunting explicitly requires a setup like this to generate mixed samples, AEs technically do not, as they can function as anomaly detectors in a fully unsupervised setting. However, even for AEs one generally needs to assume something about the signal and the background in order to enable robust, data-driven background estimation.

In this scenario, both models can be trained to exploit the information in the substructure of the two jets to gain discriminating power between the signal and background events. 
CWoLa Hunting, being able to take advantage of the weak labels, should excel in the limit of moderately high signal rate in the sample because it is able to take advantage of learnt features of the signal. It should fail however in the limit of no signal. On the other hand, an unsupervised approach like the AE is fully agnostic to the specific features of the signal, and thus should be robust in the limit of low signal statistics. While the behaviour of these strategies in the high and low signal statistics limits can be understood on general grounds, it is the intermediate regime in which the two strategies might have a `cross-over' in performance that is of most interest for realistic searches. It is therefore worth analyzing in detail for some case studies the nature of this crossover and the degree of complementary of the strategies.

In this work, we provide a detailed comparative analysis of the performance of CWoLa Hunting and AEs at anomaly detection on a fully hadronic resonance search. After evaluating the ability of both methods to identify the signal events for different cross sections, we test whether they are able to increase the significance of the signal region excess. Here we emphasize the importance of going beyond the AUC metric and consider more meaningful performance metrics such as the Significance Improvement Characteristic (SIC). Furthermore, a realistic fit procedure based on ATLAS and CMS hadronic diboson searches is implemented. We will confirm the general behavior of AE and CWoLa Hunting approaches at large and small signal strengths described in the previous paragraph, and we will demonstrate quantitatively the existence of a cross-over region in a part of parameter space that could be of practical relevance. We conclude that the approaches have complementary sensitivity to different amounts or types of signals.

This paper is organized as follows. In Section~\ref{sec:sim}, we describe the resonant hadronic new physics signal that we consider and the simulation details for the generated events. In Section~\ref{sec:ML}, we introduce the details of CWoLa Hunting and the AE and explain how they can be successfully implemented in this type of new physics searches. We present results for the two models in Section~\ref{sec:results} and discuss their performance at anomaly detection. Finally, the conclusions are presented in Section~\ref{sec:conc}.

\section{Simulation}
\label{sec:sim}

In order to investigate the performance of CWoLa Hunting and AEs in a generic hadronic resonance search, we consider a benchmark new physics signal $pp \rightarrow Z^{\prime} \rightarrow XY$, with $X \rightarrow jjj$ and $Y \rightarrow jjj$.  There is currently no dedicated search to this event topology.  The mass of the new heavy particle is set to $m_{Z^{\prime}} = 3.5 \; \TeV$, and we consider two scenarios for the masses of the new lighter particles: $m_{X}$, $m_{Y} = 500 \; \GeV$ and $m_{X}$, $m_{Y} = 300 \; \GeV$. These signals typically produce a pair of large-radius jets $J$ with invariant mass $m_{\text{JJ}} \simeq 3.5 \; \TeV$, with masses of $m_{J} = 500, 300 \; \GeV$ and a three-prong substructure. These signals are generated in the LHC Olympics framework~\cite{Kasieczka:2021xcg}.

\begin{figure}[H]
   \begin{center}
        \hspace{-10pt}
        \includegraphics[scale=0.38]{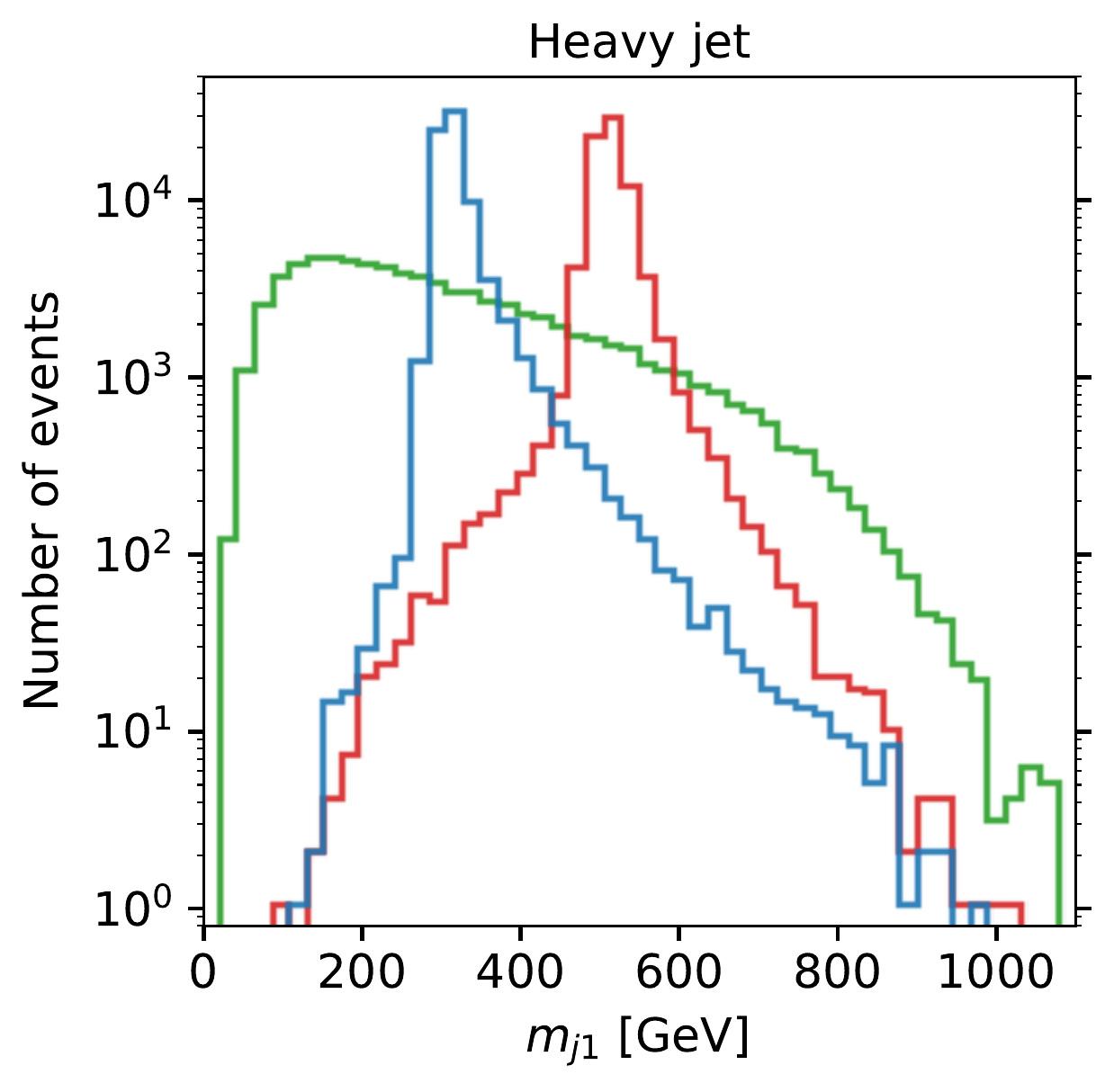}
        \hspace{1pt}
        \includegraphics[scale=0.38]{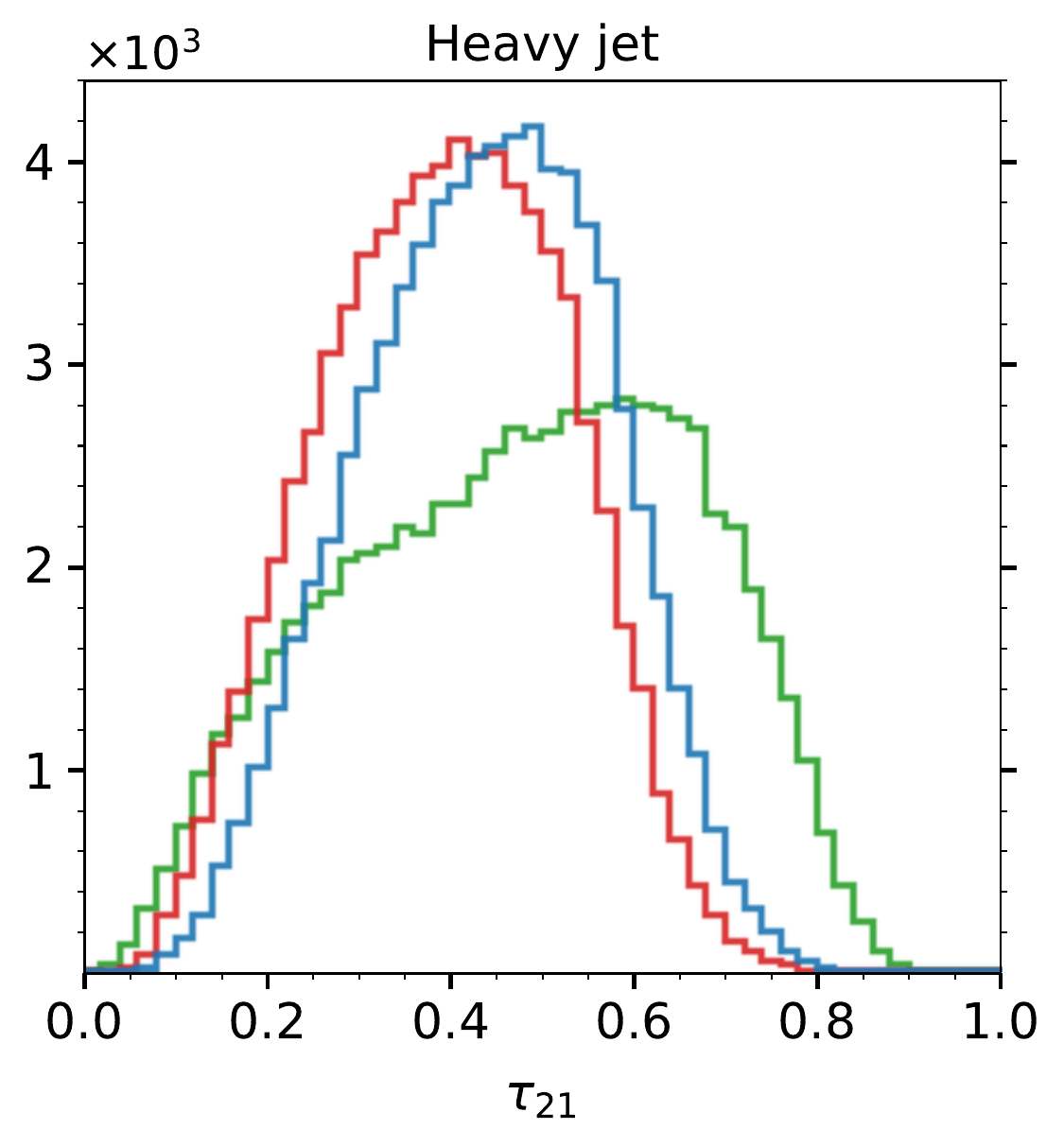}
        \hspace{1pt}
        \includegraphics[scale=0.38]{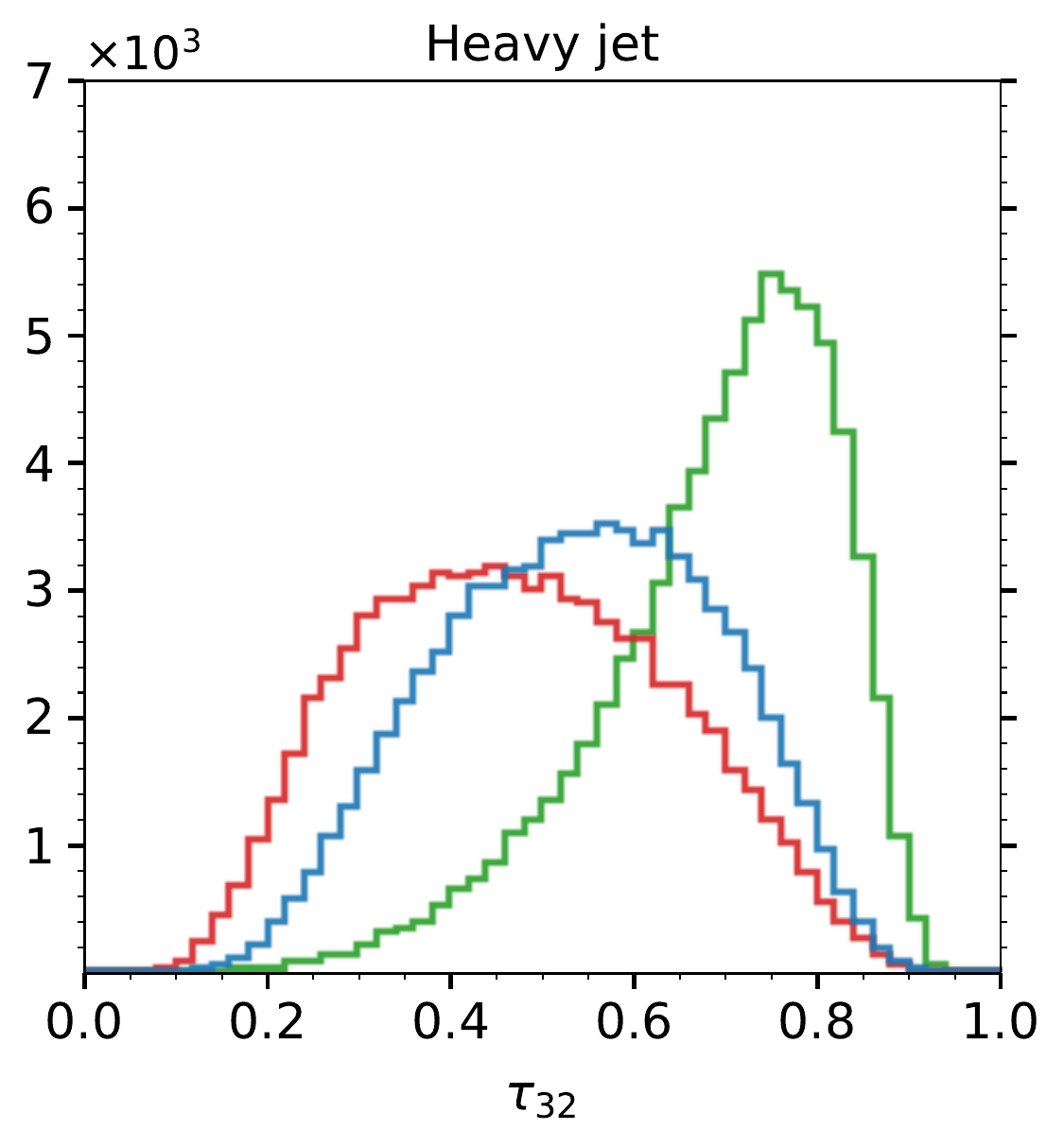} \\
        \vspace{1pt}
        \includegraphics[scale=0.38]{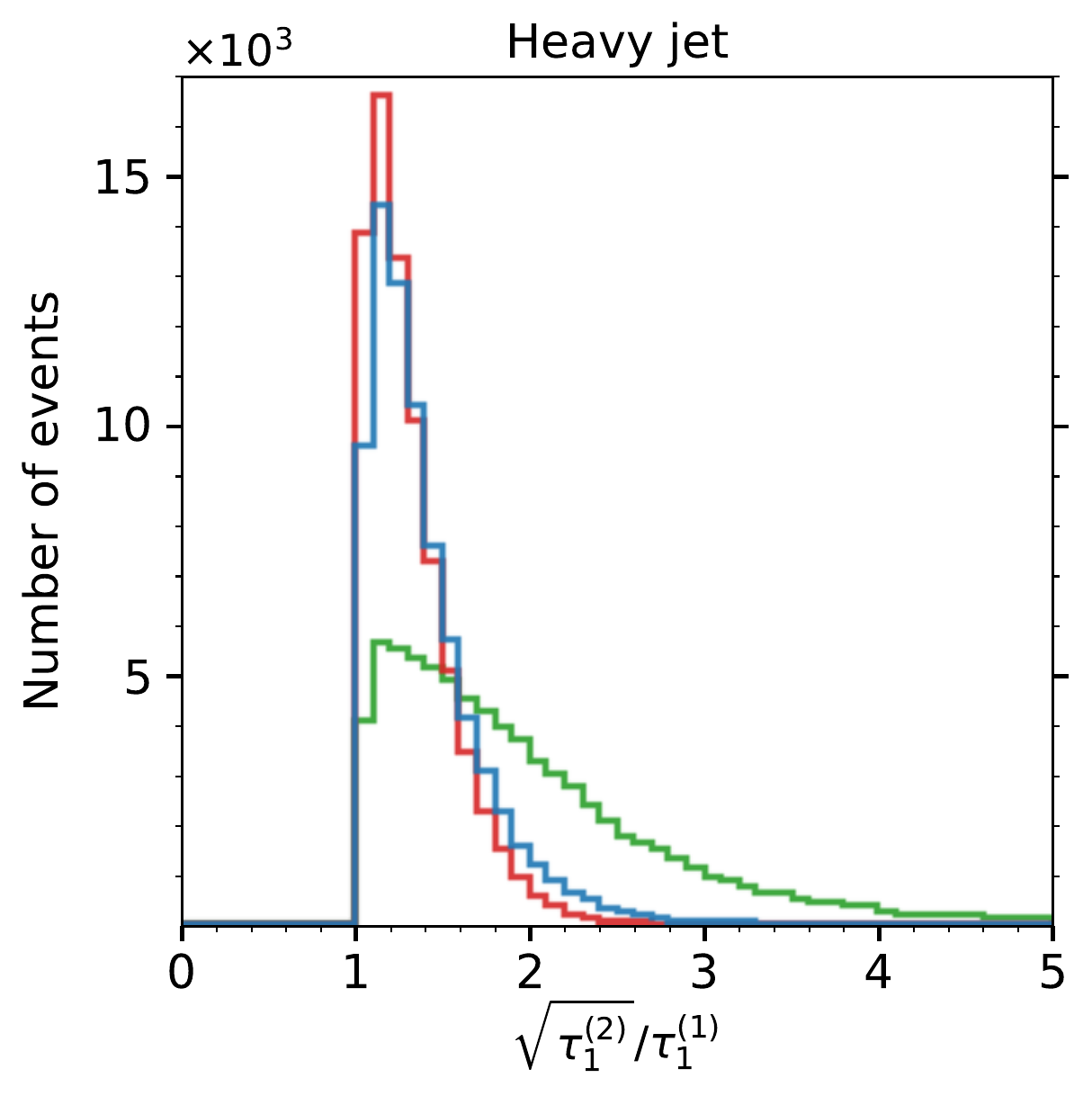}
        \hspace{1pt}
        \includegraphics[scale=0.38]{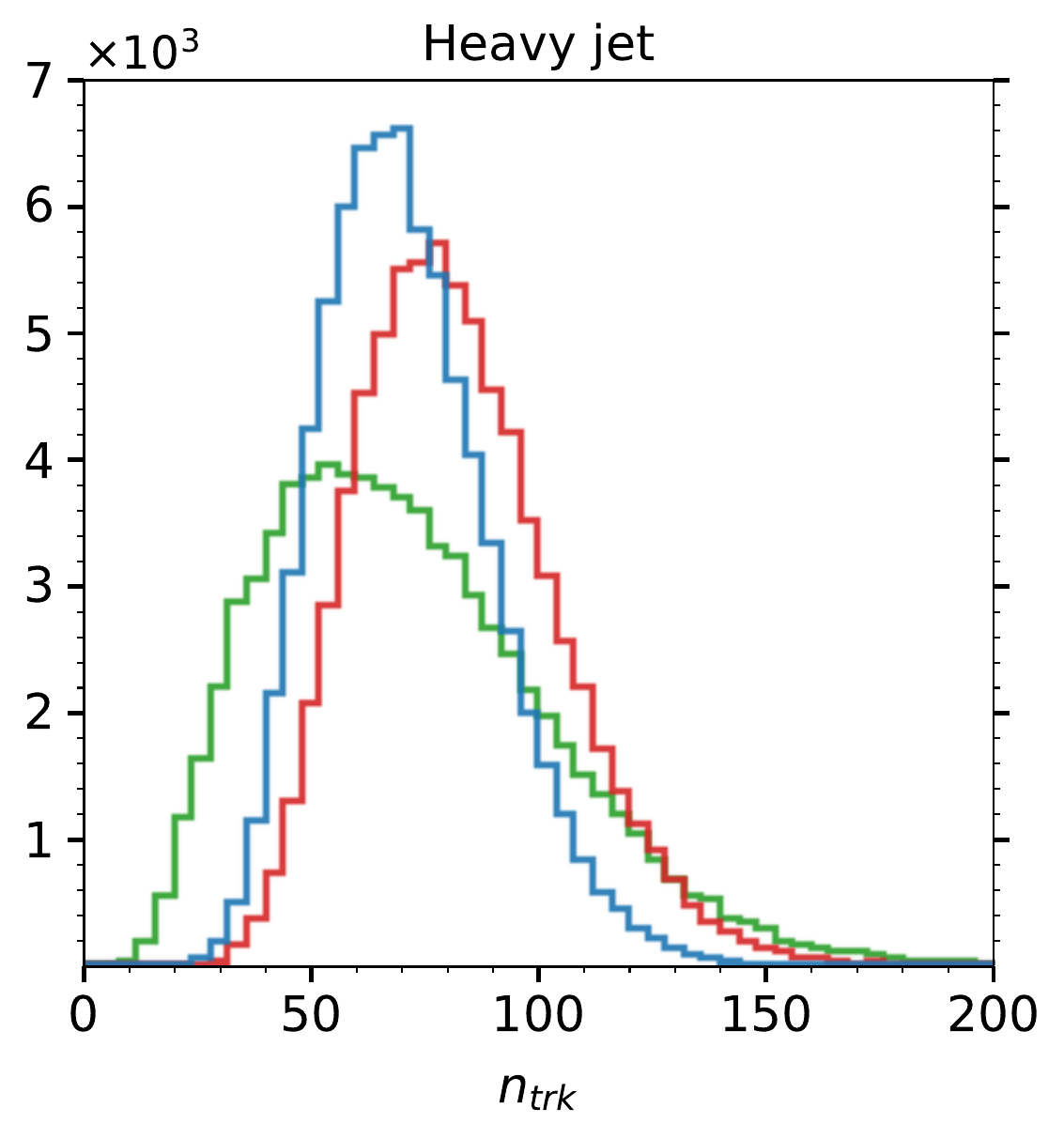}
        \hspace{1pt}
        \includegraphics[scale=0.38]{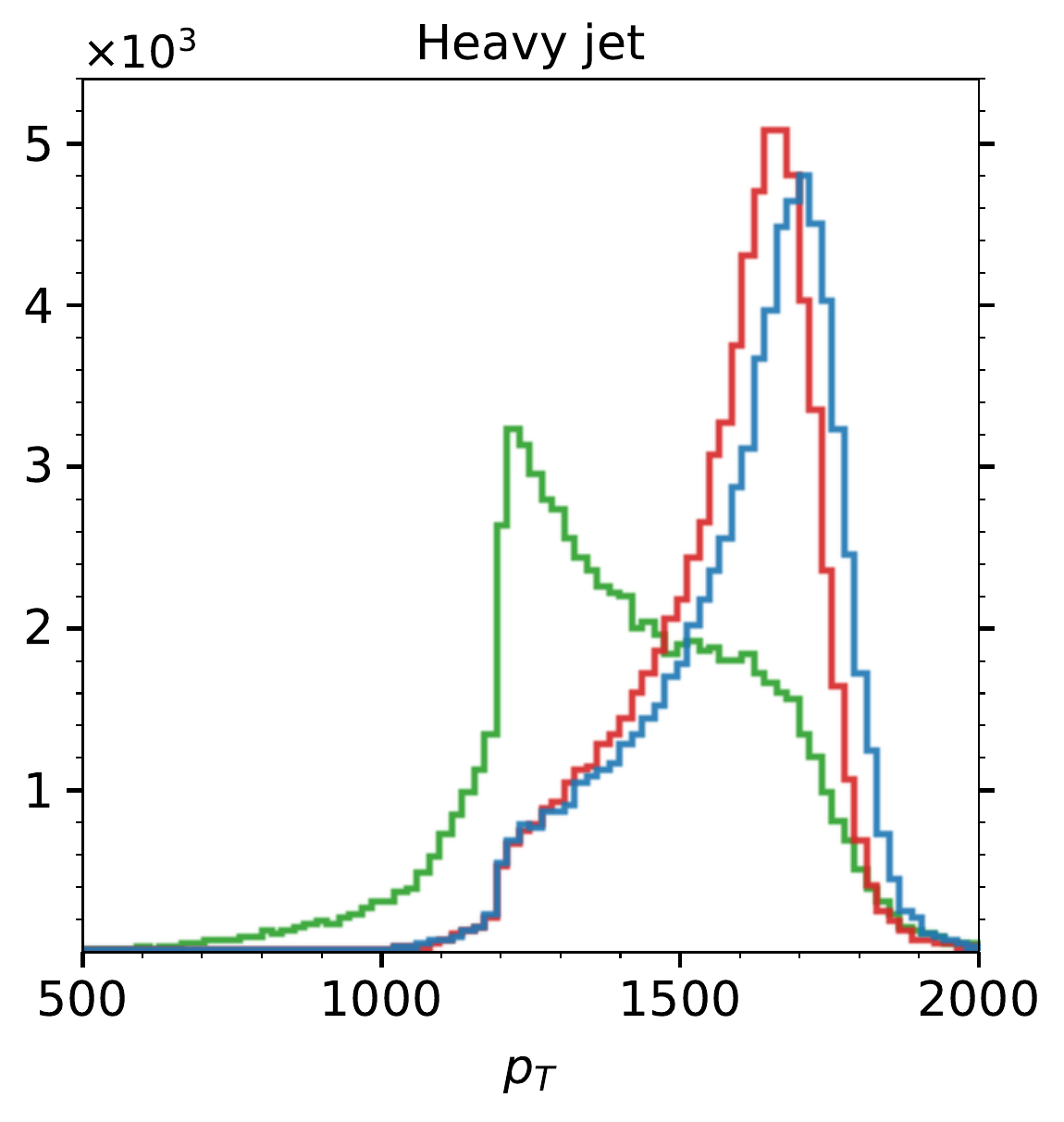} \\
        \vspace{1pt}
        \hspace{-10pt}
        \includegraphics[scale=0.38]{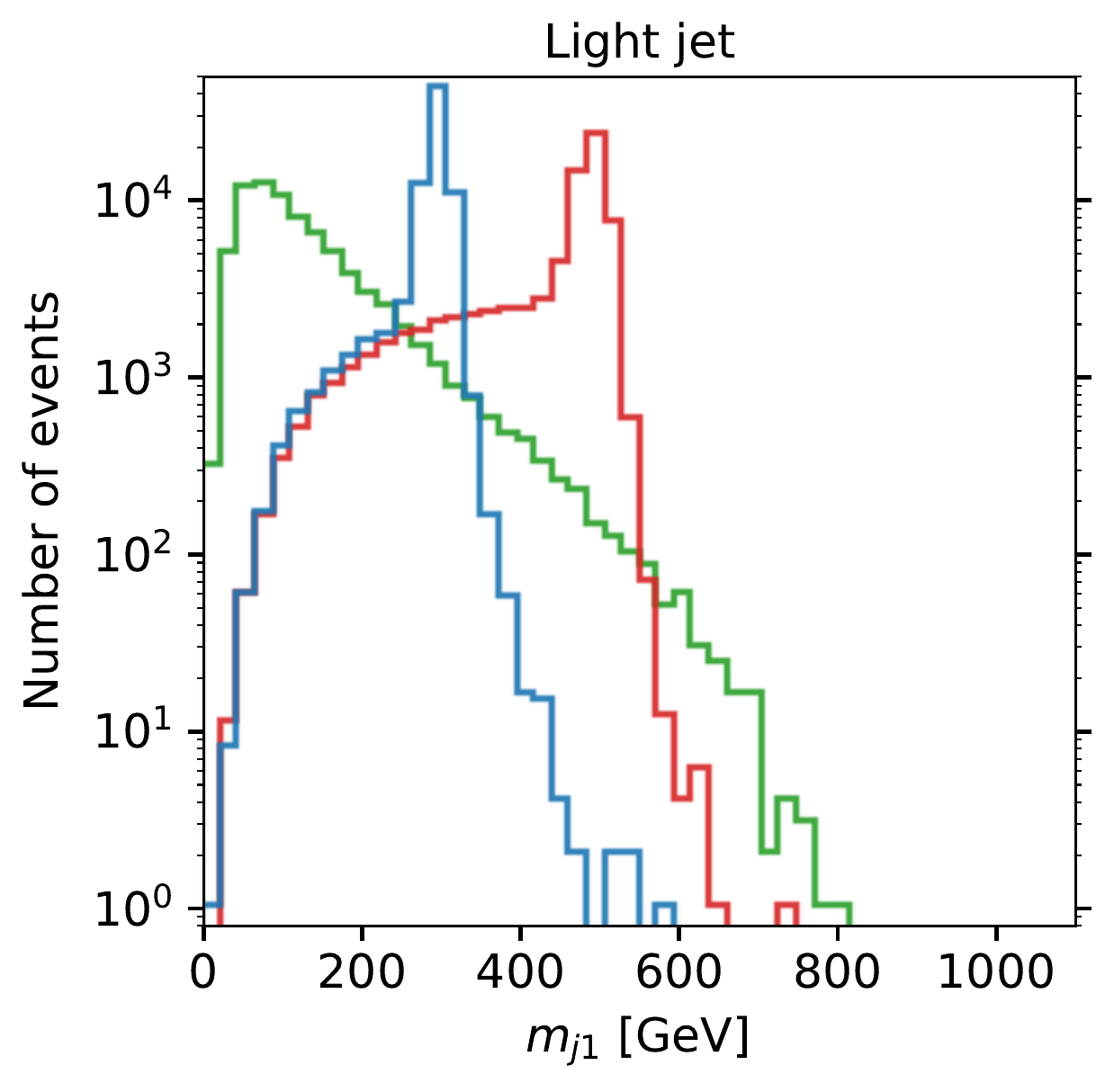}
        \hspace{1pt}
        \includegraphics[scale=0.38]{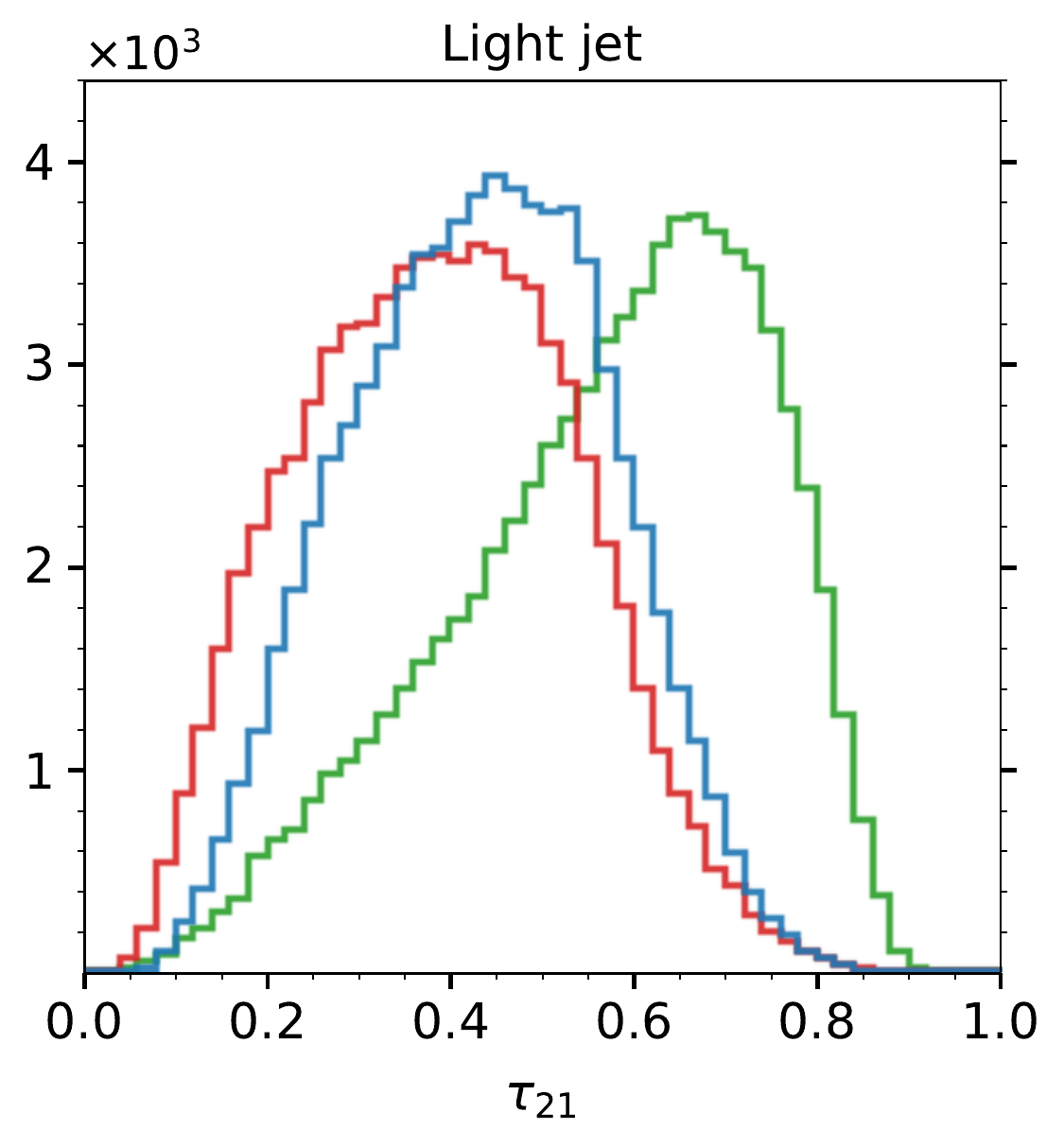}
        \hspace{1pt}
        \includegraphics[scale=0.38]{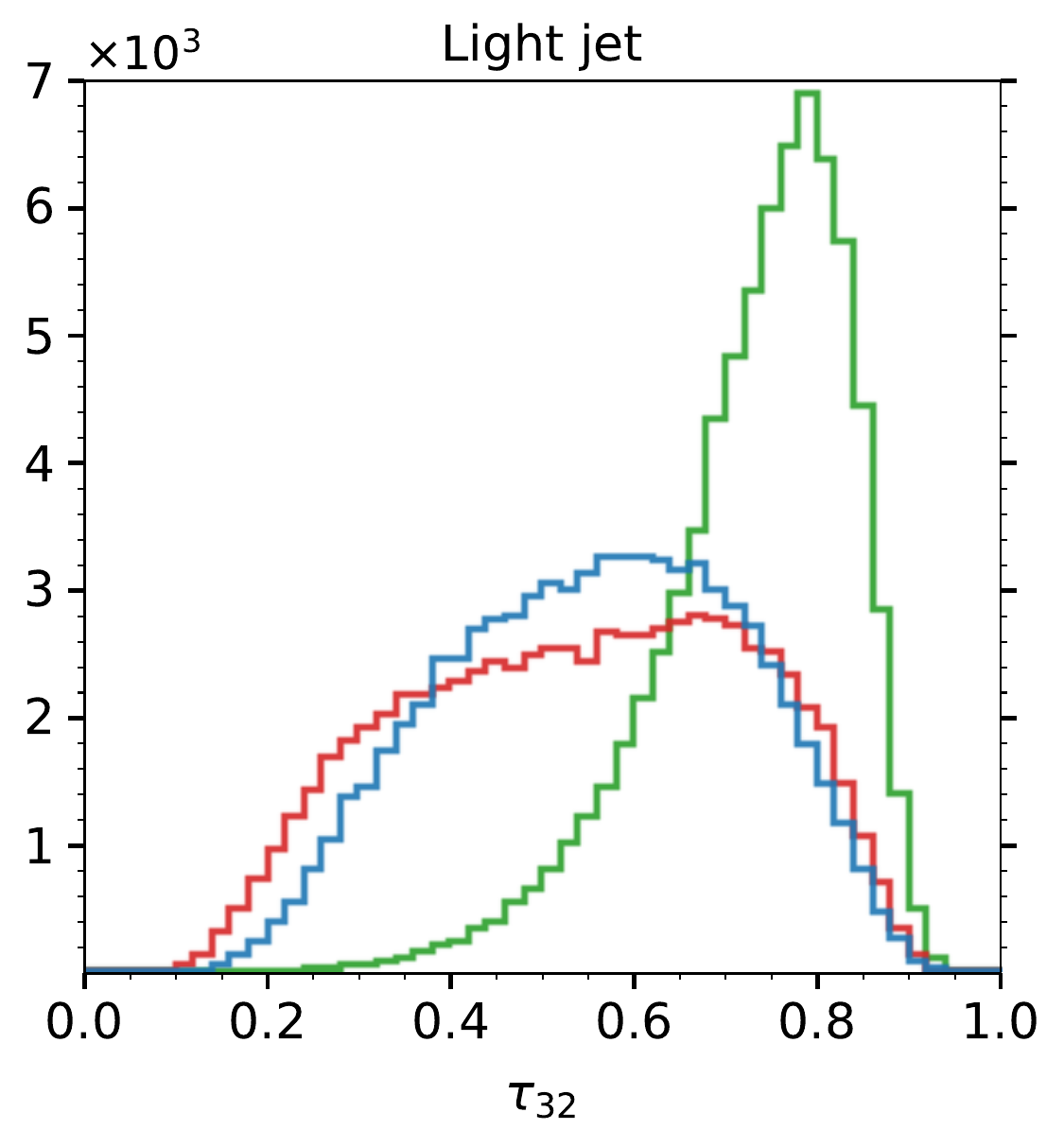} \\
        \vspace{1pt}
        \includegraphics[scale=0.38]{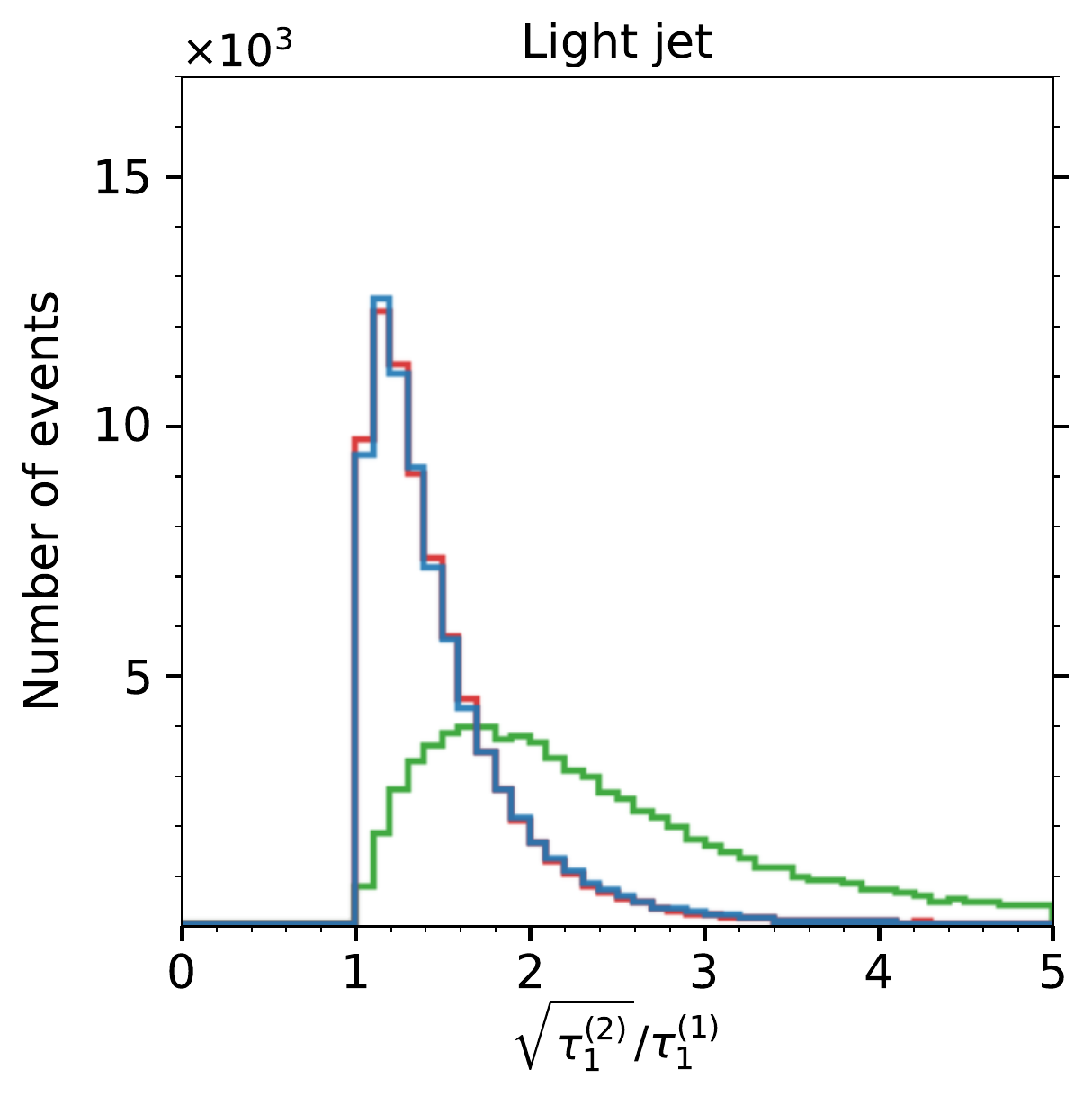}
        \hspace{1pt}
        \includegraphics[scale=0.38]{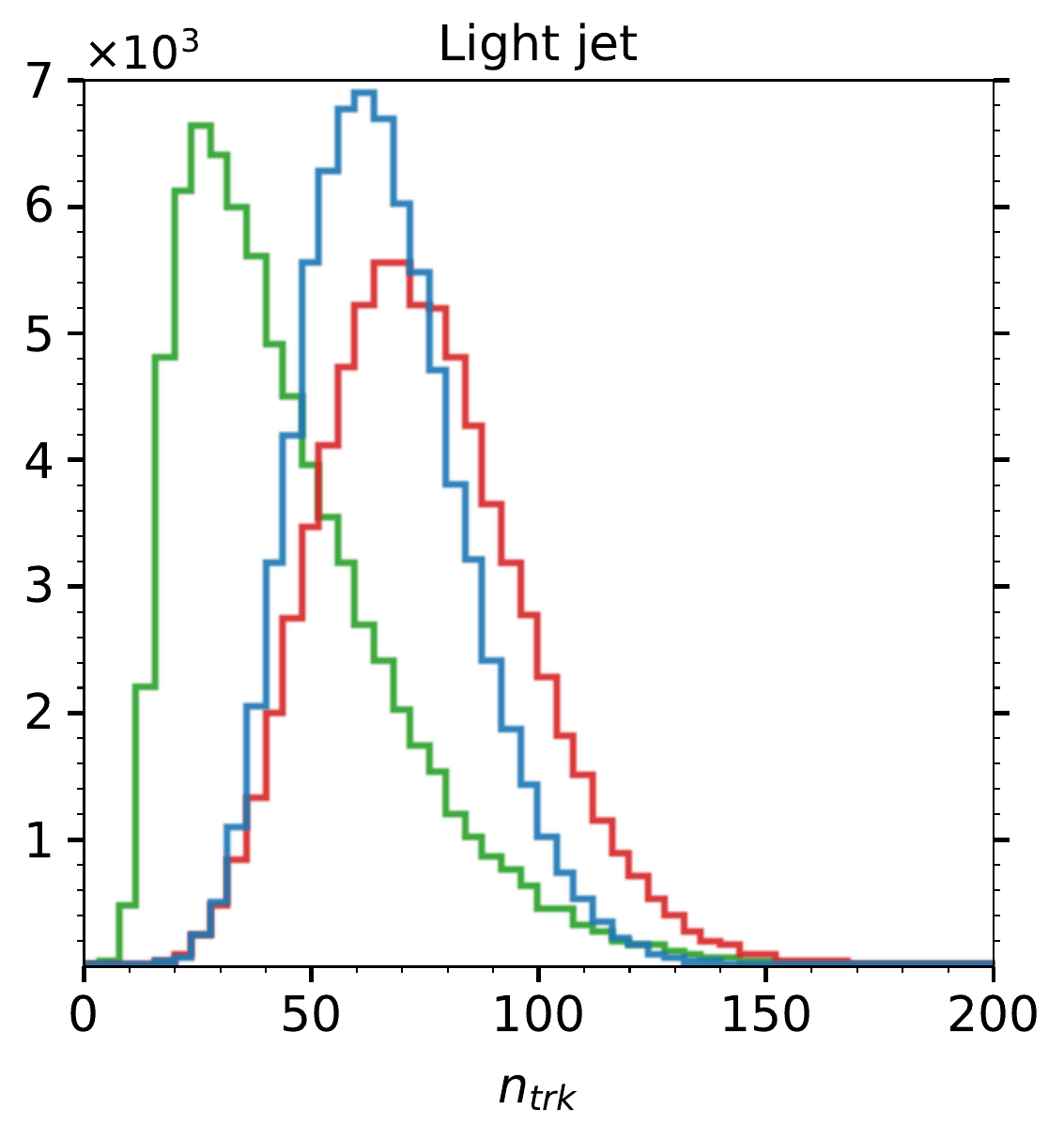}
        \hspace{1pt}
        \includegraphics[scale=0.38]{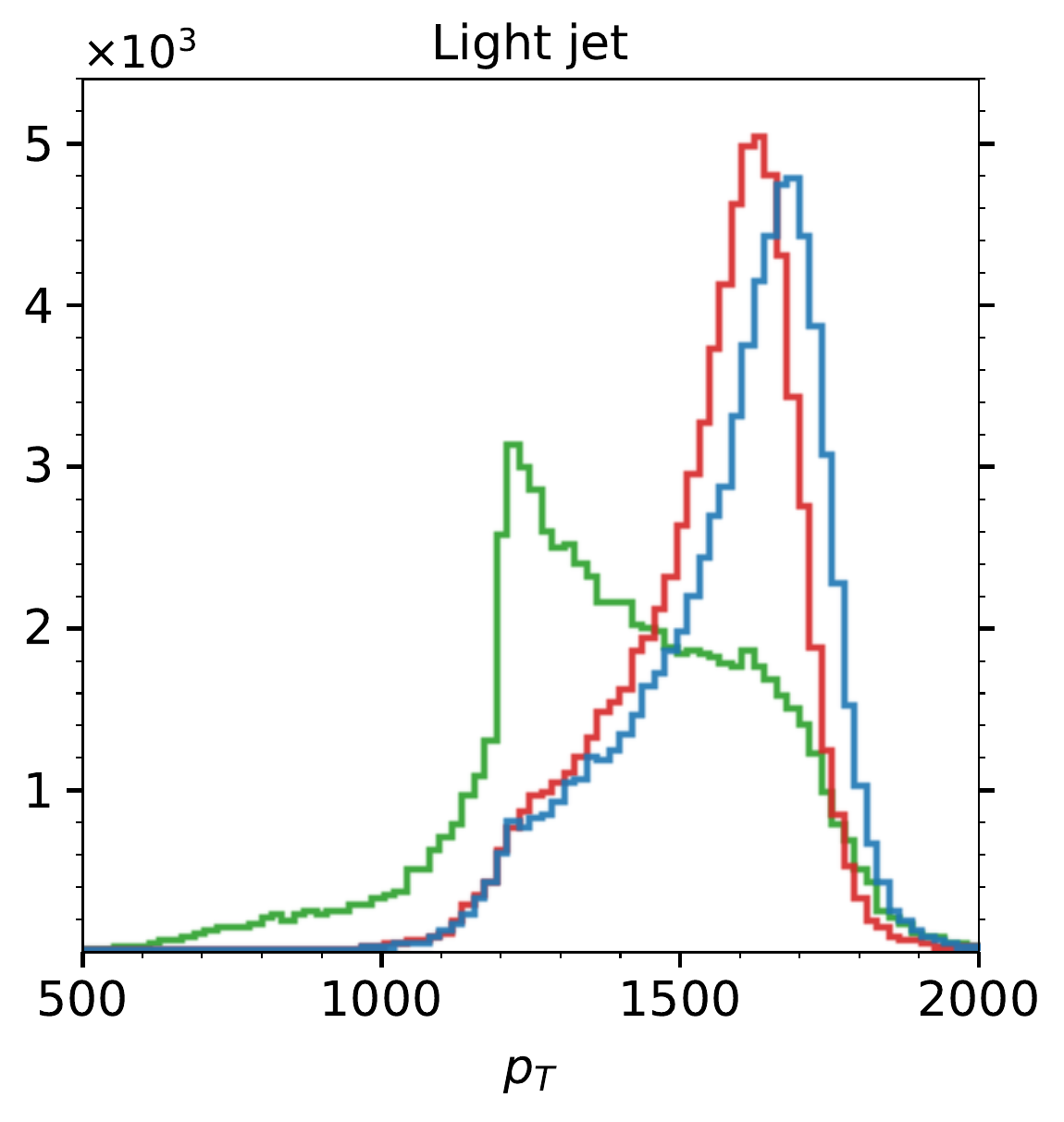}
   \end{center}
   \caption{A reduced set of the input features that we use for training the models are shown for Jet $1$ (first and second rows) and Jet $2$ (third and fourth rows) for the signals with $(m_{j_{1}}, m_{j_{2}}) = (500, 500) \; \GeV$ (red) and $(m_{j_{1}}, m_{j_{2}}) = (300, 300) \; \GeV$ (blue), and the background (green). We plot the same number of signal and background events for visualization purpose.}
   \label{fig:input_features}
\end{figure}

For both signal models, we generated $10^{4}$ events. One million QCD dijet events serve as the background and are from the LHC Olympics~\cite{Kasieczka:2021xcg} dataset. All the events were produced and showered using \texttt{Pythia 8.219} \cite{Sjostrand:2007gs} and the detector simulation was performed using \texttt{Delphes 3.4.1} \cite{deFavereau:2013fsa}, with no pileup or multiparton interactions included. All jets are clustered with \texttt{FastJet 3.3.2} \cite{Cacciari:2011ma} using the anti-$k_{t}$ algorithm \cite{Cacciari:2008gp} with radius parameter $R = 1$. We require events to have at least one large-radius jet with $p_{T} > 1.2 \; \TeV$ and pseudo-rapidity $|\eta| < 2.5$. The two hardest jets are selected as the candidate dijet and a set of substructure variables are calculated for these two jets as shown in Fig.~\ref{fig:input_features}.  In particular, the $N$-subjettiness variables $\tau_i^{\beta}$ were first proposed in Ref.~\cite{Thaler:2011gf,Thaler:2010tr} and probe the extent to which a jet has $N$ subjets.  All $N$-subjettiness variables are computed using \texttt{FastJet 3.3.2} and angular exponent $\beta = 1$ unless otherwise specified in the superscript. The observable $n_{\text{trk}}$ denotes the number of constituents in a given jet. Jets are ordered by mass in descending order.

\section{Machine Learning Setup}
\label{sec:ML}

In this section, we describe the machine learning setup and the strategies that we follow to train CWoLa Hunting and the AE approaches.

\subsection{Classification Without Labels (CWoLa)}
\label{sec:CWoLa}

The strategy closely follows Ref.~\cite{Collins:2018epr, Collins:2019jip}.  To begin, we use a set of high-level observables computed from the two leading jets.  In particular, we consider the following set of input features for each jet:
\begin{equation}
Y_{i} = \left\{ m_{J}, \, \sqrt{\tau_{1}^{(2)}} / \tau_{1}^{(1)}, \, \tau_{21}, \, \tau_{32}, \, \tau_{43}, \, n_{\text{trk}} \right\} \, .
\label{eq:input_CW}
\end{equation}
A reduced set of input features is shown in Fig.~\ref{fig:input_features}.

\begin{figure}[t!]
   \begin{center}
        \includegraphics[scale=0.65]{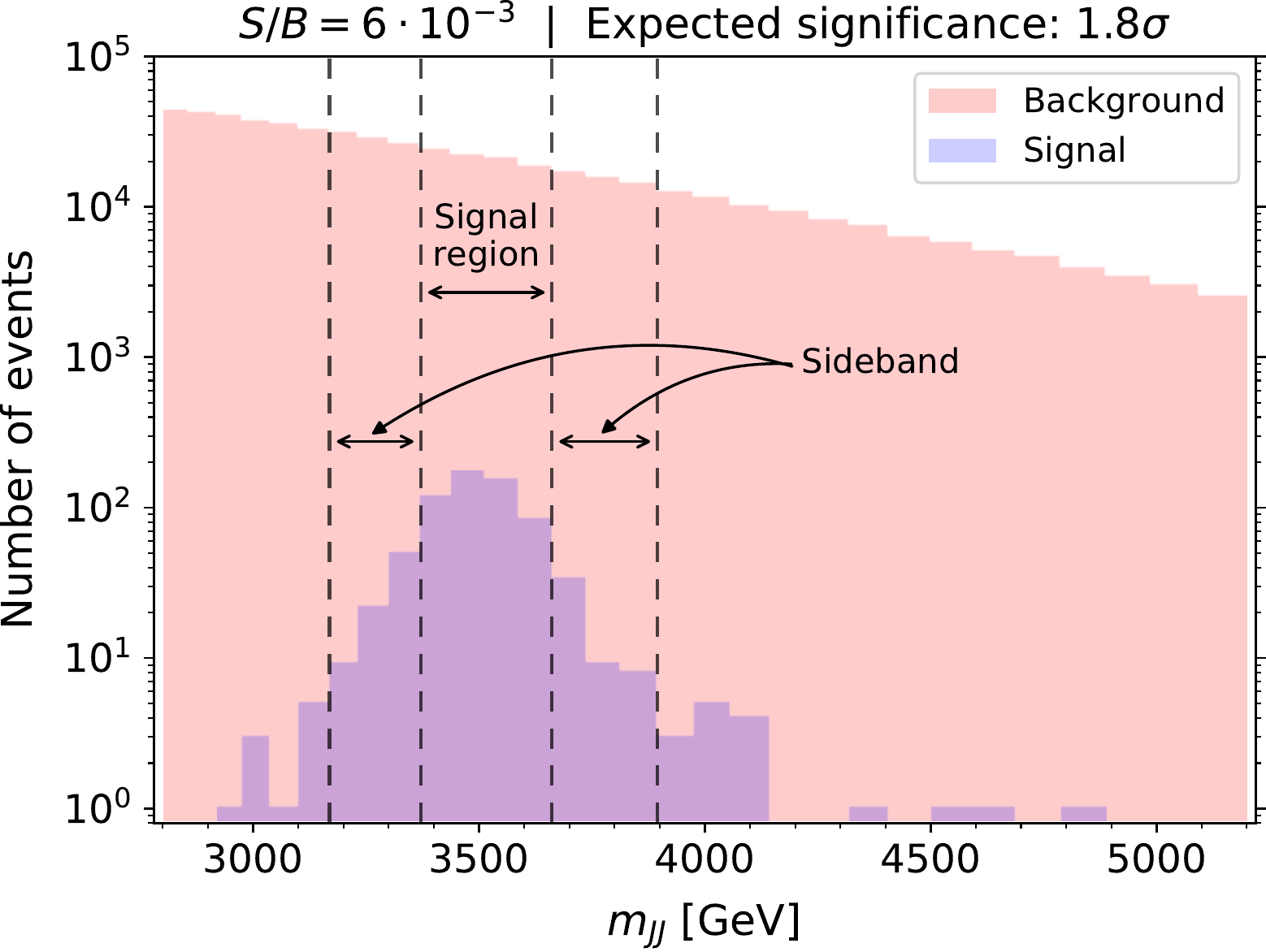}
   \end{center}
   \caption{Distribution of a fraction of signal and all background events on the $m_{JJ}$ plane. Events are divided in $30$ bins and a signal region and a sideband region are defined, as described in the text in Section~\ref{sec:CWoLa}. The amount of signal that has been injected corresponds to $S/B = 6 \cdot 10^{-3}$ and $S/ \sqrt{B} = 1.8\sigma$ in the signal region.}
   \label{fig:mJJ_distribution}
\end{figure}

We select all of the events in the range $m_{JJ} \in [2800, 5200] \; \GeV$ and split them uniformly in $\log(m_{JJ})$ in $30$ bins. After selecting this range, $537304$ background events remain in our sample. In order to test for a signal hypothesis with mass $m_{JJ} = m_{\text{peak}}$, where $m_{\text{peak}}$ is the mean mass of the injected signal, we build a signal region and a sideband region. The former contains all of the events in the four bins centered around $m_{\text{peak}}$, while the latter is built using the three bins below and above the signal region. By doing this, we obtain a signal region in the range $m_{JJ} \in (3371, 3661) \; \GeV$ with a width of $290 \; \GeV$, and a lower and upper sidebands that are $202 \; \GeV$ and $234 \; \GeV$ wide, respectively. The size of the signal region window depends on the signal width\footnote{This is dominated by detector effects; for models with a non-trivial off-shell width, this may not be optimal.} and can be scanned for optimal performance. In Fig.~\ref{fig:mJJ_distribution}, we show the binned distribution of a fraction of signal and all background events, with a signal-to-background ratio of $S/B = 6 \cdot 10^{-3}$ and a naive expected significance $S/ \sqrt{B} = 1.8\sigma$ in the signal region. Note that if a signal is present in data, the signal region will have a larger density of signal events than the mass sidebands, which are mainly populated by background events by construction. In a real search the location of the mass peak of any potential signal would be unknown, and thus the mass hypothesis must be scanned, as described in Ref. \cite{Collins:2019jip}.

After defining the signal and sideband regions, a CWoLa classifier is trained to distinguish the events of the signal region from the events of the sideband using the set of twelve input features that describe the jet substructure of each event, presented in Eq.~\eqref{eq:input_CW}. In this way, the CWoLa classifier will ideally learn the signal features that are useful to discriminate between both regions. It is important to remark that the classifier performance should be very poor when no signal is present in the signal region, but if a signal is present with anomalous jet substructure then the classifier should learn the information that is useful to distinguish the signal and sideband regions.

In this work, the classifiers that we use are fully connected neural networks with four hidden layers. The first layer has $64$ nodes and a leaky Rectified Linear Unit (ReLU)~\cite{maasrectifier} activation \cite{ReLu} (with an inactive gradient of $0.1$), and the second through fourth layers have $32$, $16$ and $4$ nodes respectively, with Exponential Linear Unit (ELU) activation \cite{clevert2015fast}. The output layer has a sigmoid activation. The first three hidden layers are followed by dropout layers with a $20 \, \%$ dropout rate \cite{JMLR:v15:srivastava14a}. We use the binary cross-entropy loss function and the Adam optimizer \cite{adam} with learning rate of $0.001$ and learning rate decay of $5 \cdot 10^{-4}$, batch size of $20480$ and first and second moment decay rates of $0.8$ and $0.99$, respectively. The training data is reweighted such that the low and high sidebands have equal total weight, the signal region has the same total weight as the sum of the sidebands, and the sum of all events weights in the training data is equal to the total number of training events. This reweighting procedure ensures that the two sideband regions have the same contribution to the training process in spite of their different event rates, and results in a classifier output peaked around $0.5$ in the absence of any signal. All classifiers are implemented and trained using \texttt{Keras} \cite{keras} with \texttt{TensorFlow} \cite{tensorflow} backend.

We implement a nested cross-validation procedure with five $k$-folds and therefore all data are used for training, validation and testing. We standardize all the input features from the training and validation sets using training information, and those from the test set using training and validation information. The full dataset is divided randomly, bin by bin, in five event samples of identical size. We set one of the samples aside for testing and perform four rounds of training and validation with the other four, using one of the subsets for validation each time. For each round, we train ten neural networks for $700$ epochs on the same training and validation data, using a different initialization each time. We measure the performance of each classifier on validation data using the metric $\epsilon_{\text{val}}$, defined as the true positive rate for the correct classification of signal region events, evaluated at a threshold with a false positive rate $f = 1 \, \%$ for incorrectly classifying events from the sideband region. Only the best out of the ten models is saved. We use an early stopping criterion to stop training if the validation performance has not improved for 300 epochs. At the end of the four rounds, we use the mean of the outputs of the four selected models to build an ensemble model which is more robust on average than any individual model. This ensemble model is used to classify the events in the test set, and the $x \, \%$ most signal-like events are selected by applying a cut on the classifier output. This procedure is repeated for all five choices of test set, and the selected most signal-like events from each are combined into a signal-like sample. If a signal is present in data and CWoLa Hunting is able to find it, it will show as a bump in the signal region of the signal-like sample on the $m_{JJ}$ plane, and standard bump-hunting techniques can be used to locate the excess.

It is worth mentioning that using an averaged model ensemble is important to reduce any potential overfitting. The cross-validation procedure ensures that even if an individual classifier learns any statistical fluctuations in the training data, each model will tend to overfit different regions of the phase space. As a result, the models will disagree in regions where overfitting has occurred, but will tend to agree in any region where a consistent excess is found.

\subsection{Autoencoder}
\label{sec:AE}

In this subsection we describe the strategy followed for the AE implementation. In the first place, we take the two leading jets in each event, ordered by mass, and consider the following set of input features for each jet:
\begin{equation}
Y_{i} = \left\{ m_{J}, \, \tau_{21}, \, \tau_{32}, \, n_{\text{trk}}, \, p_{T} \right\} \, .
\label{eq:input_AE}
\end{equation}
After analyzing different sets of input features, we found that the collection of $10$ features presented in Eq.~\eqref{eq:input_AE} led to optimal performance. All the input features are standardized for the analysis.

Unlike the CWoLa method, the AE is trained on all the available background events in the full $m_{JJ}$ range. The AE only requires a signal region and a background region for the purposes of background estimation through sideband interpolation. For the anomaly score itself (the reconstruction error), the AE is completely agnostic as to the $m_{JJ}$ range of the signal.

In this work, the AE that we consider is a fully connected neural network with five hidden layers. The AE has an input layer with $10$ nodes. The encoder has two hidden layers of $512$ nodes, and is followed by a bottleneck layer with $2$ nodes and linear activation. The decoder has two hidden layers of $512$ nodes, and is followed by an output layer with $10$ nodes and linear activation. All of the hidden layers have ReLU activation, and the first hidden layer in the encoder is followed by a Batch Normalization layer. We use the Minimum Squared Error (MSE) loss function and the Adam optimizer with learning rate of $10^{-4}$, first and second moment decay rates of $0.9$ and $0.999$, respectively, and a mini-batch size of $128$. In Appendix~\ref{sec:AEmodel} we describe our quasi-unsupervised model-selection procedure. We use \texttt{Pytorch} \cite{NEURIPS2019_9015} for implementing and training the AE.

In order to achieve a satisfactory generalization power, we decided to build an AE ensemble. For this purpose, we train fifty different models (i.e. the ensemble components) with random initialization on random subsamples of $50000$ background events. Each model is trained for only $1$ epoch. It is important to note that the training sample size and number of training epochs had a significant impact in the AE performance. When these are too large, the AE learns too much information and losses both generalization power and its ability to discriminate between signal and background events. For this reason, our training strategy gives the AE more generalization power and makes it more robust against overfitting.

The autoencoder ensemble is evaluated on the full dataset. The final MSE reconstruction loss of an event is obtained by computing the mean over the fifty different ensemble components. The optimal anomaly score is derived from the SIC curve as described in Appendix~\ref{sec:AEmodel}. The results presented in this paper are for an AE trained on $S = 0$. We have verified that including relevant amounts of signal $S$ do not significantly change the results. Therefore, for the sake of computational efficiency, we choose to present the AE trained with $S=0$ everywhere.

\section{Results}
\label{sec:results}

\subsection{Signal benchmarks}

Now we are ready to test the performance of CWoLa Hunting and the AE for different amounts of injected signal. Importantly, we will quantify the performance of CWoLa Hunting and the AE not using the full $m_{JJ}$ range, but using a narrower slice $m_{JJ}\in (3371, 3661)$~GeV, the signal region defined in Sec.~\ref{sec:CWoLa}. This way, all performance gains from the two methods will be measured relative to the naive significance obtained from a simple dijet resonance bump hunt.

We define a set of eight benchmarks with a different number of injected signal events. For this purpose, to the current sample of $537304$ background events in the range $m_{JJ} \in [2800, 5200] \; \GeV$, we add from $175$ to $730$ signal events. This results in a set of benchmarks distributed over the range $S/B \in [1.5 \cdot 10^{-3}, 7 \cdot 10^{-3}]$ in the signal region, corresponding to an expected naive significance in the range $S/\sqrt{B} \in [0.4, 2.1]$. To test the consistency of both models when no signal is present in data, we add a final benchmark with no signal events which allows us to evaluate any possible biases. For each $S/B$ benchmark, the performance of CWoLa Hunting is evaluated across ten independent runs to reduce the statistical error using a random subset of signal events each time. After exploring a large range of cross sections, we decided to examine this range in $S/B$ because it is sufficient to observe an intersection in the performance of the two methods. The observed trends continue beyond the limits presented here.

\begin{figure}[t!]
   \begin{center}
        \includegraphics[scale=0.51]{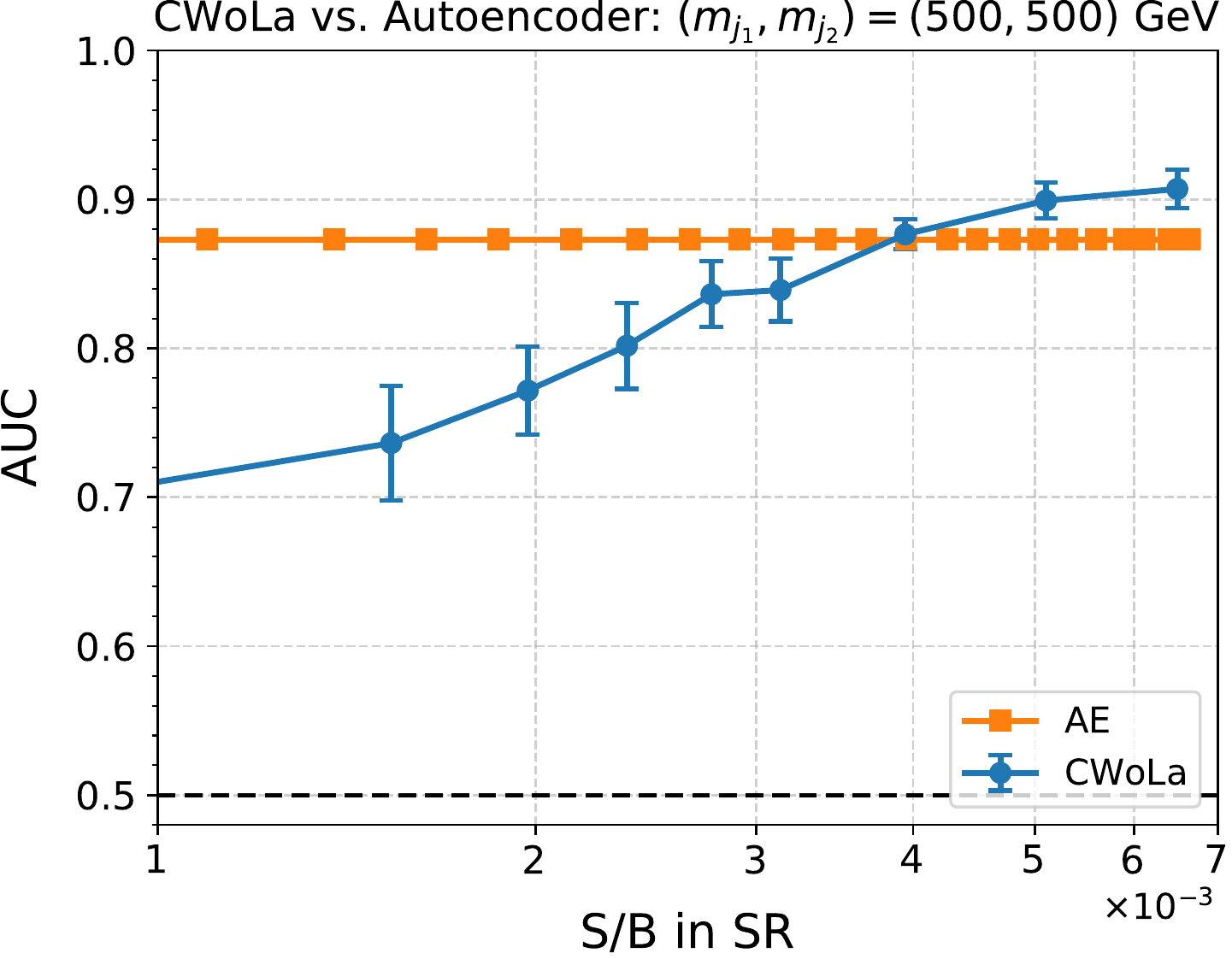}
        \hspace{10pt}
        \includegraphics[scale=0.51]{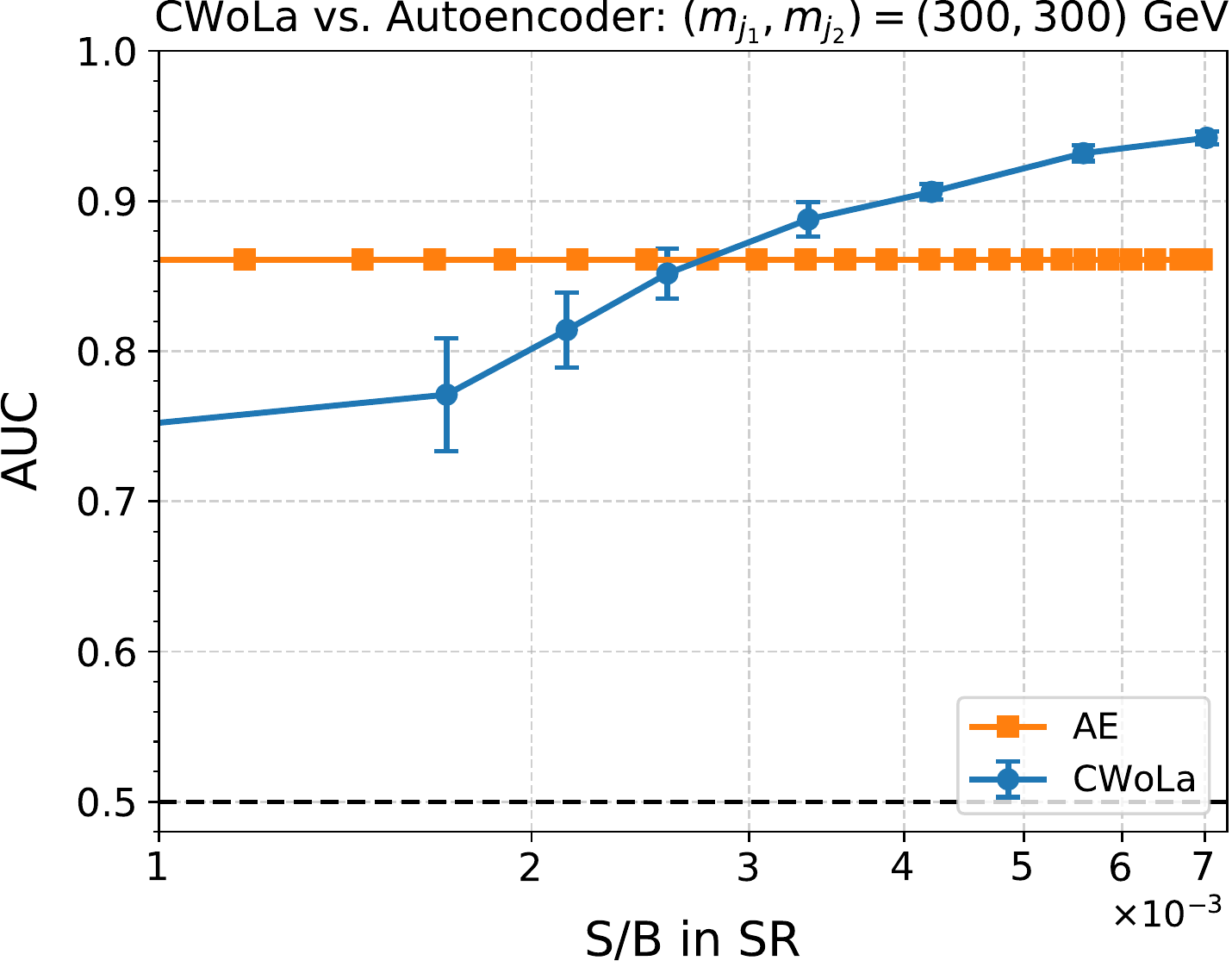}
   \end{center}
   \caption{Performance of CWoLa Hunting (blue) and the AE (orange) as measured by the AUC metric on the signal with $(m_{j_{1}}, m_{j_{2}}) = (500, 500) \; \GeV$ (left plot) and $(m_{j_{1}}, m_{j_{2}}) = (300, 300) \; \GeV$ (right plot). The error bars denote the standard deviation on the AUC metric from statistical uncertainties.}
   \label{fig:CWoLa_vs_AE_AUC}
\end{figure}

\subsection{Supervised metrics}

The performance of CWoLa Hunting and the AE in the signal region for different $S/B$ ratios as measured by the Area Under the Curve (AUC) metric is presented in Fig.~\ref{fig:CWoLa_vs_AE_AUC} for the two signal hypotheses considered in this work. Even though only a small fraction of signal events is used for training, the AUC metric is computed using all the available signal to reduce any potential overfitting. The results in both cases show that CWoLa Hunting achieves excellent discrimination power between signal and background events in the large $S/B$ region, reaching AUC scores above $0.90$ and approaching the $0.98$ score from the fully supervised case. As the number of signal events in the signal region decreases, the amount of information that is available to distinguish the signal and sideband regions in the training phase becomes more limited. As a result, learning the signal features becomes more challenging and performance drops in testing. When the $S/B$ ratio in the signal region is close to zero, the signal and sideband regions become nearly identical and the classifier should not be able to discriminate between both regions. For the benchmark with no signal events, the AUC scores are only $0.43$ and $0.59$ for the signals with larger and smaller jet masses, respectively\footnote{For visualization purpose, this benchmark is not shown in the plot.}. It is interesting to note that, in the absence of signal, the AUC should converge to $0.5$. However, we will see that the presence of background events (from a statistical fluctuation) with a feature distribution that partially overlaps with the one from signal events, located in a region of the phase space with low statistics, allows the classifier to learn some information that turns out to be useful to discriminate between signal and background. Importantly, this does not imply that the information learnt by the classifier will be useful for enhancing the signal excess, as we discuss in detail below. By contrast, the AE performance is solid and stable across the whole $S/B$ range. The reason is that, once the AE learns to reconstruct background events, its performance is independent of the number of signal events used for training as long as the contamination ratio is not too large. Interestingly, the AUC curves from CWoLa Hunting and the AE cross at $S/B \sim 3 \cdot 10^{-3}$.

The most standard way of measuring the performance of a given model is through the Receiver Operating Characteristic (ROC) curve, and the area under this curve, the AUC metric. These two metrics are useful to compare the overall performance of different models in many classification tasks. However, the goal of a resonant anomaly detection search is to find a localized signal over a large background. For this purpose, the most important variables to consider are the signal-to-background ratio ($S/B$) and the naive expected significance ($S/\sqrt B$). With this in mind, we will consider the Significance Improvement Characteristic (SIC) \cite{Gallicchio:2010dq} to measure the performance of CWoLa Hunting and the AE at enhancing the significance of the signal excess. The SIC metric measures the significance improvement after applying a cut in the classifier output. In particular, any given cut will keep a fraction $\epsilon_{S}$ of signal events and a fraction $\epsilon_{B}$ of background events, which are defined as the signal and background efficiencies of the cut. The significance improvement for this cut is thus given by $\text{SIC} = \epsilon_{S}/\sqrt{\epsilon_{B}}$.

\begin{figure}[t!]
   \begin{center}
        \includegraphics[scale=0.48]{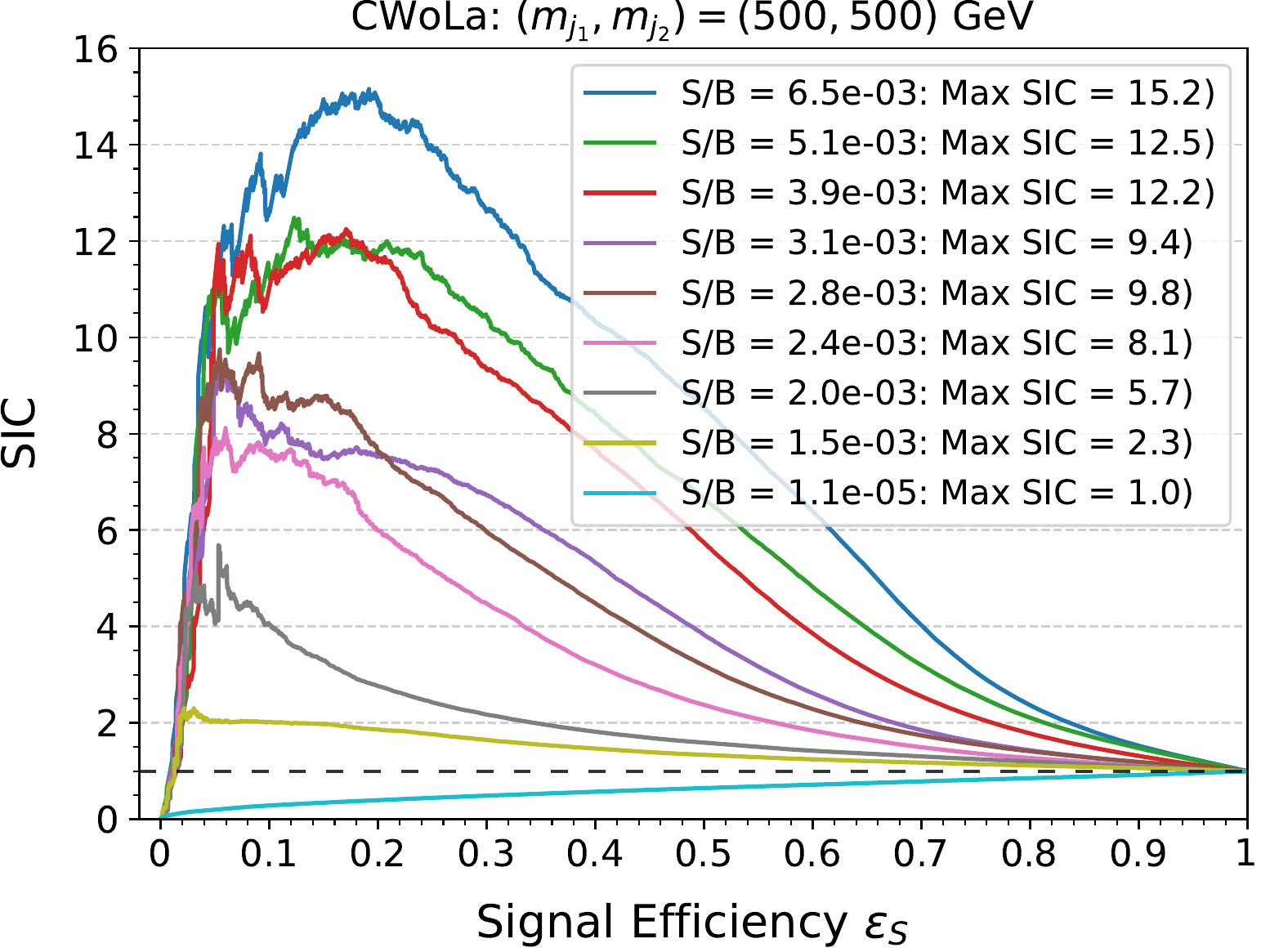}
        \hspace{3pt}
        \includegraphics[scale=0.48]{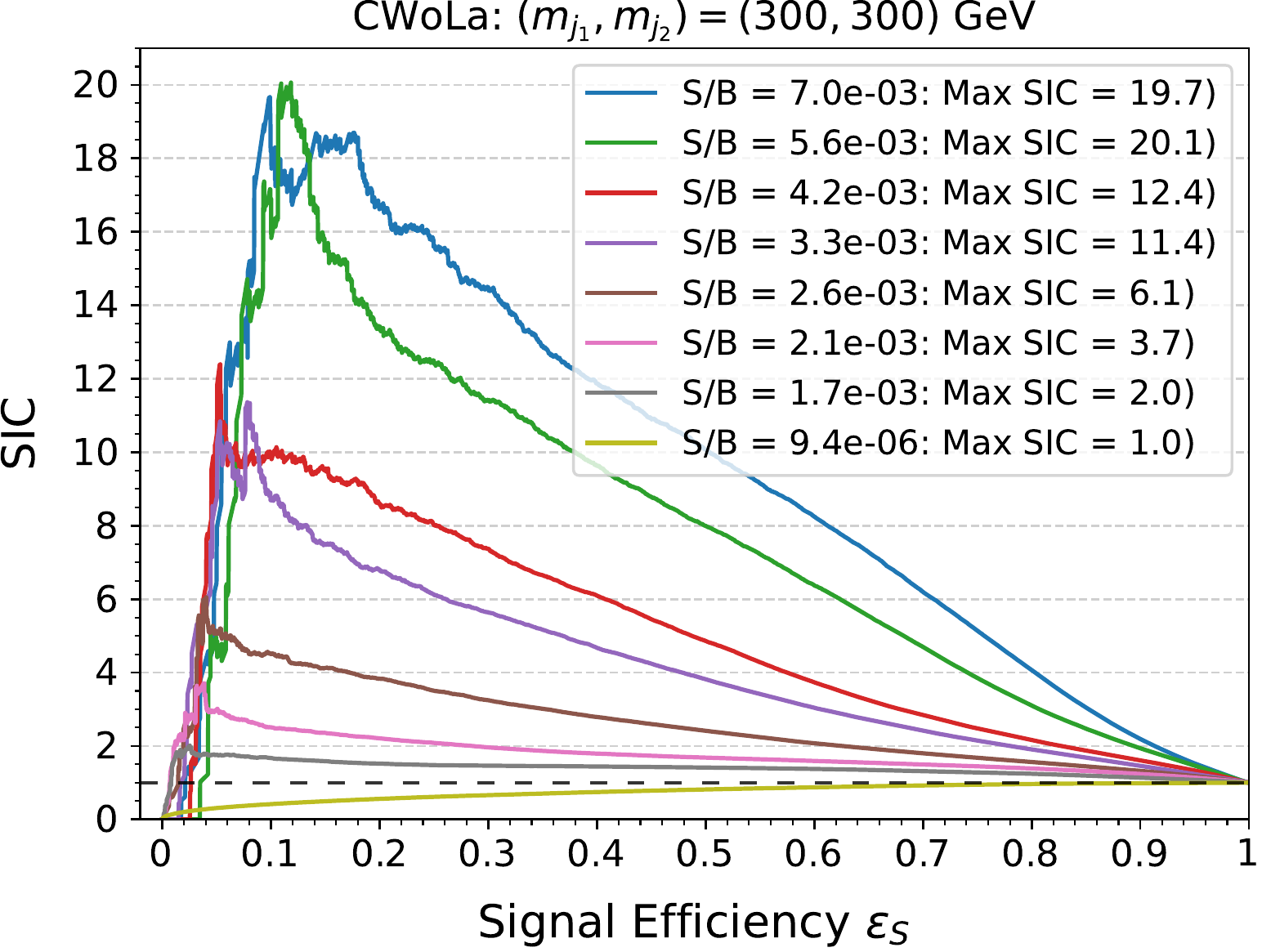} \\
        \vspace{10pt}
        \hspace{1.5pt}
        \includegraphics[scale=0.48]{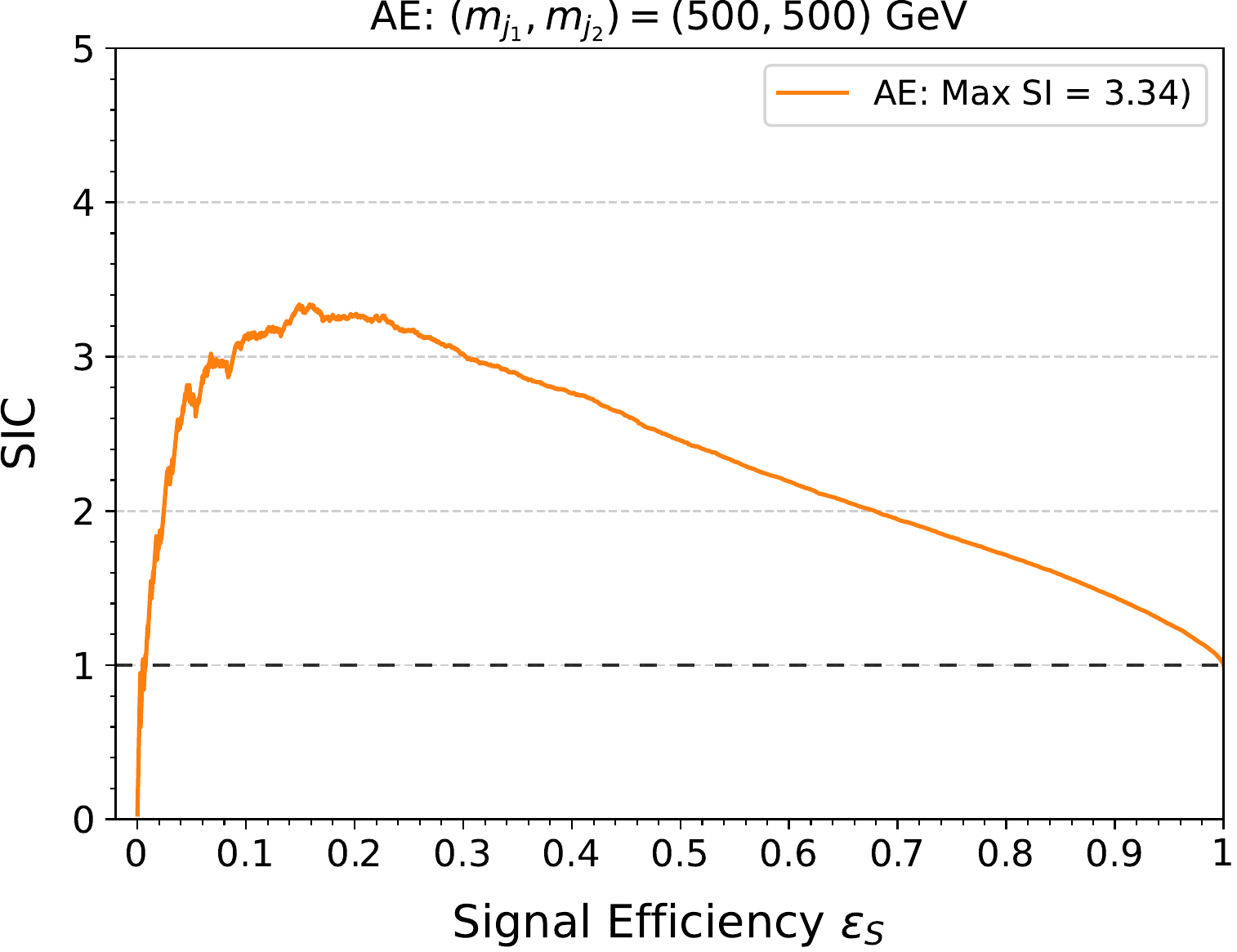}
        \hspace{6.9pt}
        \includegraphics[scale=0.48]{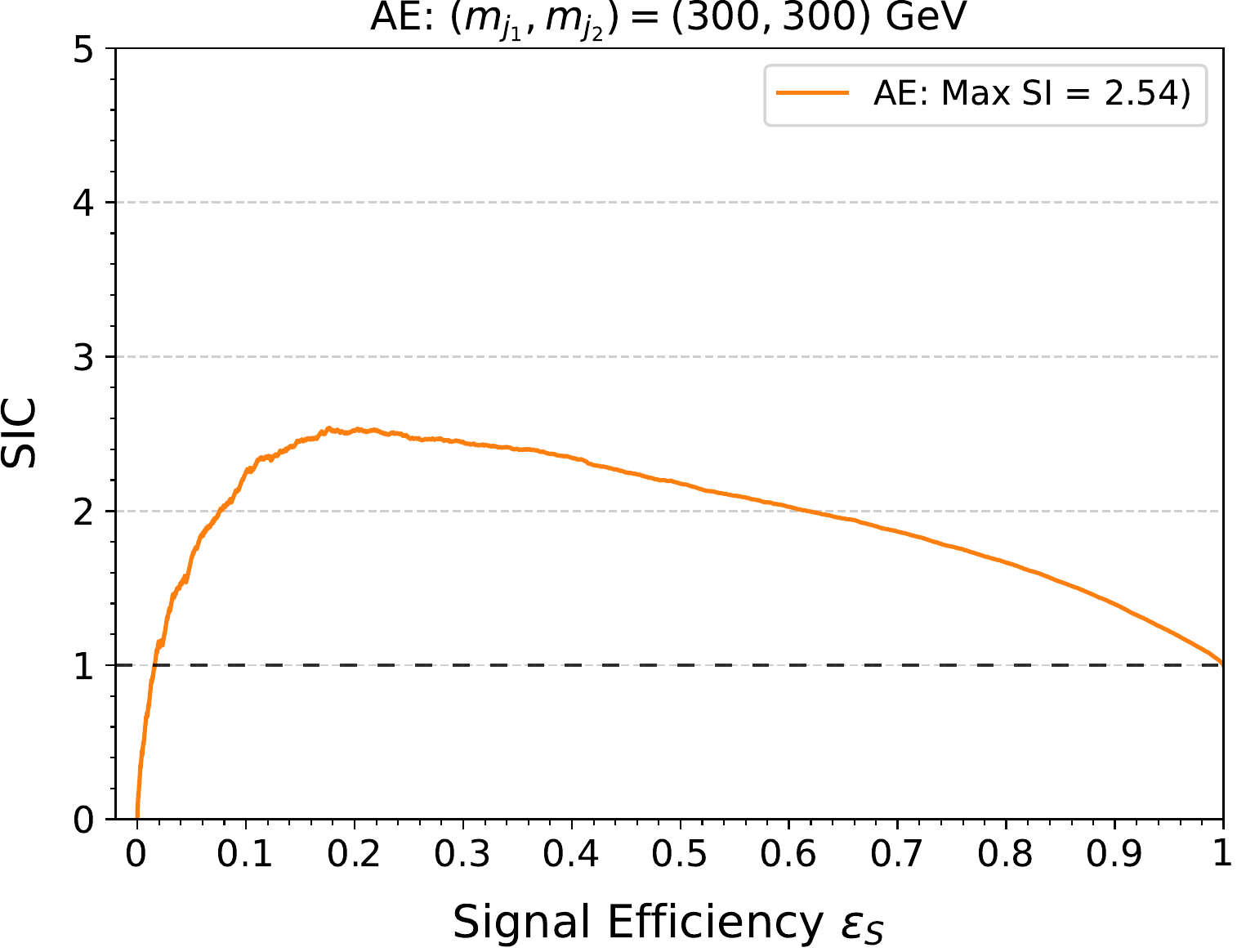}
   \end{center}
   \caption{The SIC curves for CWoLa Hunting (top row) and the AE (bottom row) are shown for the signals with $(m_{j_{1}}, m_{j_{2}}) = (500, 500) \; \GeV$ and $(m_{j_{1}}, m_{j_{2}}) = (300, 300) \; \GeV$ in the left and right plots, respectively. For CWoLa Hunting, a SIC curve is shown for each of the classifiers that were trained on mixed samples with different amounts of injected signal.}
   \label{fig:CWoLa_and_AE_SIC_curves}
\end{figure}

\begin{figure}[t!]
   \begin{center}
        \includegraphics[scale=0.48]{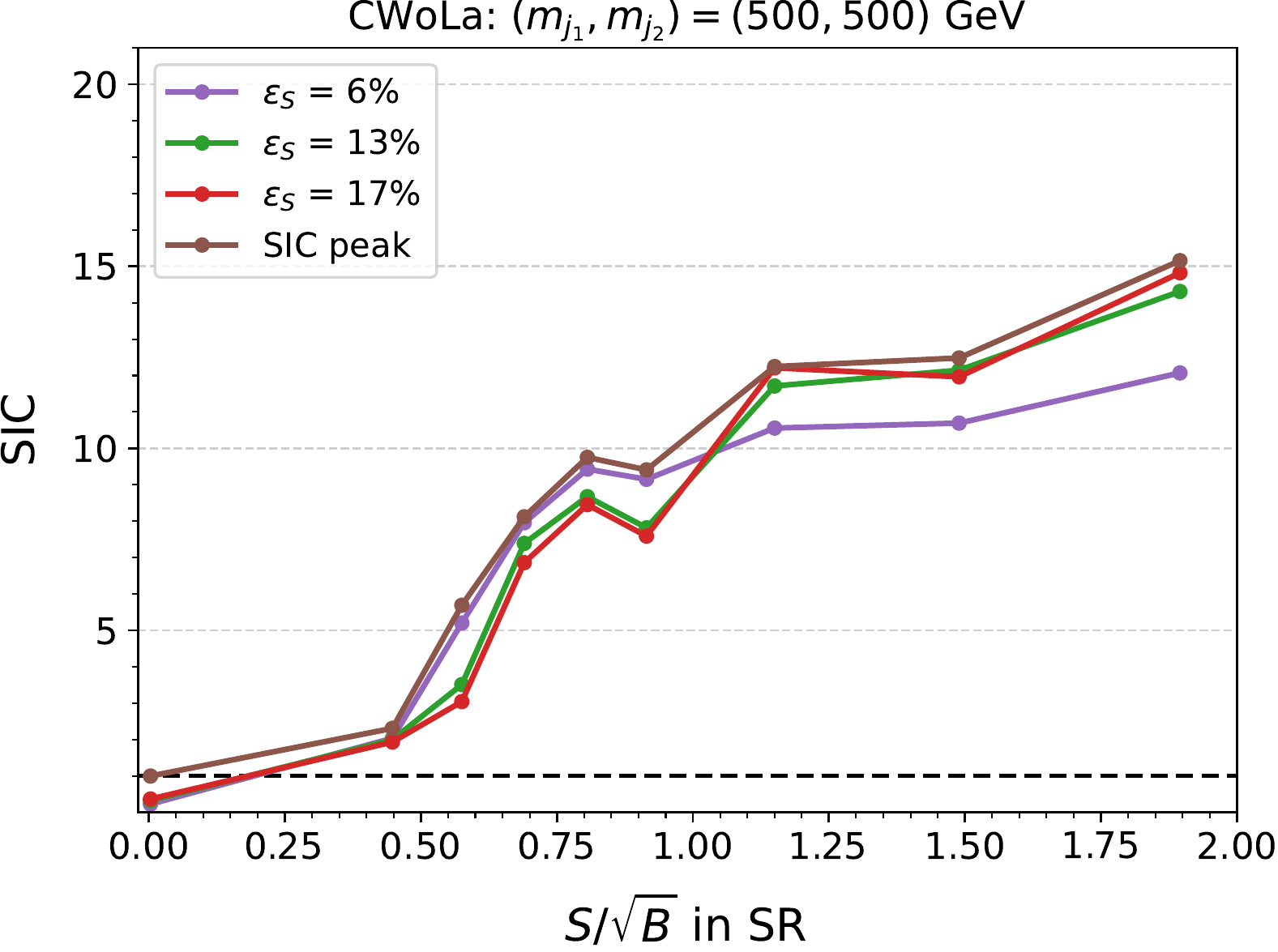}
        \hspace{3pt}
        \includegraphics[scale=0.48]{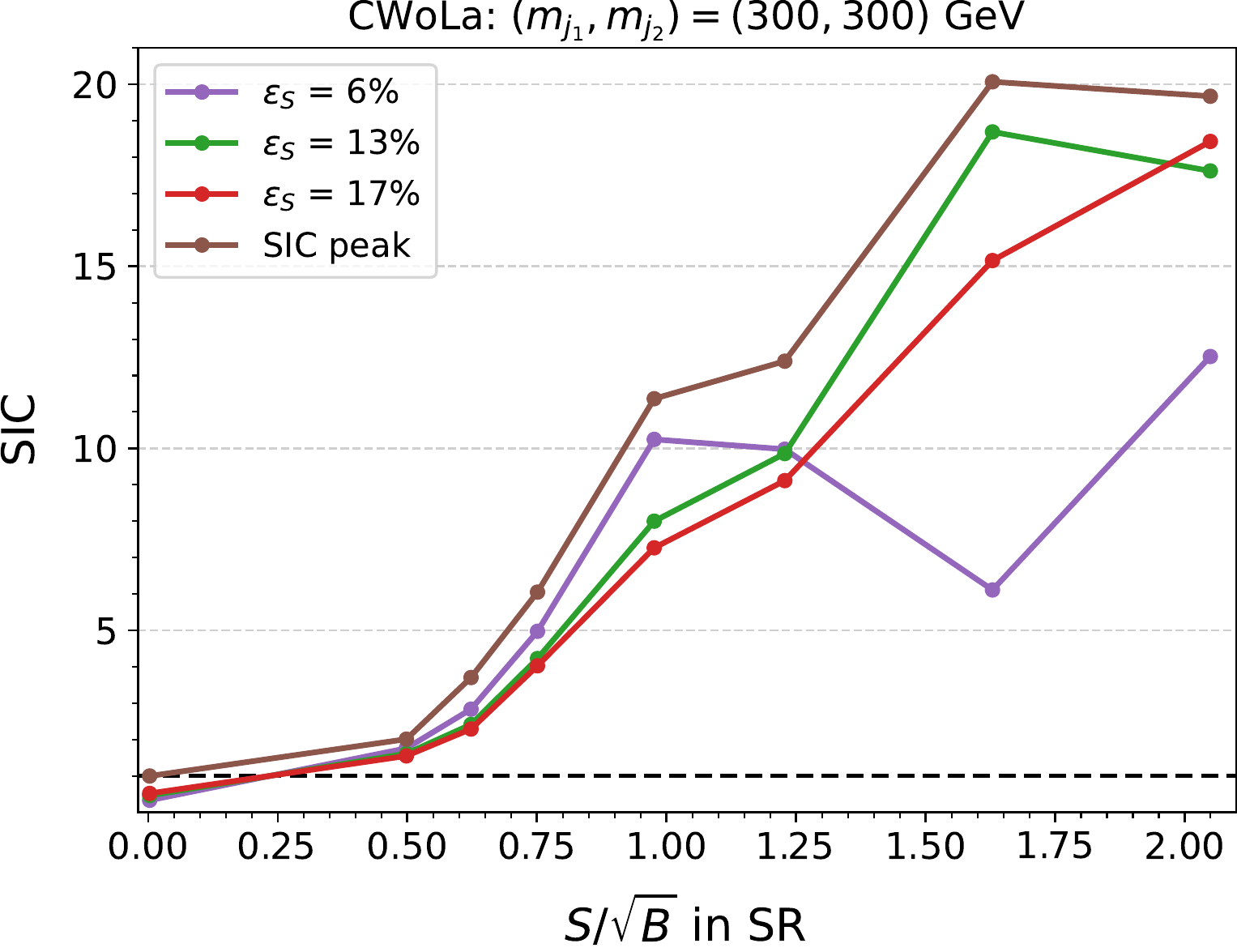} \\
        \vspace{10pt}
        \includegraphics[scale=0.48]{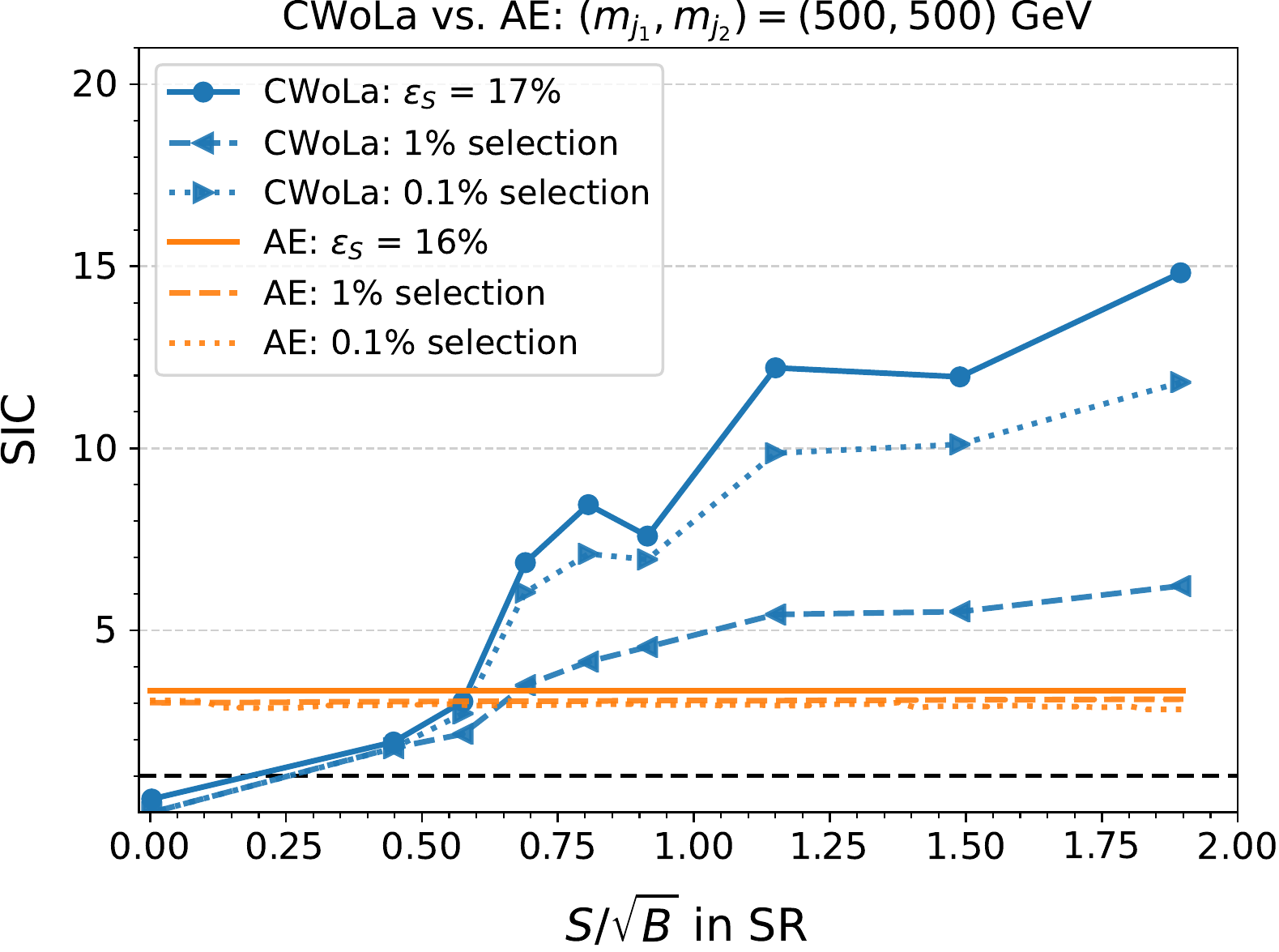}
        \hspace{3pt}
        \includegraphics[scale=0.48]{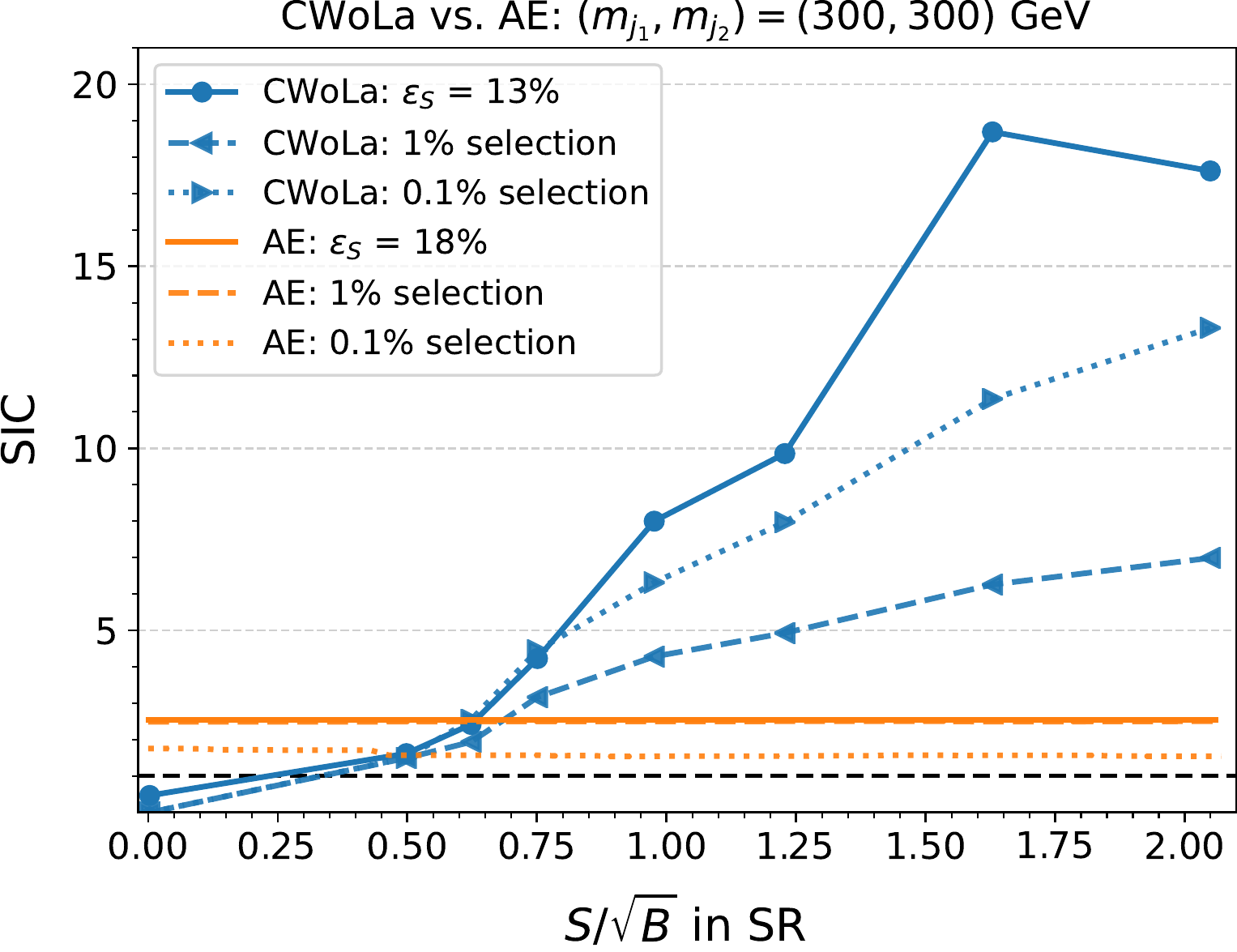}
   \end{center}
   \caption{\textbf{Top row:} The SIC value as a function of $S/\sqrt{B}$ for a set of fixed signal efficiencies is shown for the signals with $(m_{j_{1}}, m_{j_{2}}) = (500, 500) \; \GeV$ (left plot) and $(m_{j_{1}}, m_{j_{2}}) = (300, 300) \; \GeV$ (right plot). The $\epsilon_{S} = 17 \, \%$ and $\epsilon_{S} = 13 \, \%$ signal efficiencies, respectively, maximize the overall significance improvement for all $S/B$ benchmarks. \textbf{Bottom row:} The signal efficiencies are chosen such that the SIC values are maximized for CWoLa Hunting and the AE. The SIC values associated to the $1 \, \%$ and $0.1 \, \%$ are also shown for comparison. These values are calculated using only the fraction of signal that defines each $S/B$ benchmark.}
   \label{fig:CWoLa_and_AE_SIC_values}
\end{figure}

In order to find the localized signal over the large background, which we presented in Fig.~\ref{fig:mJJ_distribution}, we will use the SIC metric to find the optimal cut in the classifiers output that leads to the maximal enhancement in $S/\sqrt B$ in the signal region. The SIC curves for CWoLa Hunting and the AE are shown in Fig.~\ref{fig:CWoLa_and_AE_SIC_curves}. The SIC curves are calculated using all the available signal and background events in the signal region. For CWoLa Hunting, the results show that the shape and the location of the peak of the SIC curve depend on the amount of injected signal used during training. In order to find the signal efficiency that leads to a maximal overall significance improvement for all $S/B$ benchmarks, we analyze how the SIC value changes as a function of $S/\sqrt{B}$ for a set of fixed signal efficiencies in the top row of Fig.~\ref{fig:CWoLa_and_AE_SIC_values}. We find that the signal efficiencies that yield the maximum overall significance improvement for CWoLa Hunting are $\epsilon_{S} = 17 \, \%$ and $\epsilon_{S} = 13 \, \%$ for the high and low jet mass signals, respectively. For the AE, the optimal signal efficiencies are $\epsilon_{S} = 16 \, \%$ and $\epsilon_{S} = 18 \, \%$, respectively. Now we will use these optimal signal efficiencies to set an anomaly score threshold that maximizes the significant improvement in the signal region for each model.  In practice, model independence would prevent picking a particular value and so we will later compare these optimized values with fixed values at round logarithmically spaced efficiencies.

\subsection{Sideband fit and $p$-values}

After evaluating the quality of the two methods at identifying the signal events among the background, we compare how they perform at increasing the significance of the signal region excess. For this purpose, we performed a parametrized fit to the $m_{JJ}$ distribution in the sideband region.
We then interpolate the fitted background distribution into the signal region and evaluate the $p$-value of the signal region excess. 

For the CWoLa method, we used the following $4$-parameter function to fit the background:
\begin{equation}
\frac{d\sigma}{dm_{JJ}} = \frac{p_{0}(1-x)^{p_{1}}}{x^{p_{2}+p_{3}\ln(x)}} \, ,
\label{eq:fit}
\end{equation}
where $x = m_{JJ}/\sqrt{s}$. We use the previous function to estimate the background in the range $m_{JJ} \in [2800, 5200] \; \GeV$. This function has been previously used by both ATLAS \cite{Aad:2019hjw} and CMS \cite{Sirunyan:2018xlo} collaborations in hadronic heavy resonance searches.

For the AE, we find that this function does not fit well the distribution of surviving events on $m_{JJ}$ after applying a cut on the reconstruction error. Instead, we found that a simple linear fit (on a narrower sideband region) is able to describe the background distribution on the sideband with good accuracy and it is sensitive to an excess on the signal region for the cuts that we considered. For the cut based on the SIC curve and the $1 \%$ cut, the fit is implemented on the range $m_{JJ} \in [3000, 4000] \; \GeV$. For the $0.1 \%$ cut, we need to extend this range to $m_{JJ} \in [2800, 4400] \; \GeV$.  This range extension produces a better fit $\chi^2$ in the sideband and mitigates a small bias in the predicted signal at $S=0$.

The validity of sideband interpolation relies on the assumption that the $m_{JJ}$ distribution for background events surviving a cut can still be well modelled by the chosen functional forms. This is likely to be the case so long as the selection efficiency of the tagger on background events is smooth and monotonic in $m_{JJ}$, and most simply\footnote{Complete decorrelation is sufficient, but not necessary to prevent bump-sculpting~\cite{2010.09745}.} if it is constant in $m_{JJ}$ (which would require signal features uncorrelated with $m_{JJ}$).

\begin{figure}[t!]
   \begin{center}
        \includegraphics[scale=0.59]{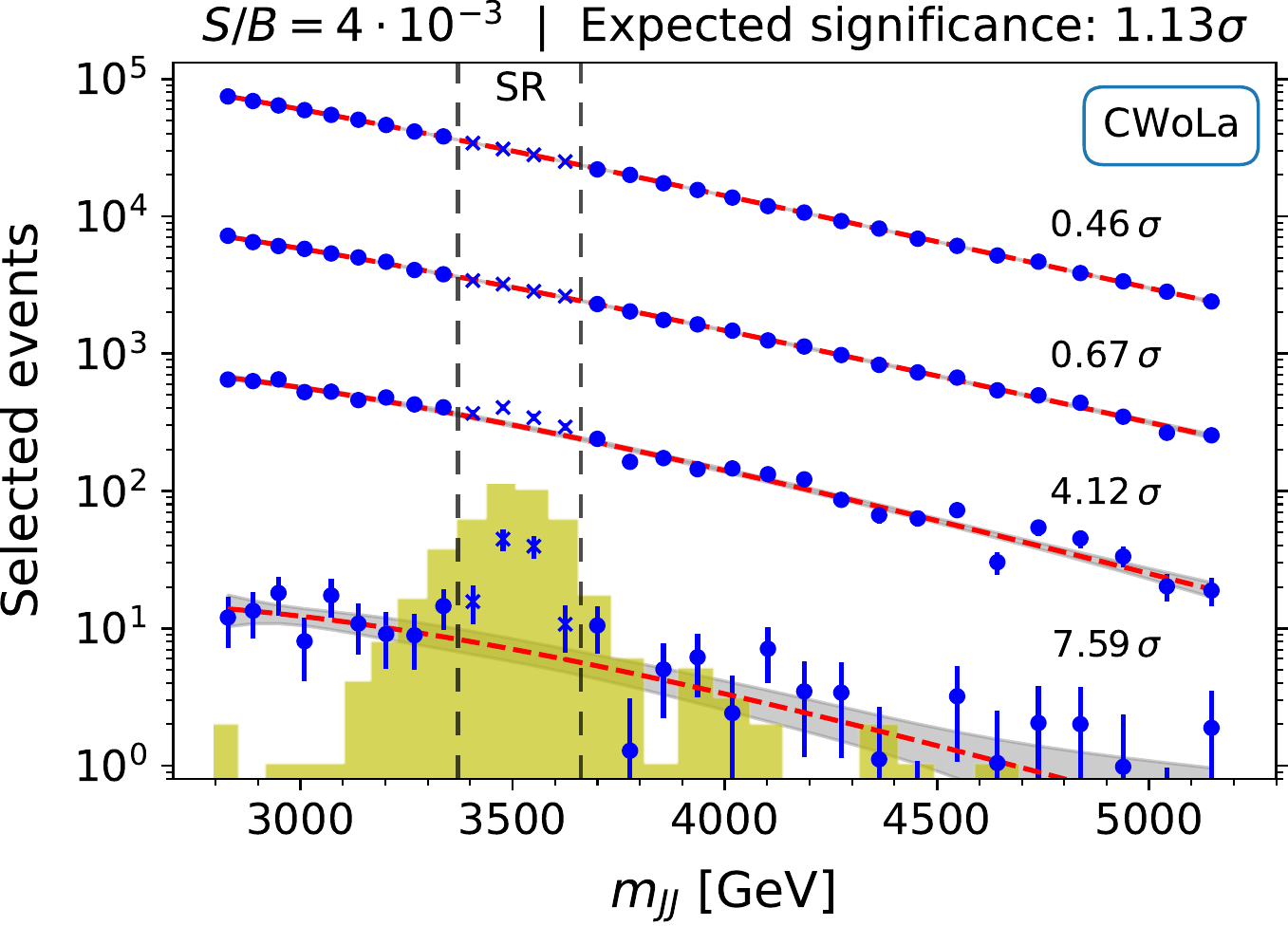}
        \hspace{0pt}
        \includegraphics[scale=0.59]{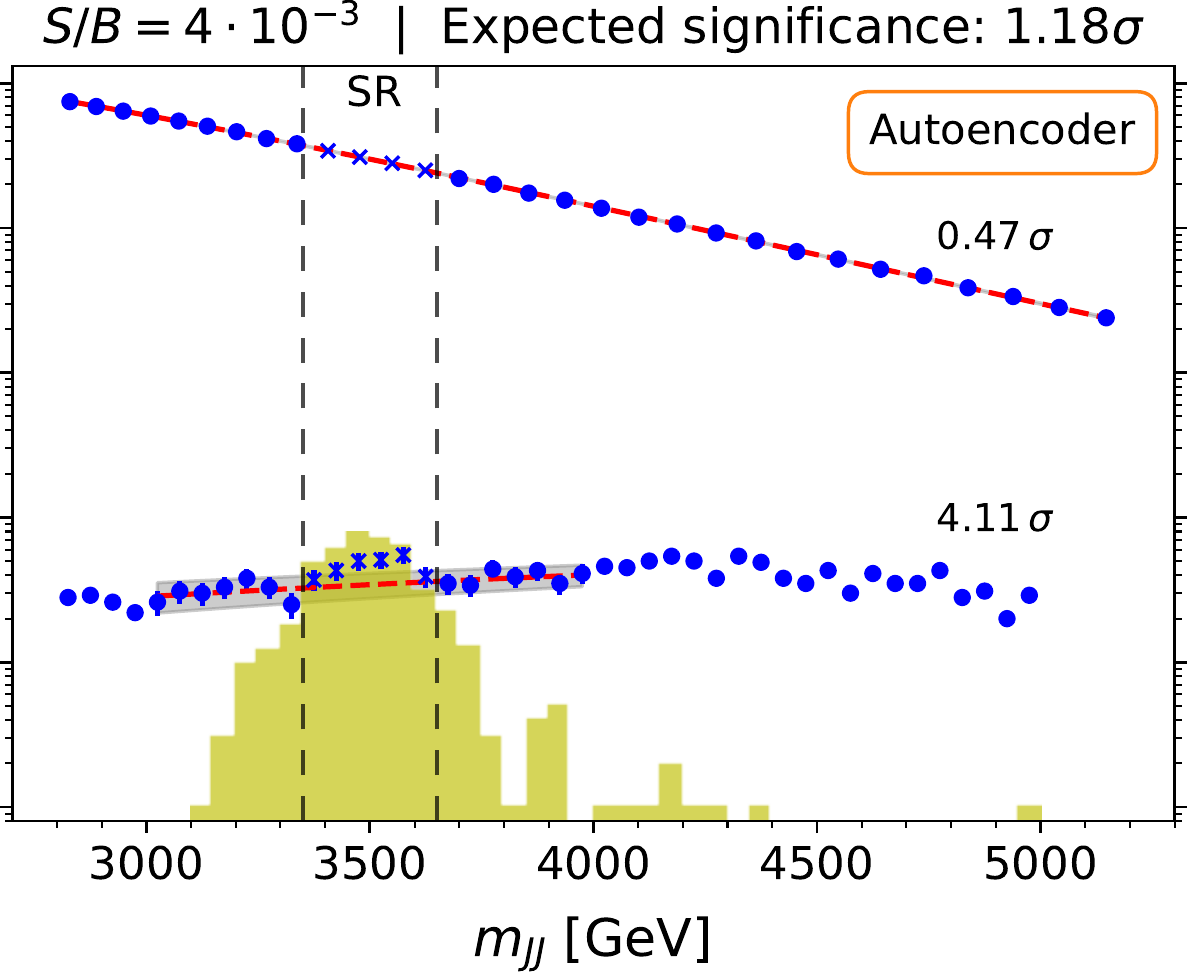} \\
        \vspace{10pt}
        \includegraphics[scale=0.59]{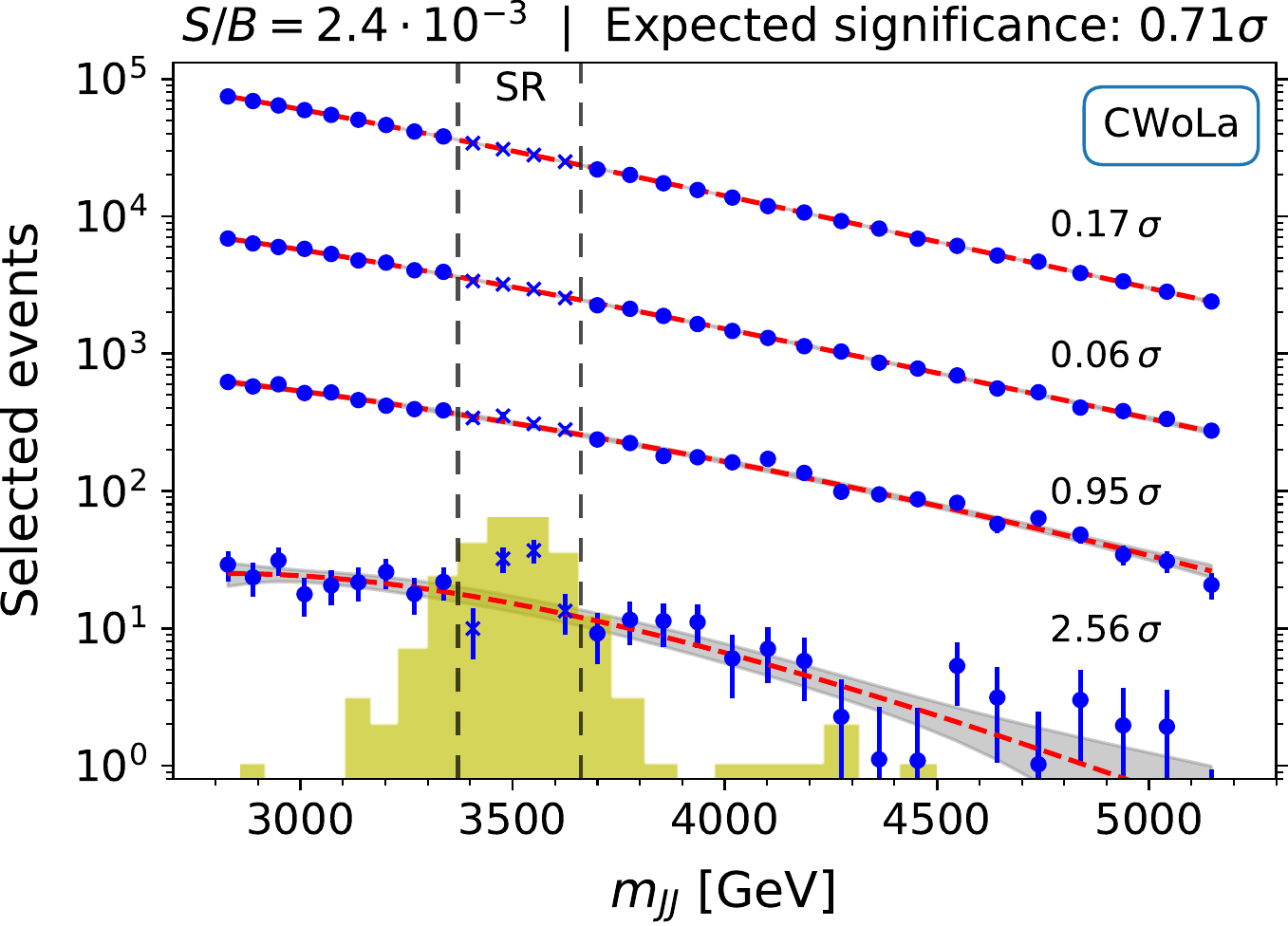}
        \hspace{0pt}
        \includegraphics[scale=0.59]{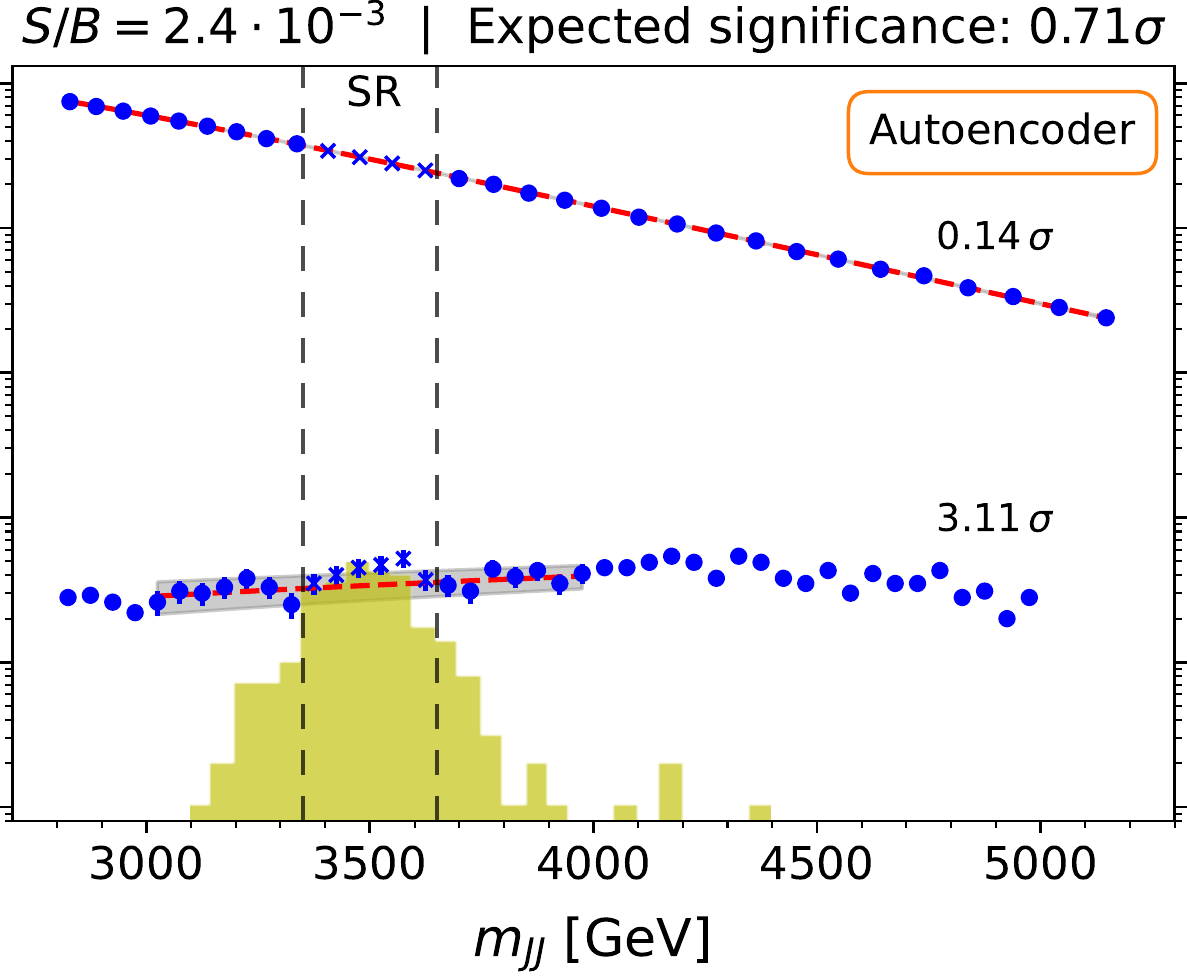}
   \end{center}
   \caption{Significance of the signal region excess after applying different cuts using the classifier output for CWoLa Hunting (left plots) and the AE (right plots), for one of the runs corresponding to the benchmarks with $S/B \simeq 4 \cdot 10^{-3}$ (top row) and $S/B \simeq 2.4 \cdot 10^{-3}$ (bottom row) on the signal with $(m_{j_{1}}, m_{j_{2}}) = (500, 500) \; \GeV$. For CWoLa, we show the $100 \, \%, 10 \, \%, 1 \, \%, 0.04 \, \%$ most signal-like events. For the AE, we show the $100 \, \%$ and $0.6 \, \%$ event selections. In both cases, the smallest cut corresponds to the optimal cut according to the SIC curve. The blue crosses denote the event selection in each signal region bin, while the blue circles represent the event selection in each bin outside of the signal region. The dashed red lines indicate the fit to the events outside of the signal region, the grey band indicates the fit uncertainty and the injected signal is represented by the green histogram.}
   \label{fig:CWoLa_vs_AE_fits}
\end{figure}

In Fig.~\ref{fig:CWoLa_vs_AE_fits}, we show the fit results for CWoLa Hunting and the AE for one of the runs corresponding to the benchmarks with $S/B \simeq 4 \cdot 10^{-3}$ and $S/B \simeq 2.4 \cdot 10^{-3}$ on the signal with $(m_{j_{1}}, m_{j_{2}}) = (500, 500) \; \GeV$. After applying different cuts using the classifiers outputs, the significance of signal region bump is significantly increased. For the benchmark with more injected signal, CWoLa Hunting yields a substantial significance increase of up to $7.6 \sigma$, while the AE is able to increase the bump significance by up to $4.1 \sigma$. When the amount of injected signal is reduced, the results show that CWoLa Hunting becomes weaker and it rises the excess significance up to only $2.6 \sigma$. However, in this case the AE performs better than CWoLa Hunting, increasing the bump significance up to $3.1 \sigma$. This is an important finding because it suggests that CWoLa Hunting and the AE may be complementary techniques depending on the cross section. Note that the event distribution from the AE is clearly shaped due to some correlations between the input features and $m_{JJ}$. In particular, since the jet $p_{T}$ is very correlated with $m_{JJ}$. However, the average jet $p_T$ scales monotonically (and roughly linearly) with $m_{JJ}$, which means that no artificial bumps are created and the distribution post-selection is still well modelled by the chosen fit function. Finally, note that the fit to the raw distribution (i.e. no cut applied) is lower than the naive expected significance $S/\sqrt B$ due to a downward fluctuation in the number of background events in the signal region, as discussed in Appendix~\ref{sec:fit}.

\begin{figure}[t!]
   \begin{center}
        \includegraphics[scale=0.46]{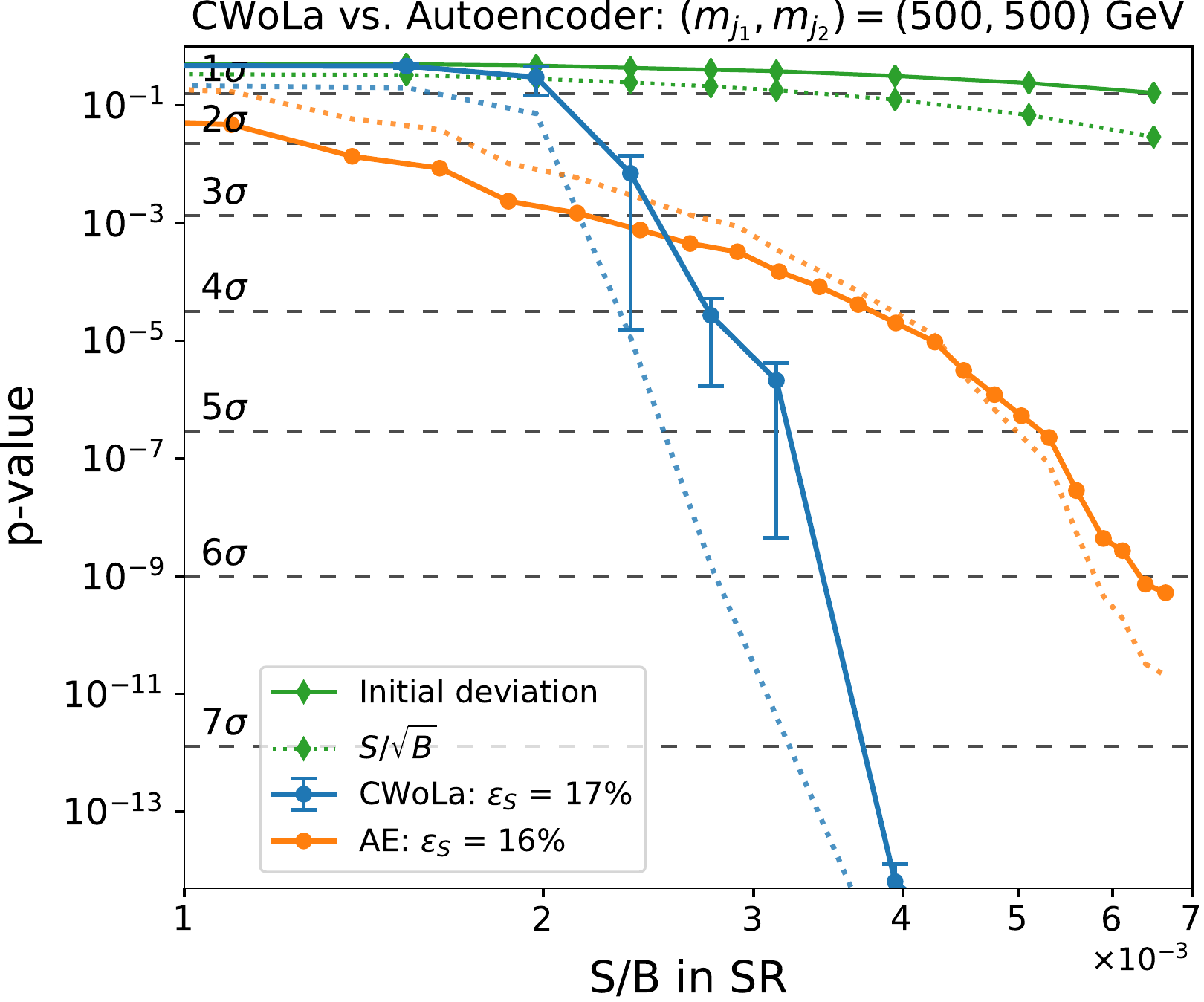}
        \hspace{3pt}
        \includegraphics[scale=0.46]{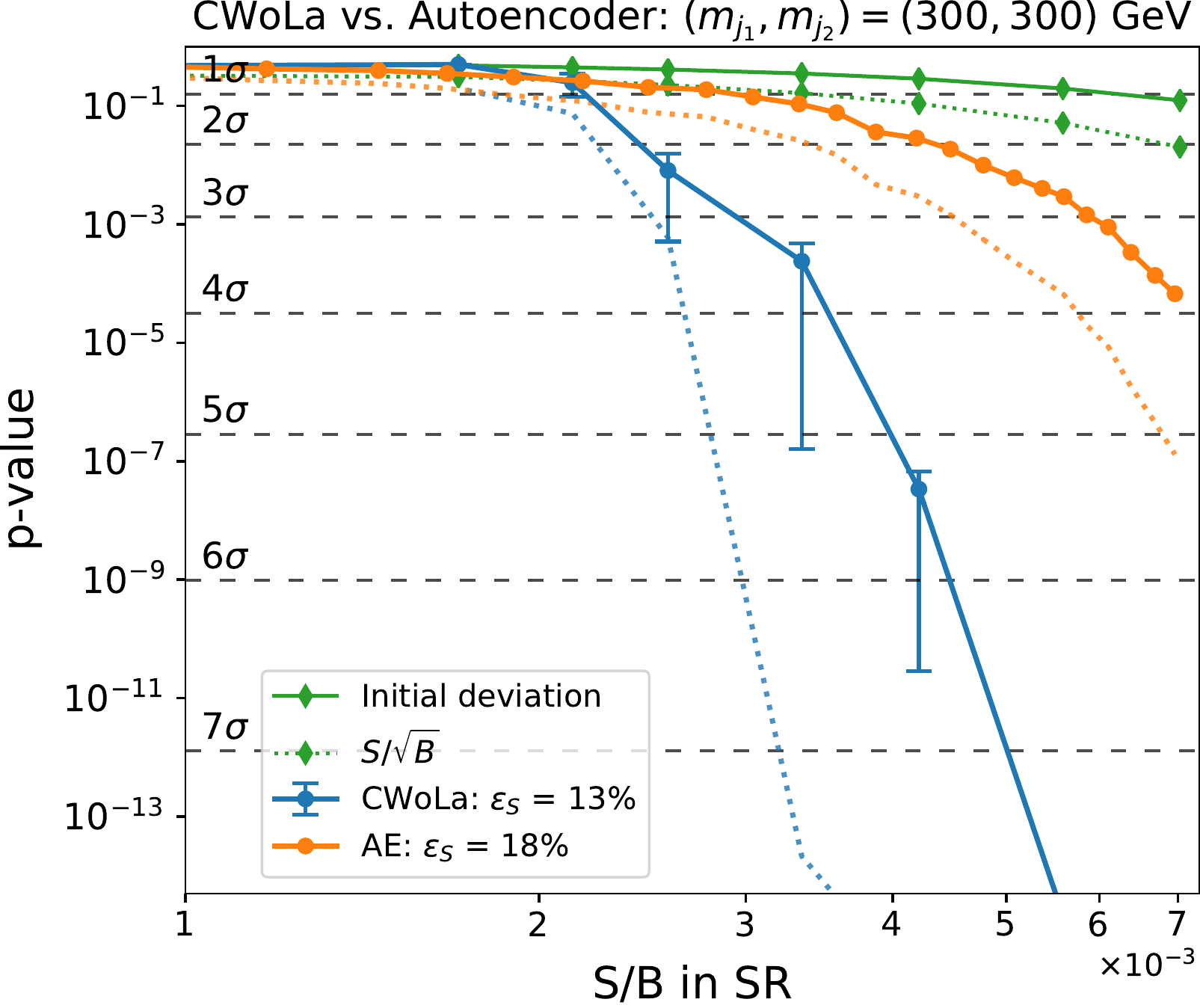} \\
        \vspace{10pt}
        \includegraphics[scale=0.46]{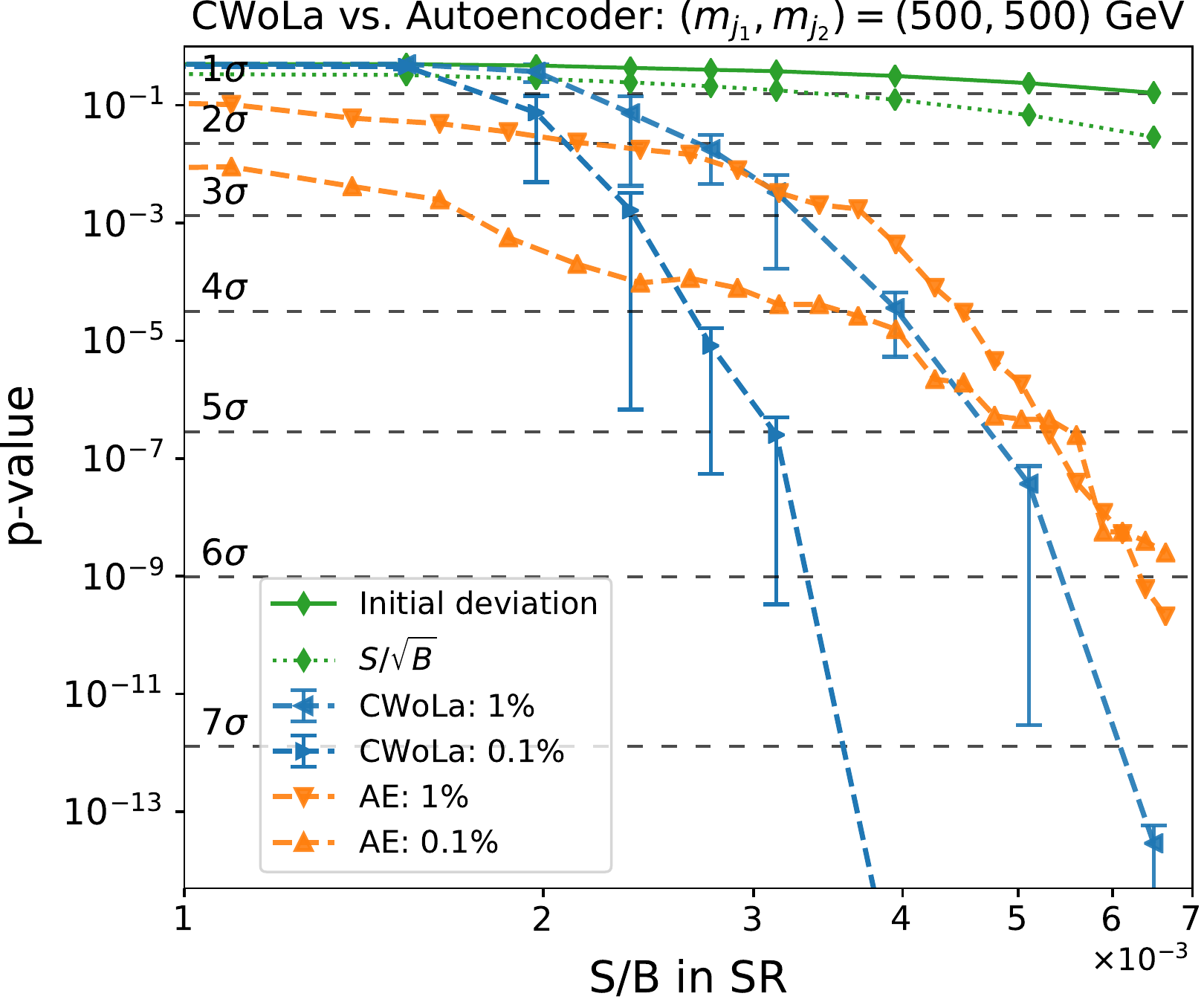}
        \hspace{3pt}
        \includegraphics[scale=0.46]{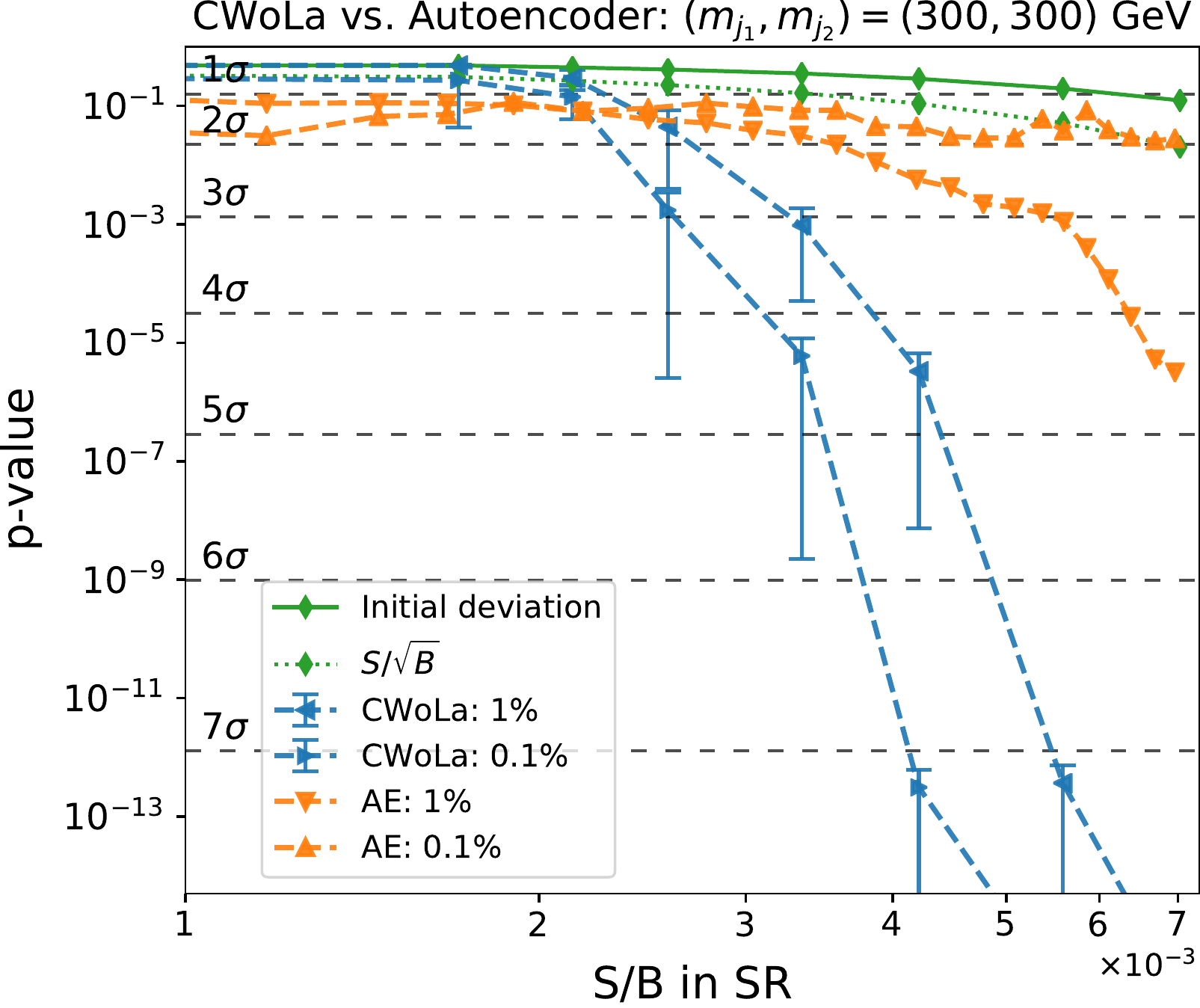}
   \end{center}
   \caption{The significance of the signal region excess after applying different cuts for CWoLa Hunting and the AE, for the signals with $(m_{j_{1}}, m_{j_{2}}) = (500, 500) \; \GeV$ and $(m_{j_{1}}, m_{j_{2}}) = (300, 300) \; \GeV$, is shown in the left and right plots, respectively. The plots in the top row show the cuts that maximize the overall significance improvement for all benchmarks according to the SIC curve, while the bottom row plots show results for fixed, predetermined cuts. The best cuts for CWoLa Hunting (blue) correspond to the $17 \%$ and $13 \%$ signal efficiencies for the signals with high and low jet masses, respectively. For the AE (orange), the best cuts correspond to the $16 \%$ and $18 \%$ signal efficiencies, respectively. The dotted lines denote the naive expected significance, $S/\sqrt B$. The round cuts from the bottom plots show the $1 \%$ and $0.1 \%$ event selections for CWoLa Hunting and the AE. The initial significance of the bump ($100 \, \%$ selection) is shown in green.
   }
   \label{fig:CWoLa_vs_AE_pvalues}
\end{figure}

In order to systematically study if CWoLa Hunting and the AE could be complementary techniques depending on the cross section, we analyze their performance at increasing the significance of the signal region excess for different $S/B$ benchmarks and the two signal hypotheses in Fig.~\ref{fig:CWoLa_vs_AE_pvalues}. The top two plots show the cuts on the classifier output that lead to the largest overall significance improvement according to the SIC curve. For CWoLa Hunting, we show the median $p$-values from the ten independent runs for every benchmark corresponding to the $17 \, \%$ (top left) and $13 \, \%$ (top right) signal efficiencies, which correspond to fractions of signal-like events between $0.04 \, \%$ and $1.7 \, \%$ depending on the benchmark. The error bars represent the Median Absolute Deviation. Note that the fit result does not always agree with the naive expected significance, $S/\sqrt B$, due to the high uncertainties among the ten independent classifiers and the small fractions of events considered in some cases. For the AE, we show the $p$-values associated to the $16 \, \%$ (top left) and $18 \, \%$ (top right) signal efficiencies, which correspond to the $0.36 \, \%$ and $0.63 \, \%$ most signal-like events, respectively. 

Importantly, there are other cuts that enhance the significance of the signal region excess, as shown in the bottom plots of Fig.~\ref{fig:CWoLa_vs_AE_pvalues}. In a real experimental search, with no previous knowledge about any potential new physics signal, the two models would be able to find the signal for fixed round cuts of $1 \%$ and $0.1 \%$. For the AE, these cuts are applied in the signal region to derive an anomaly score above which all the events in the full $m_{JJ}$ range are selected. However, note that for these cuts the AE seems to sculpt small bumps on the $m_{JJ}$ distribution even when no signal is present on data. We find that the excess significance at $S/B = 0$ is $0.89 \sigma$, $0.56 \sigma$ and $1.06 \sigma$ for the SIC-based, $1 \%$ and $0.1 \%$ cuts, respectively. We checked that this is caused by the shaping of the $m_{JJ}$ distribution and the small statistical fluctuations that appear for such tight cuts. We remark that this effect is not produced by the signal.

The statistical analysis demonstrates two things. First, CWoLa Hunting is able to increase the significance of the signal region excess up to $3 \sigma - 8 \sigma$ for $S/B$ ratios above $\sim 3 \cdot 10^{-3}$ for both signal hypotheses, even when the original fit shows no deviation from the background-only hypothesis. By contrast, the AE shows a superior performance below this range for the signal with $(m_{j_{1}}, m_{j_{2}}) = (500, 500) \; \GeV$, boosting the significance of the excess up to $2 \sigma - 3 \sigma$ in the low $S/B$ region where CWoLa Hunting is not sensitive to the signal. Importantly, there is again a crossing point in the performance of the two methods as measured by their ability to increase the significance of the excess. Therefore, our results show that the two methods are complementary for less-than-supervised anomaly detection.  Second, it is clear that the AE is not able to increase the bump excess for the signal with $(m_{j_{1}}, m_{j_{2}}) = (300, 300) \; \GeV$ below $S/B \sim 3 \cdot 10^{-3}$, even when it reaches a fairly solid AUC score, as shown in Fig.~\ref{fig:CWoLa_vs_AE_AUC}. This means that even though the AE is able to classify a sizeable fraction of signal events correctly, there is a significant fraction of background events that yield a larger reconstruction error than the signal events. In other words, the AE does not consider the signal events as sufficiently anomalous and finds more difficult to reconstruct part of the background instead. Therefore, cutting on the reconstruction error does not result in a larger fraction of signal in the selected events. By construction, this is the main limitation of the AE: it focuses its attention in anything that seems anomalous, whether it is an exciting new physics signal or something that we consider less exotic.

Finally, it is important to analyze the performance of CWoLa Hunting and the AE when training on no signal. For consistency, both models should not sculpt any bumps on the $m_{JJ}$ distribution when no signal is present on data. For CWoLa Hunting, the AUC scores for the benchmark with $S/B = 0$ are $0.43$ and $0.59$ for the signal hypotheses with larger and smaller jet masses, respectively. These numbers are slightly different to the expected value of $0.5$ due to the presence in the background of a real, low-significance statistical excess in the high mass region of phase space in the signal region, which we have checked does not appear in repeated background simulations. Indeed, even after this selection the signal region shows an overall deficit, leading to overall significance of 0.

\begin{figure}[t!]
   \begin{center}
        \includegraphics[scale=0.335]{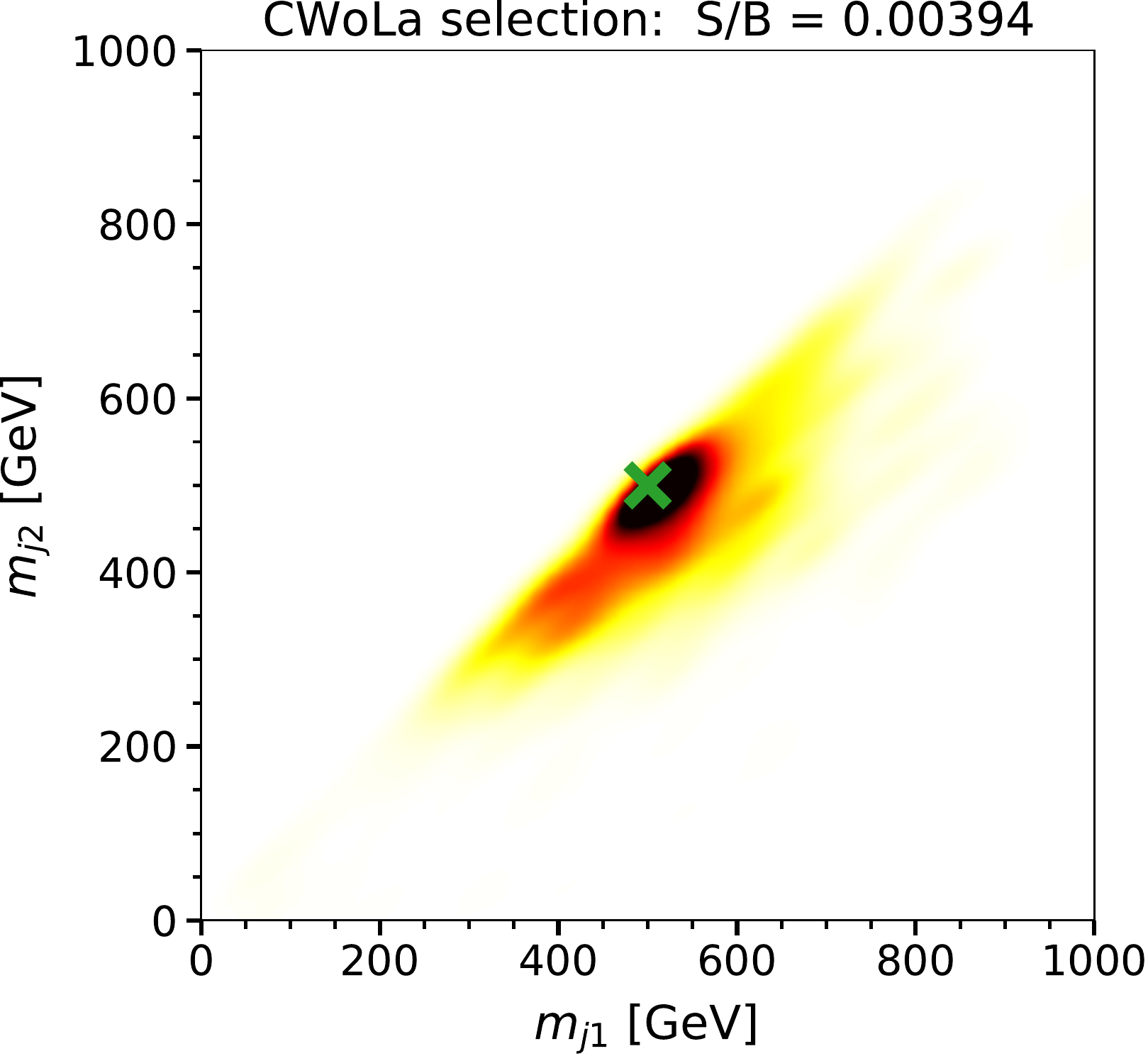}
        \includegraphics[scale=0.335]{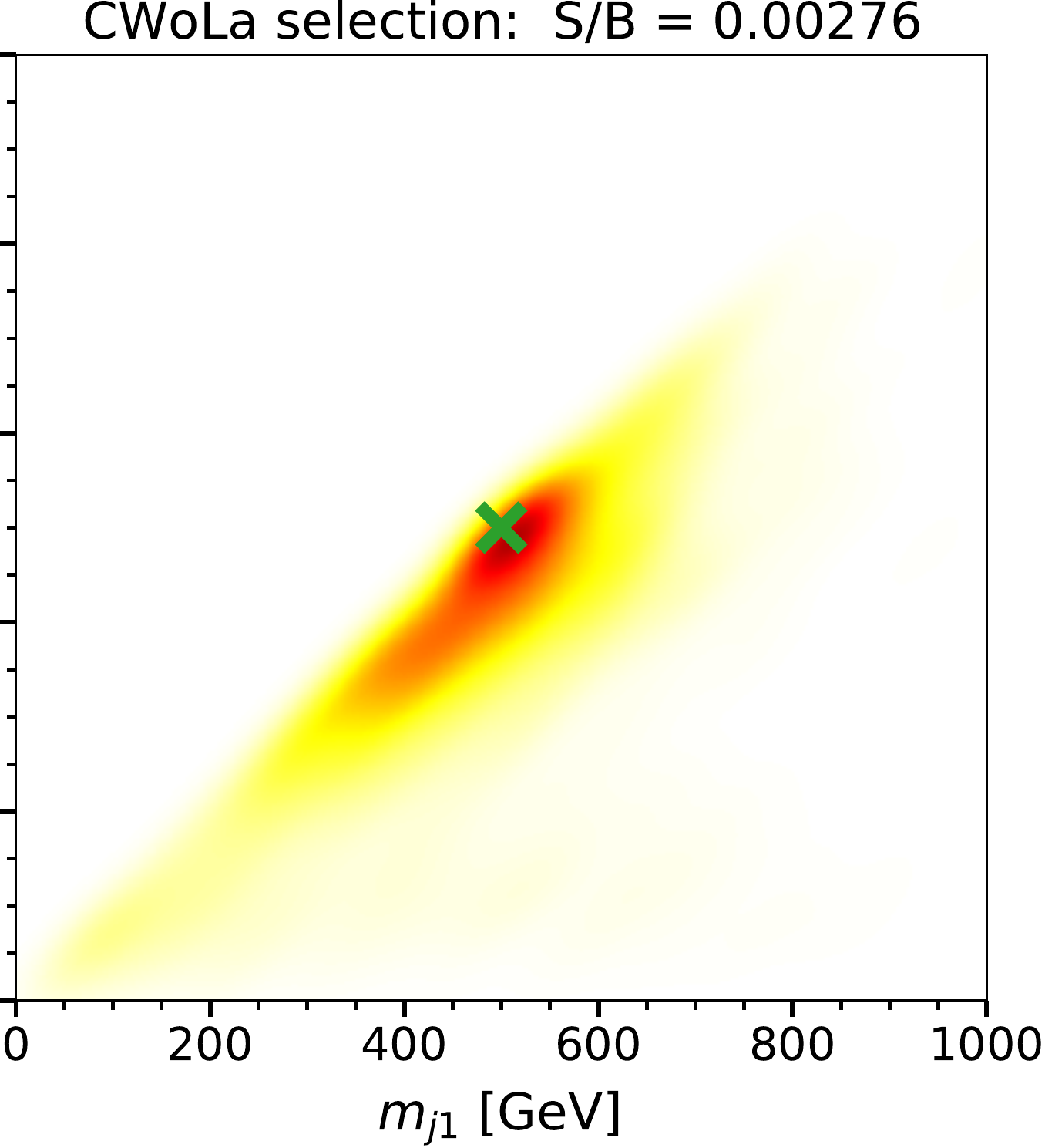}
        \includegraphics[scale=0.335]{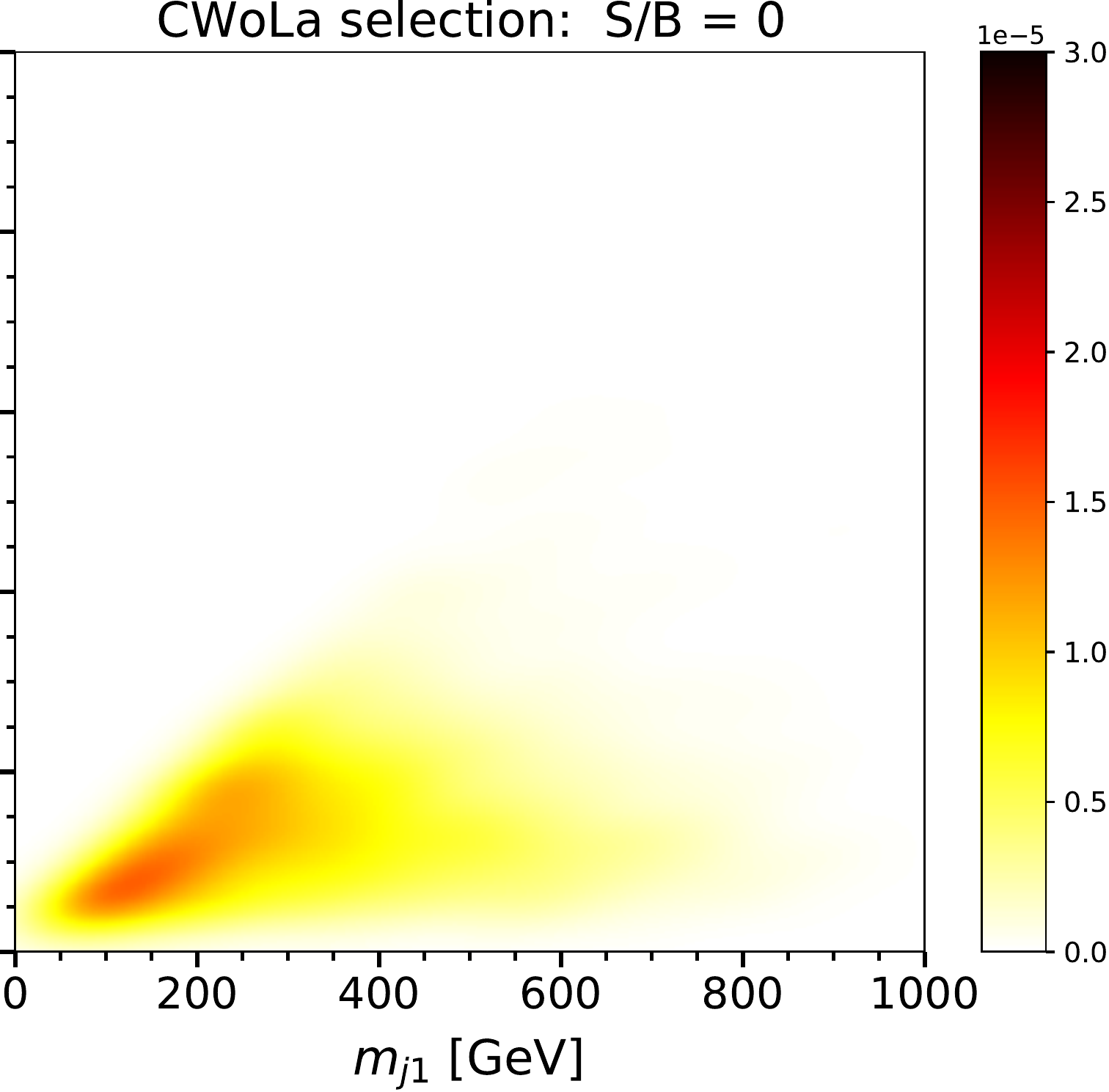} \\
        \vspace{10pt}
        \includegraphics[scale=0.335]{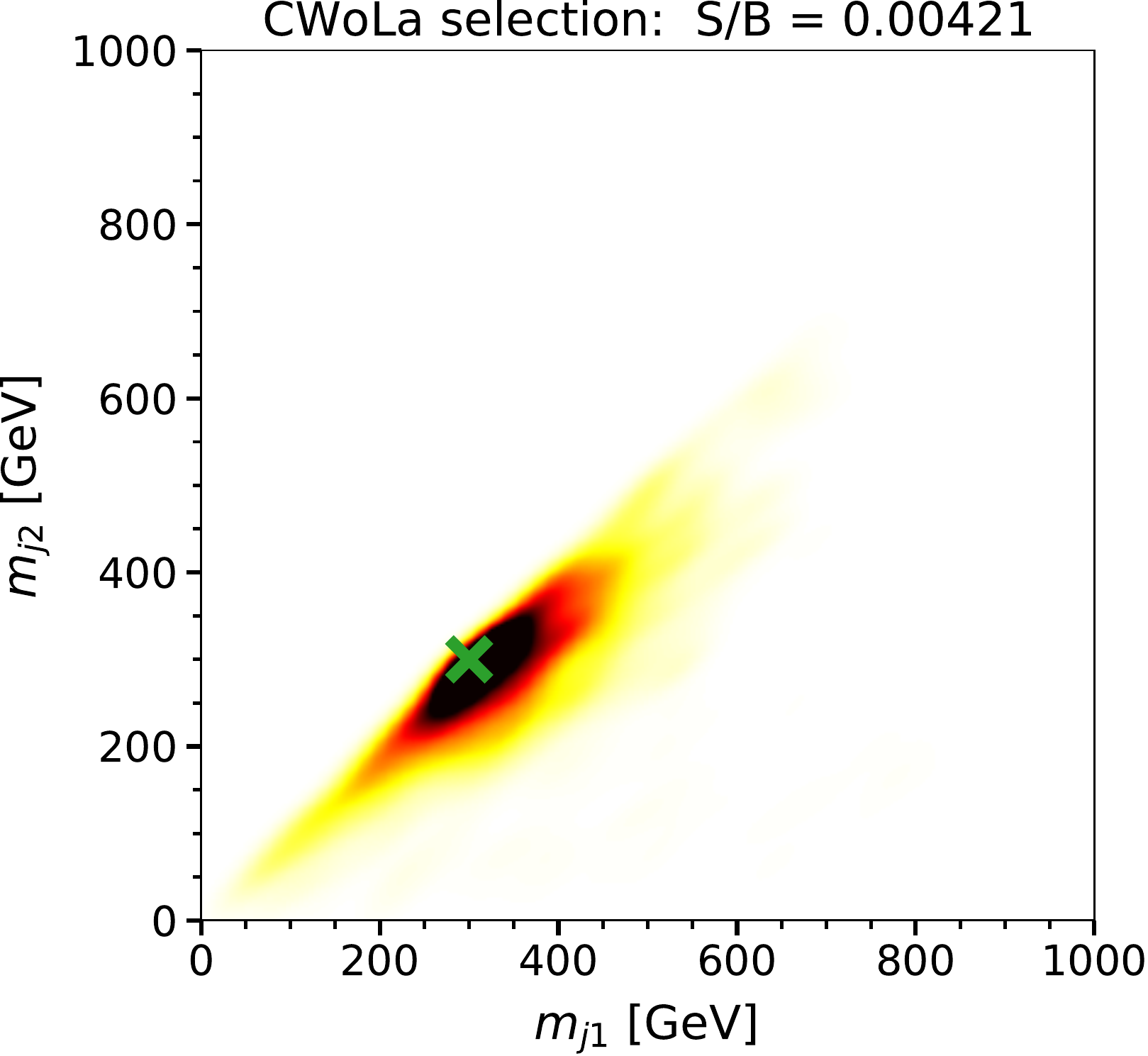}
        \includegraphics[scale=0.335]{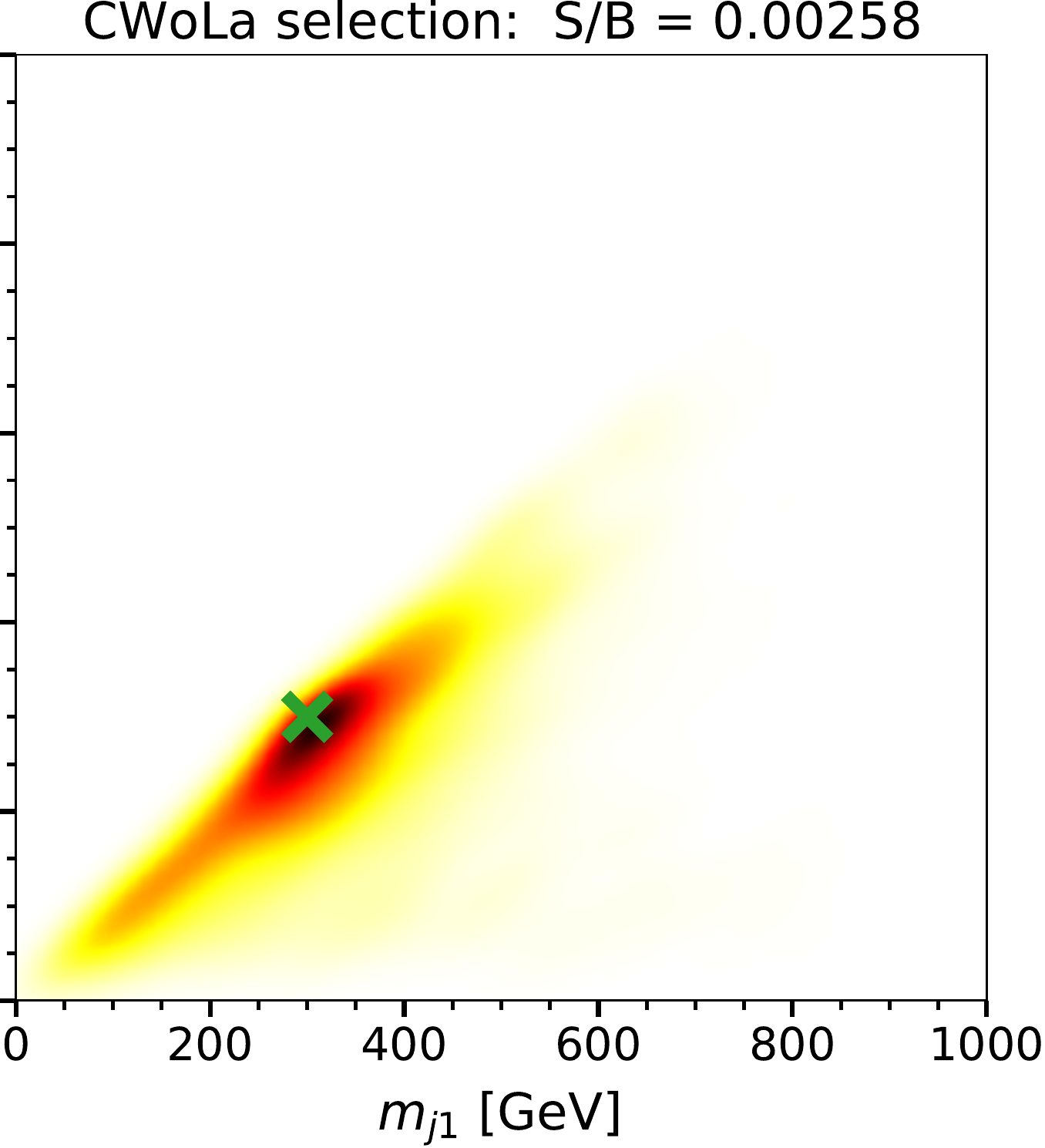}
        \includegraphics[scale=0.335]{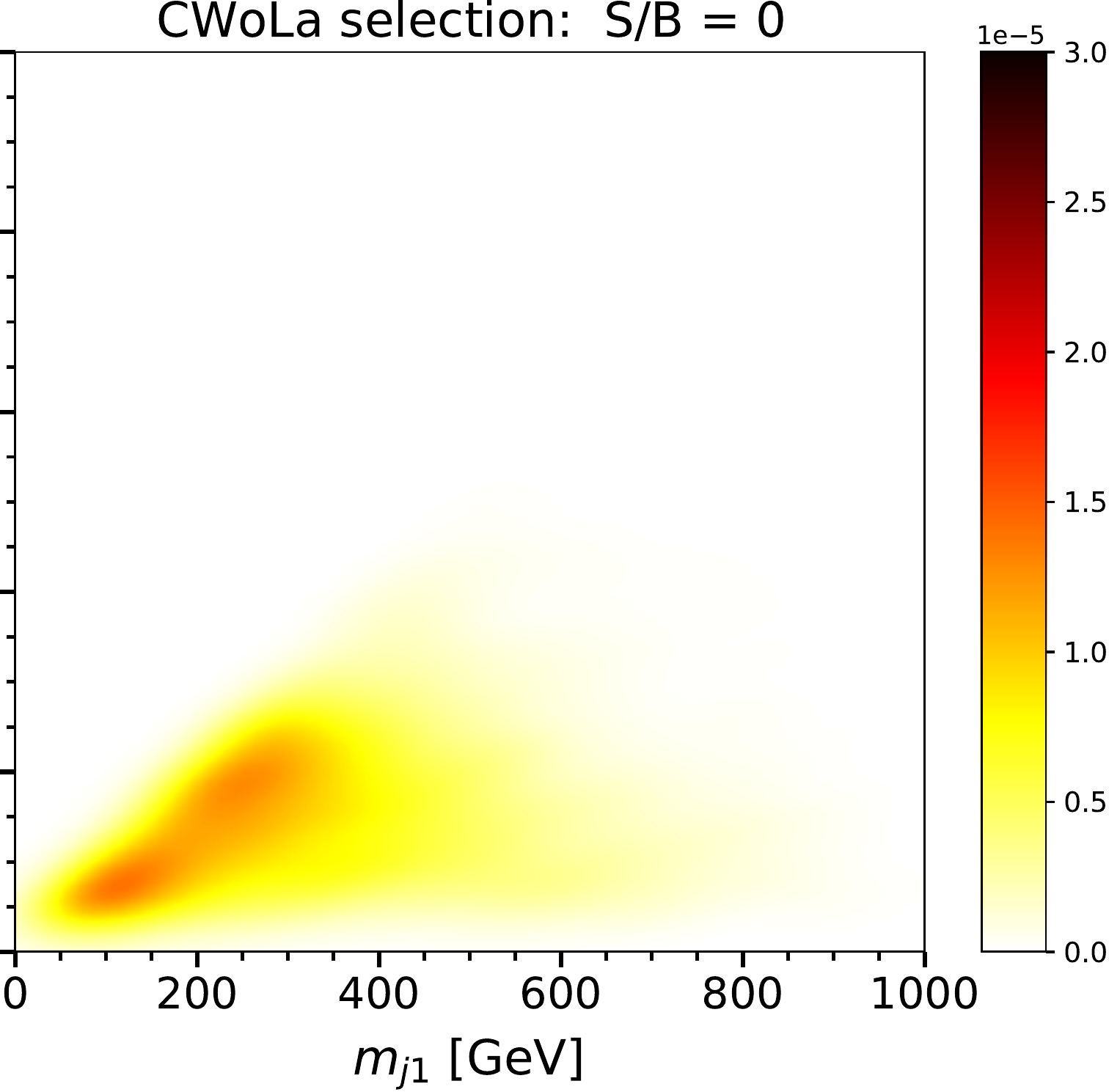}
   \end{center}
   \caption{Density of events on the $(m_{j_{1}}, m_{j_{2}})$ plane for the most signal-like events selected by CWoLa Hunting for the signal hypothesis $(m_{j_{1}}, m_{j_{2}}) = (500, 500) \; \GeV$ (top row) and  $(m_{j_{1}}, m_{j_{2}}) = (300, 300) \; \GeV$ (bottom row). From left to right, we show results for three benchmarks with $S/B \simeq 4 \cdot 10^{-3}, 2.8 \cdot 10^{-3}, 0$. The location of the injected signal is indicated by a green cross. Note that the upper right plot shows a small statistical fluctuation that disappears when averaging over a larger number of simulations.}
   \label{fig:CWoLa_selection_01_percent}
\end{figure}

\begin{figure}[t!]
   \begin{center}
        \includegraphics[scale=0.335]{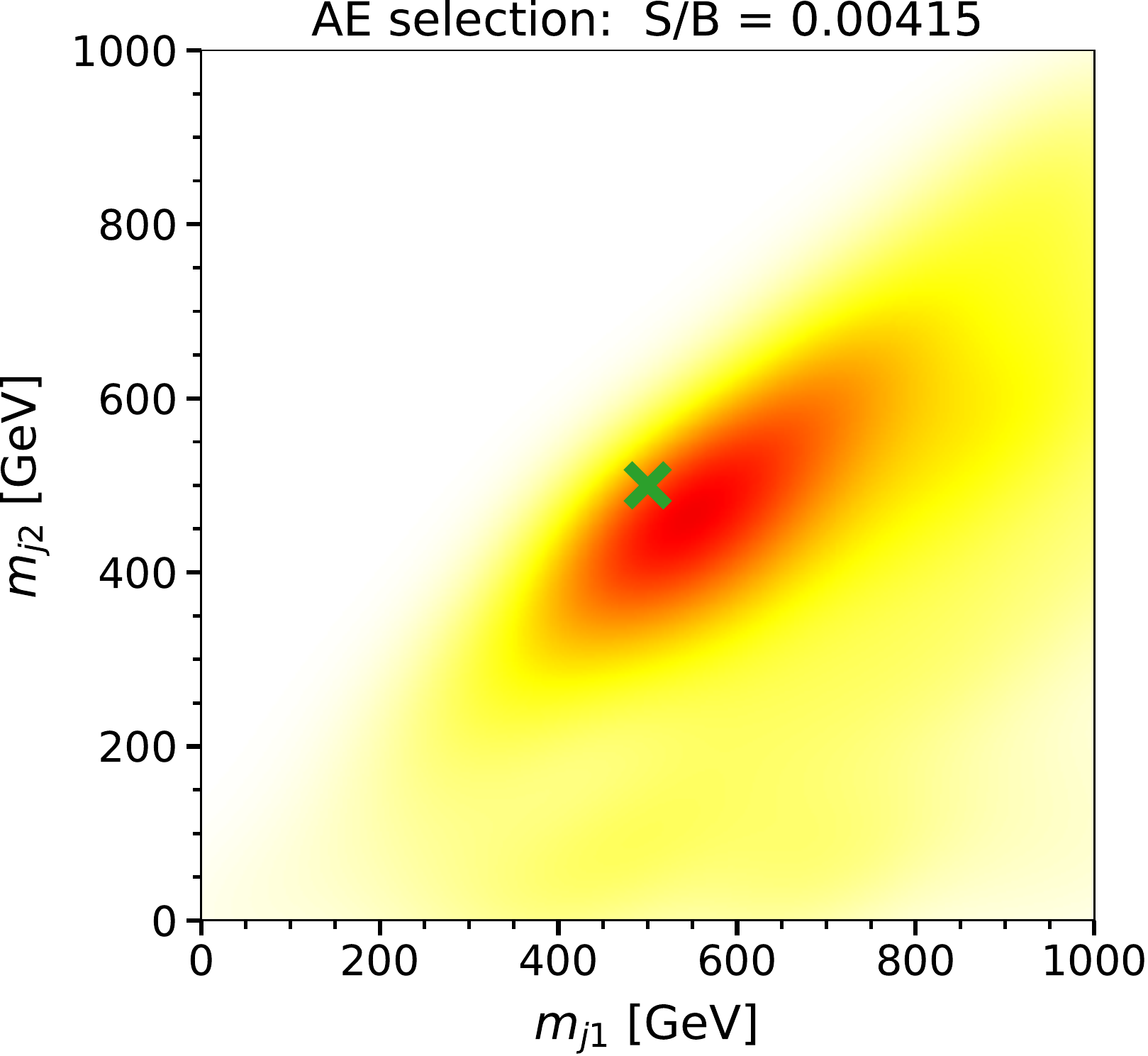}
        \includegraphics[scale=0.335]{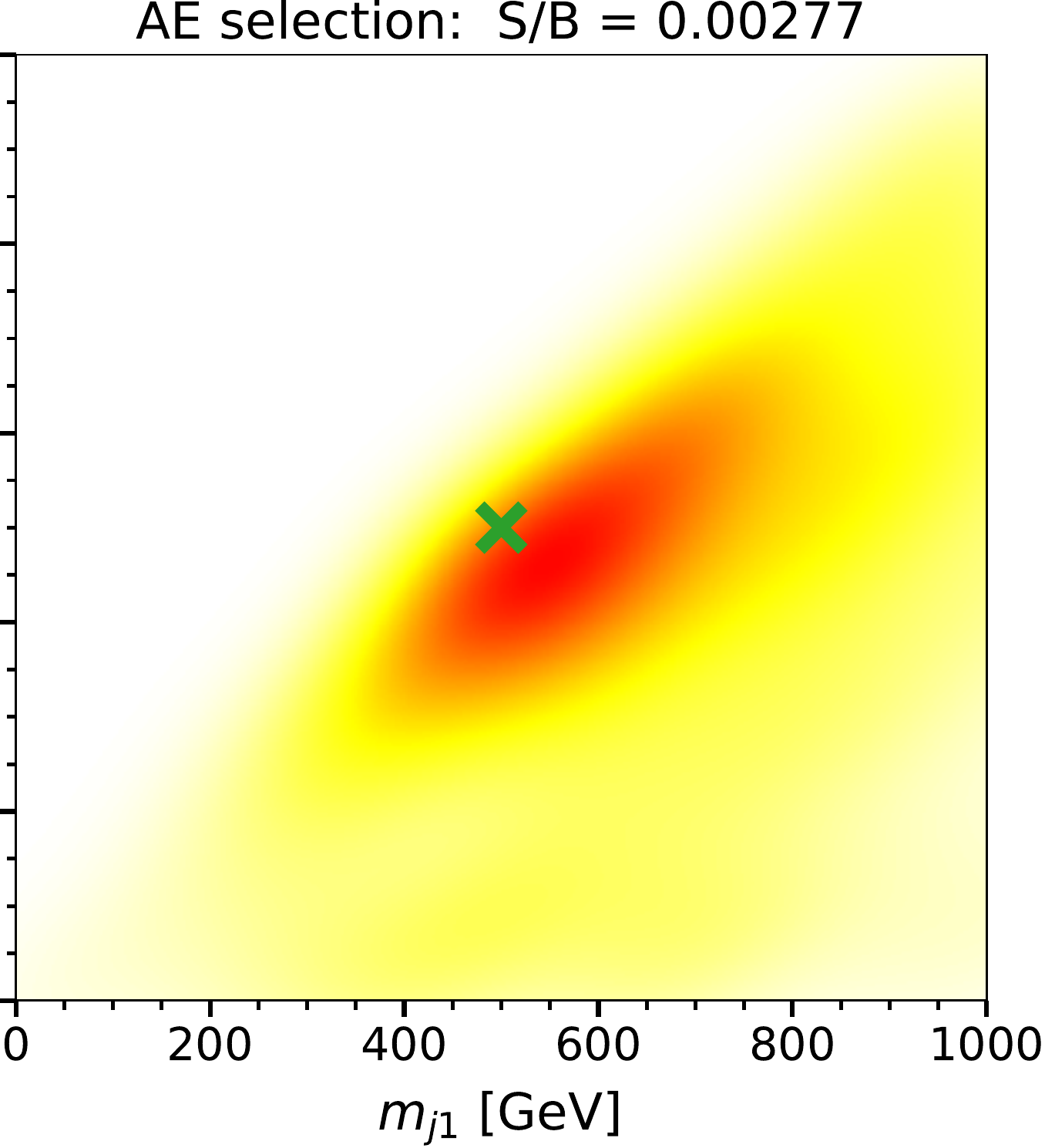}
        \includegraphics[scale=0.335]{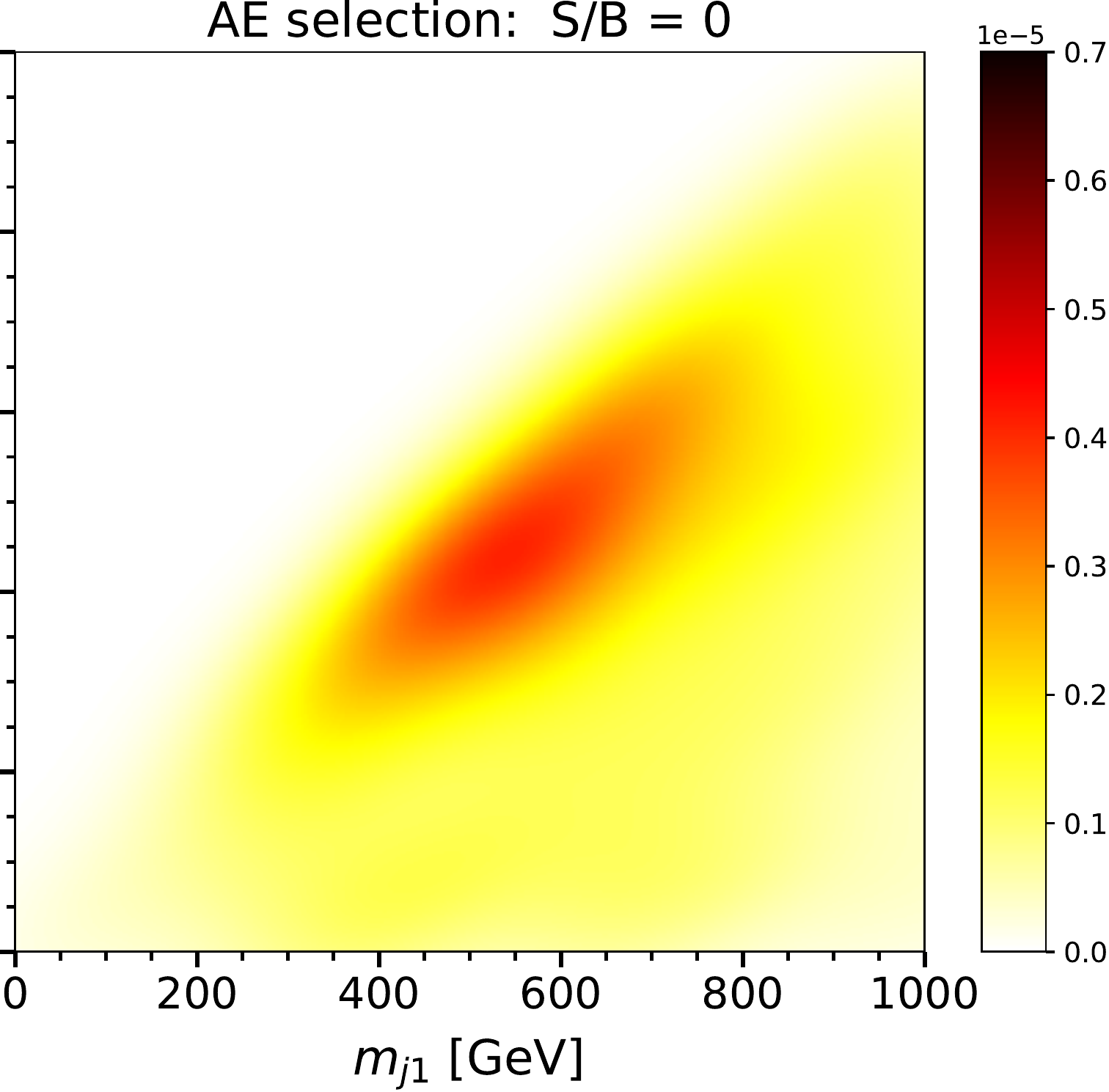} \\
        \vspace{10pt}
        \includegraphics[scale=0.335]{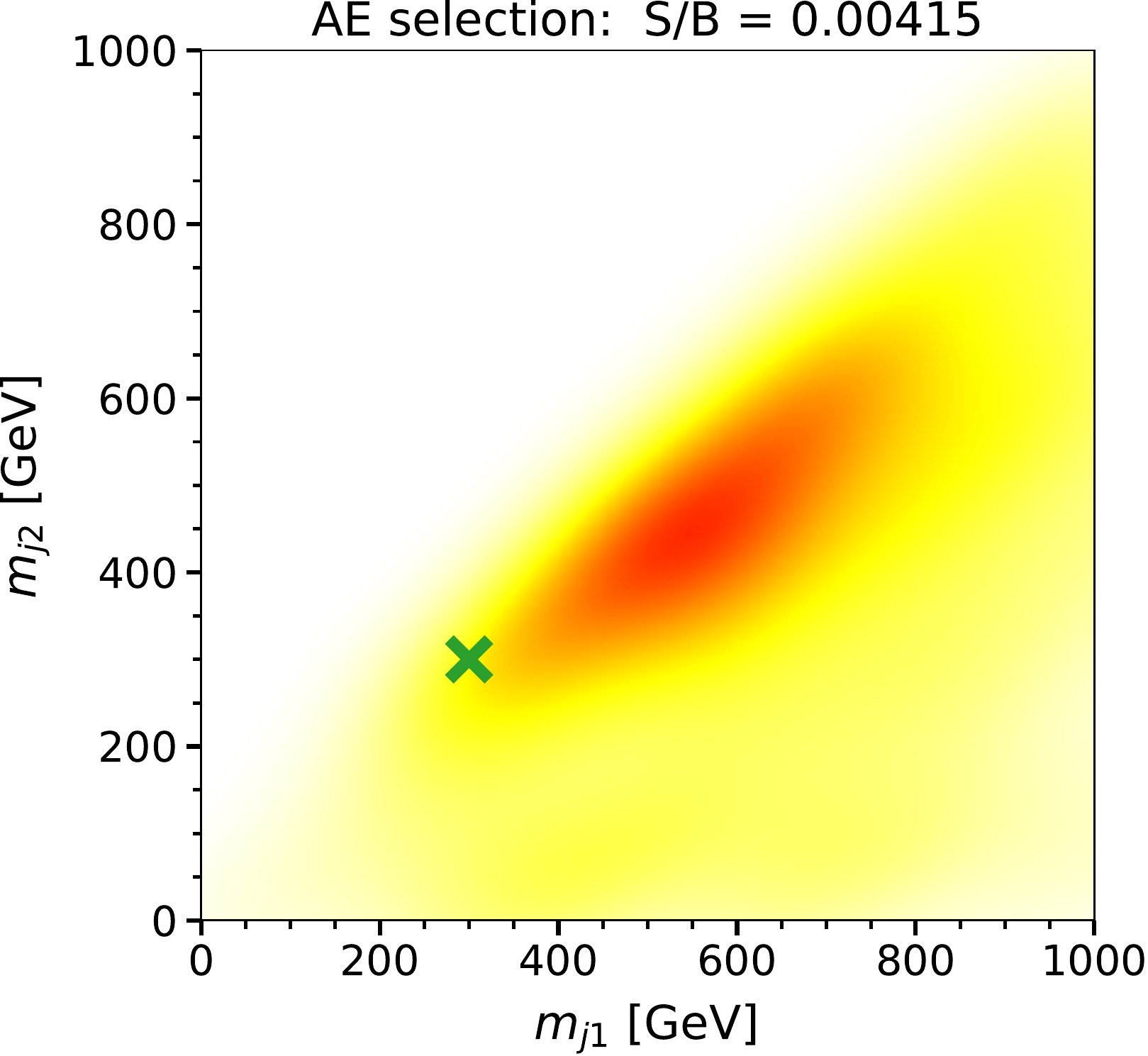}
        \includegraphics[scale=0.335]{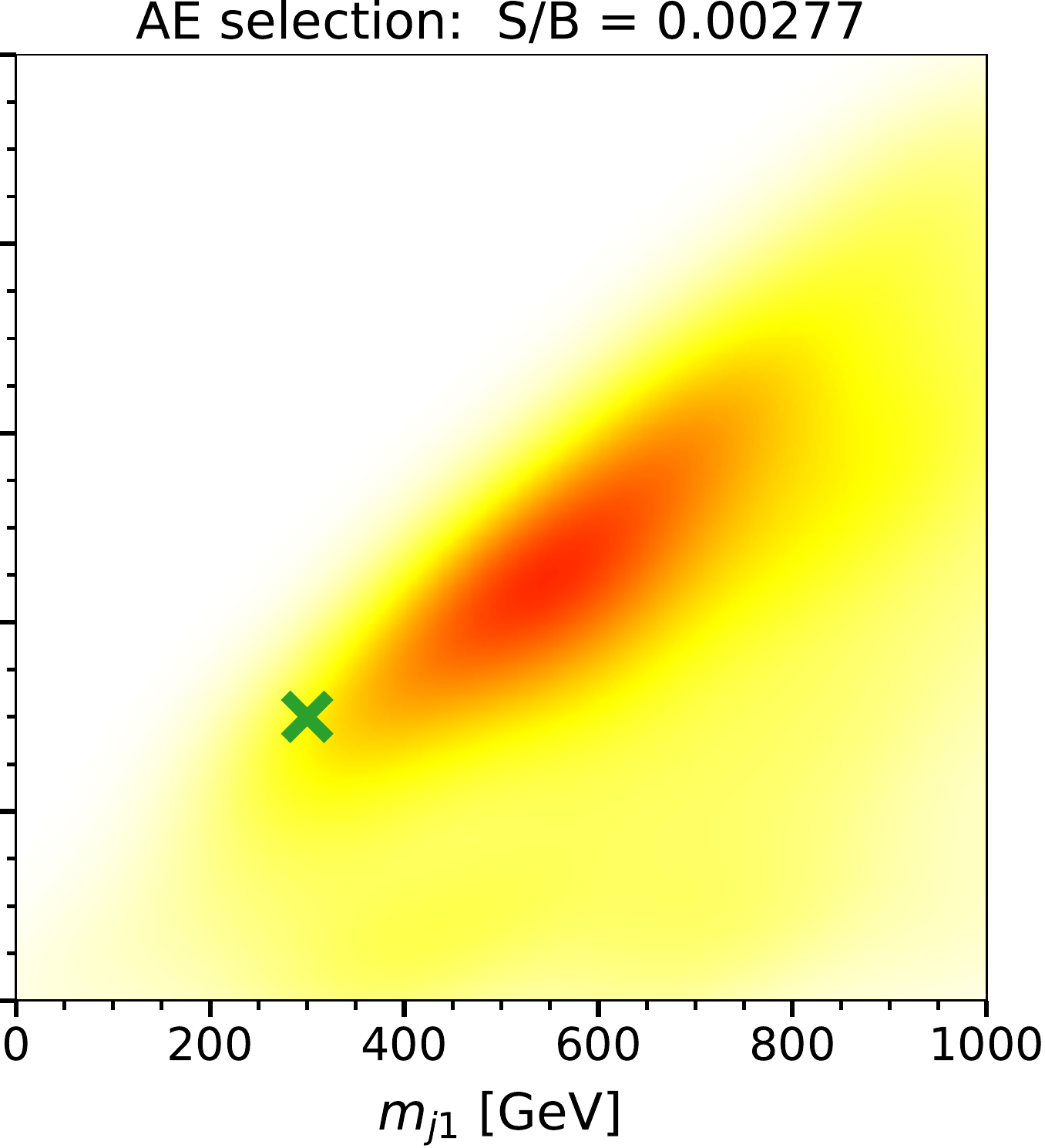}
        \includegraphics[scale=0.335]{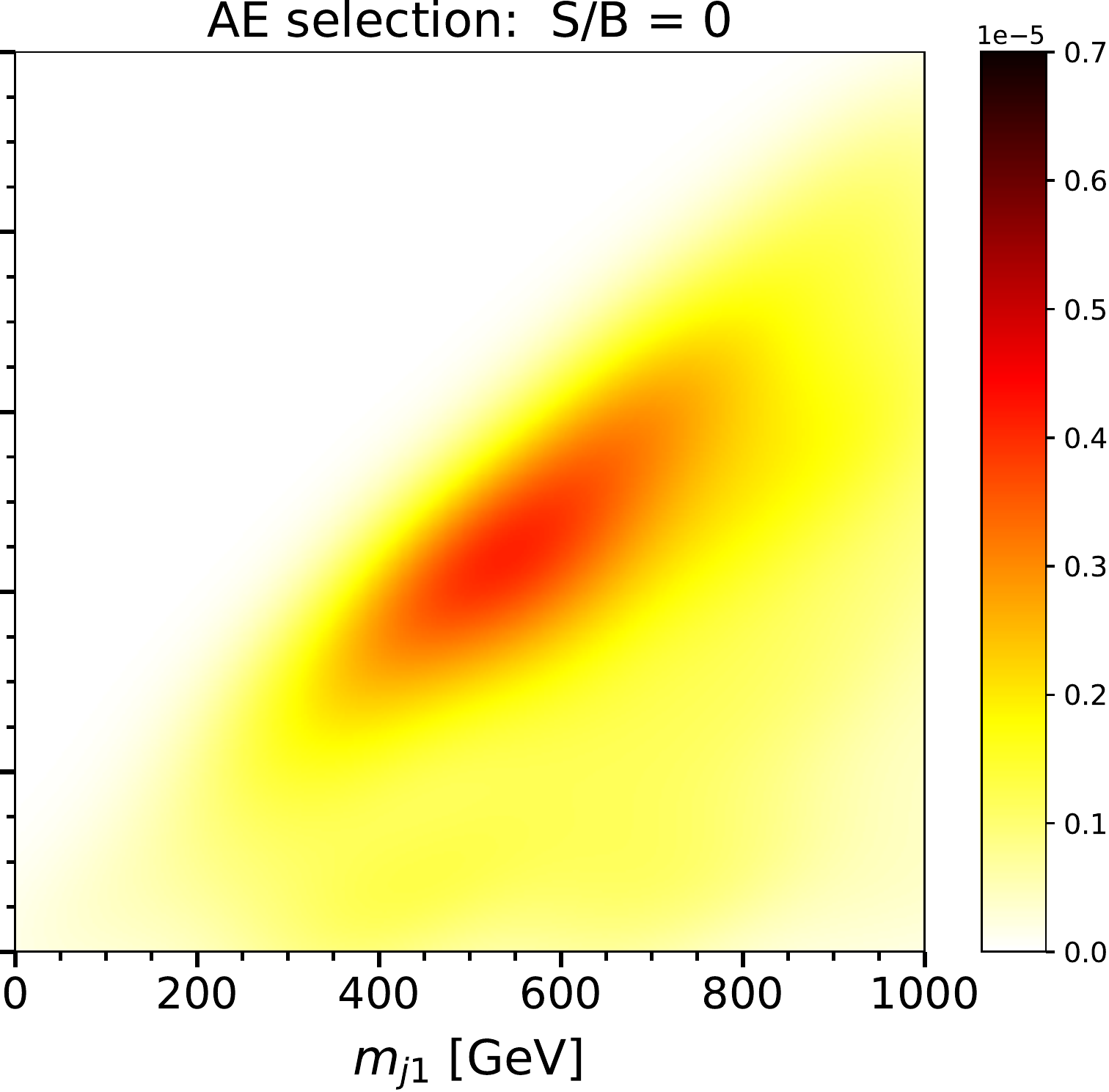}
   \end{center}
   \caption{Density of events on the $(m_{j_{1}}, m_{j_{2}})$ plane for the most signal-like events selected by the AE for the signal hypothesis $(m_{j_{1}}, m_{j_{2}}) = (500, 500) \; \GeV$ (top row) and  $(m_{j_{1}}, m_{j_{2}}) = (300, 300) \; \GeV$ (bottom row). From left to right, we show results for three benchmarks with $S/B \simeq 4 \cdot 10^{-3}, 2.8 \cdot 10^{-3}, 0$. The location of the injected signal is indicated by a green cross.}
   \label{fig:AE_selection_01_percent}
\end{figure}

\subsection{What did the machine learn?}

In order to illustrate this point, we can examine what the classifiers have learnt by looking at the properties of the events which have been classified as signal-like for three benchmarks with $S/B \simeq 4 \cdot 10^{-3}, 2.8 \cdot 10^{-3}, 0$. In Fig.~\ref{fig:CWoLa_selection_01_percent} and Fig.~\ref{fig:AE_selection_01_percent} we show the density of events on the $(m_{j_{1}}, m_{j_{2}})$ plane for the most signal-like events selected by CWoLa Hunting and the AE, respectively. The cuts applied in each case correspond to the $0.1 \, \%$ cut. For CWoLa Hunting, it is clear that the classifier is able to locate the signal for the two mass hypotheses. In addition, note that the upper and lower right plots show a small statistical fluctuation that is produced by the different fractions of signal-like events represented in each plot, which disappears when averaging over a larger number of simulations.

\begin{figure}[t!]
   \begin{center}
        \includegraphics[scale=0.31]{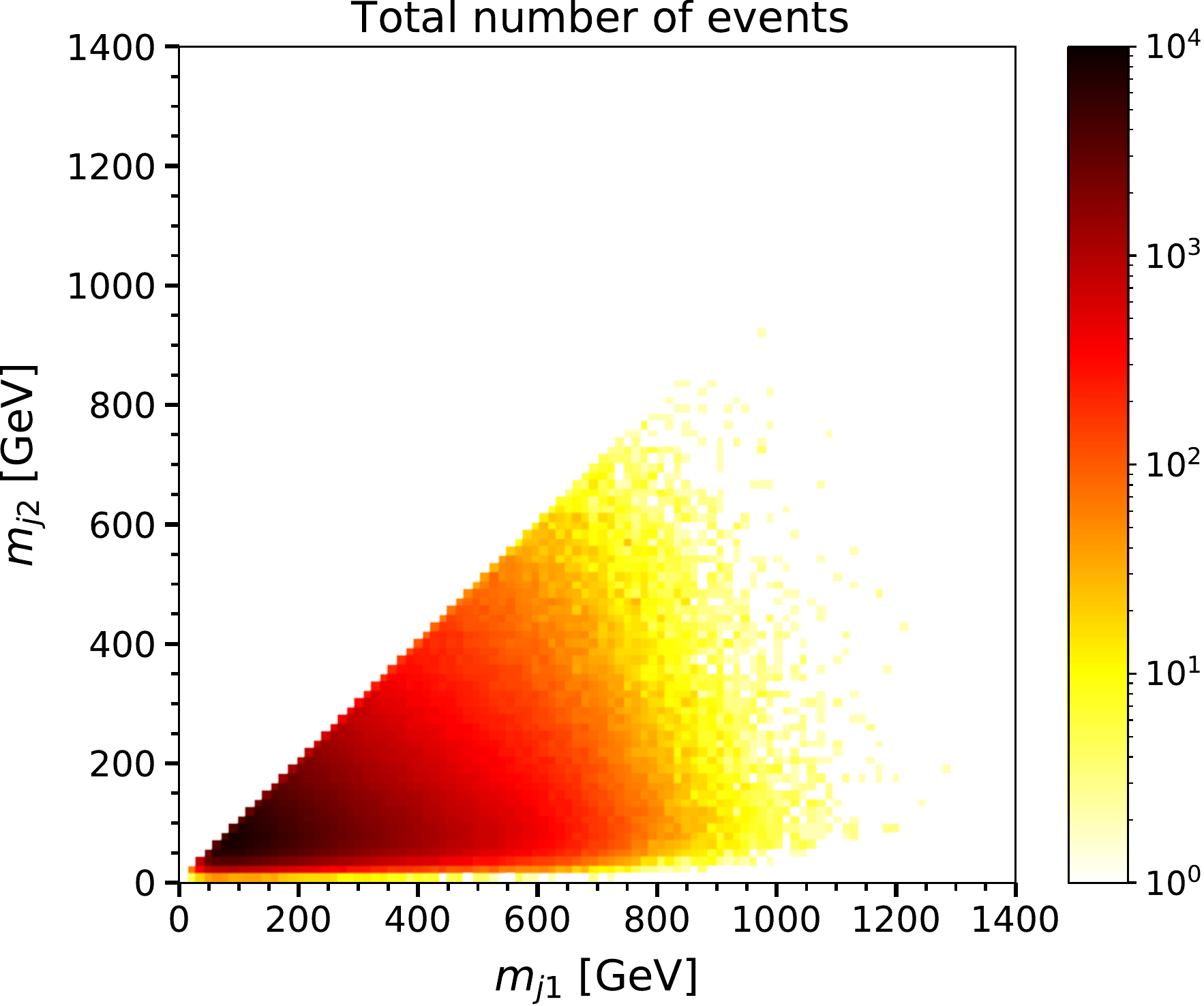}
        \hspace{-5pt}
        \includegraphics[scale=0.31]{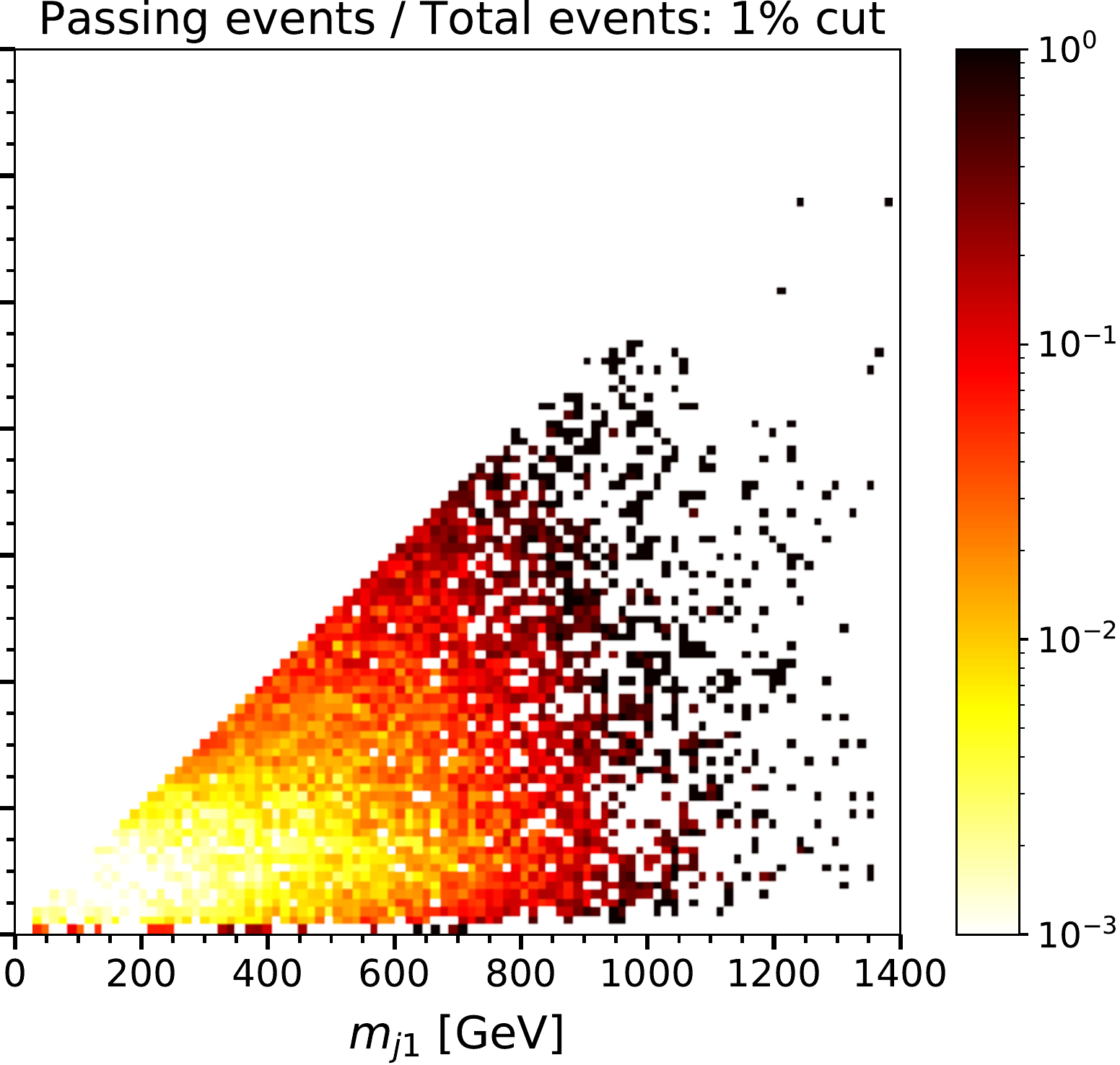}
        \hspace{-5pt}
        \includegraphics[scale=0.31]{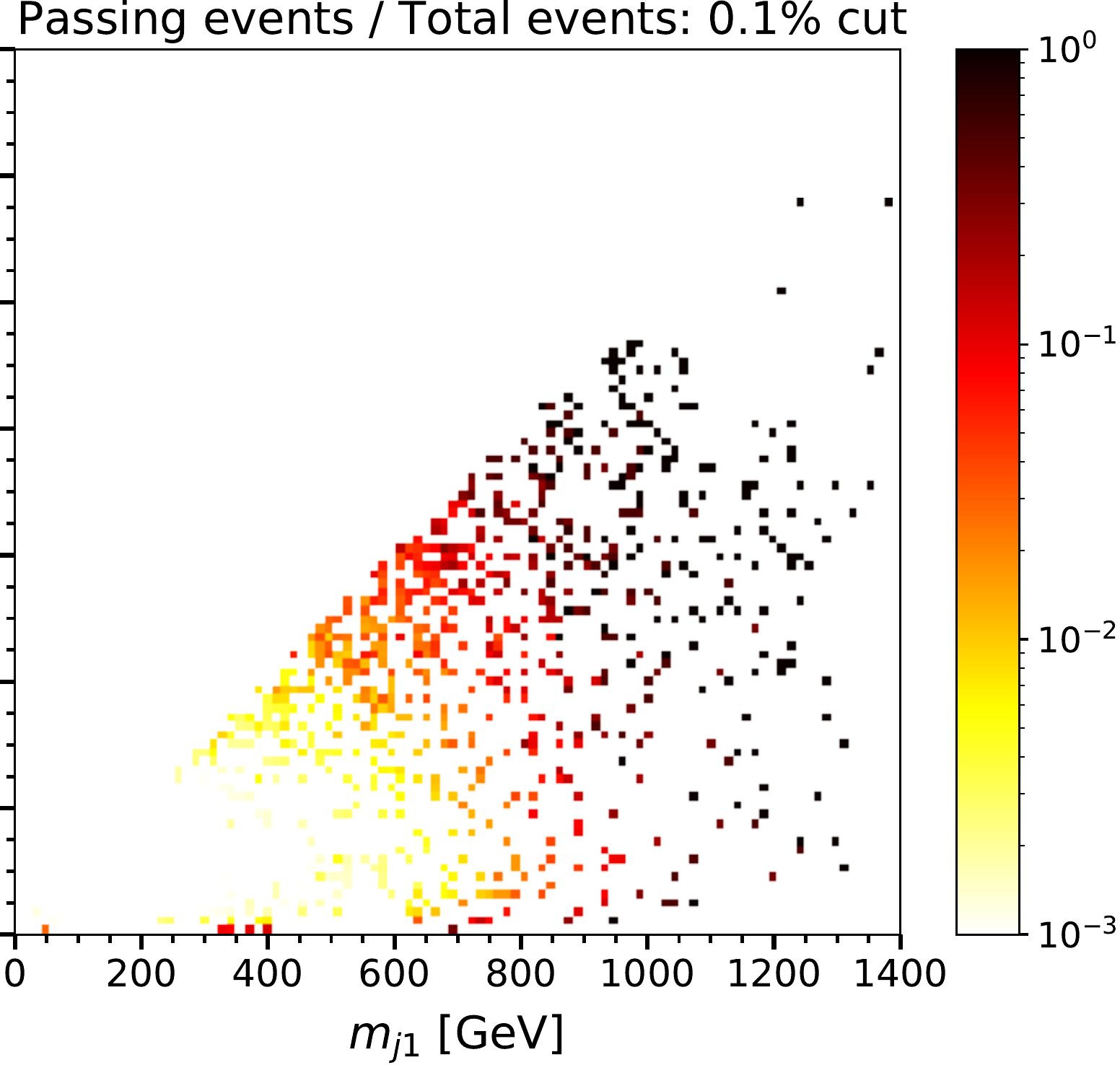} \\
        \vspace{10pt}
        \includegraphics[scale=0.31]{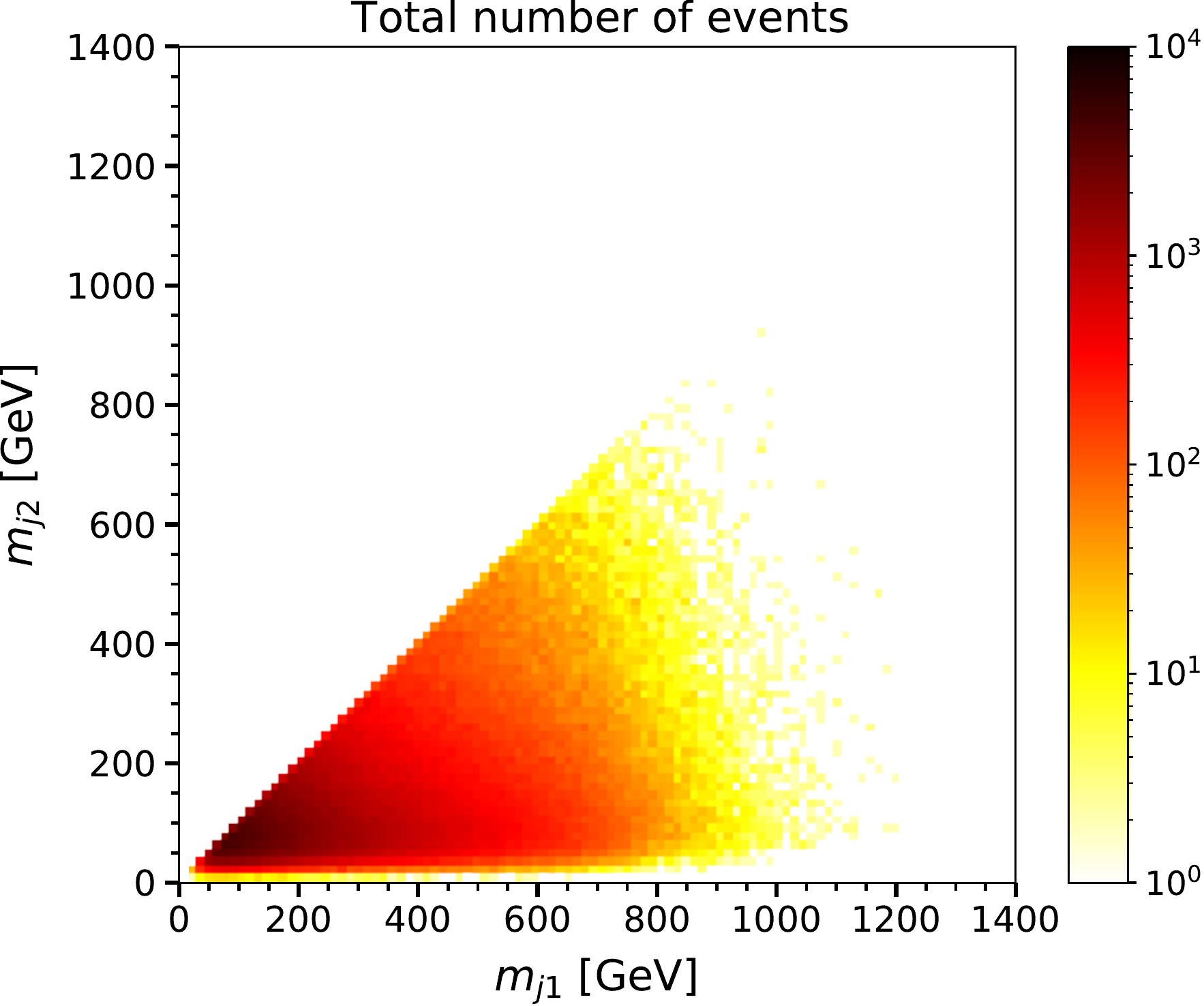}
        \hspace{-5pt}
        \includegraphics[scale=0.31]{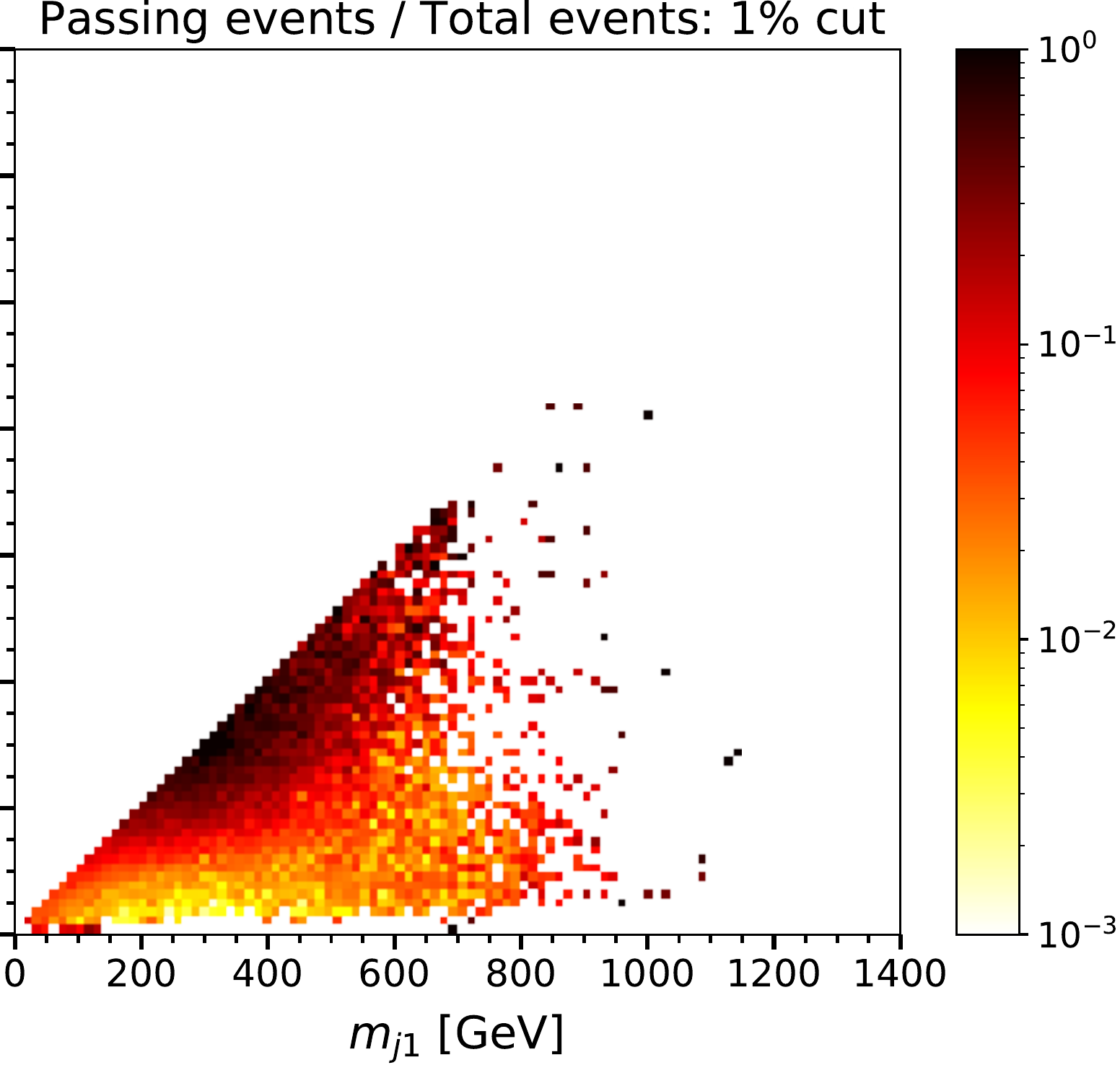}
        \hspace{-5pt}
        \includegraphics[scale=0.31]{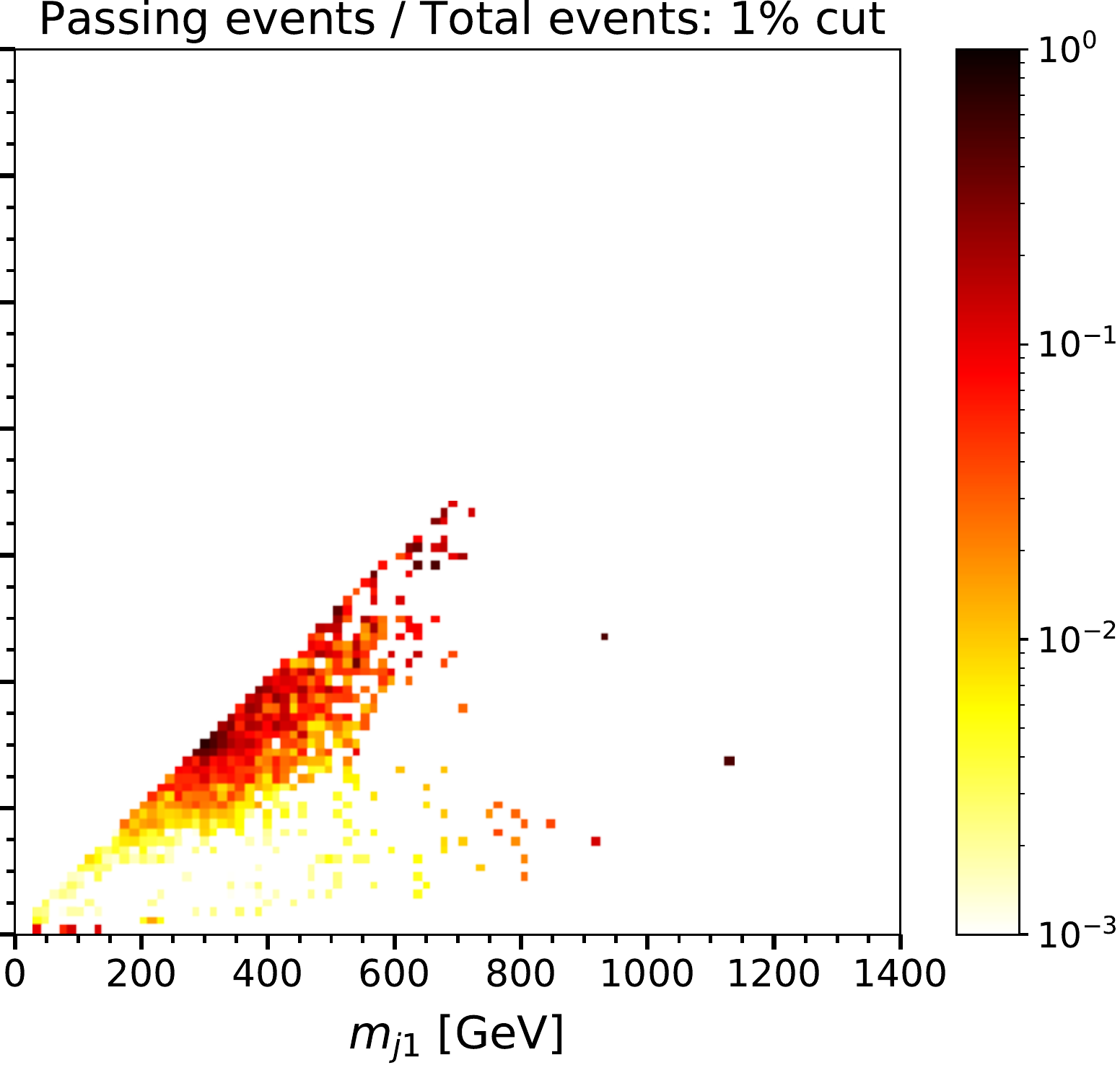}
   \end{center}
   \caption{The total density of events on the $(m_{j_{1}}, m_{j_{2}})$ plane is plotted on the left. The $1 \, \%$ and $0.1 \, \%$ selection efficiencies for the AE and CWoLa are plotted on the middle and right images, respectively. The top row shows results for the AE, while the bottom row shows results for CWoLa. The selection efficiency in a given bin is defined as the number of events passing the $x \, \%$ cut divided by the total number of events in that bin.  These results correspond to the signal with $(m_{j_{1}}, m_{j_{2}}) = (300, 300) \; \GeV$ and $S/B \simeq 4 \cdot 10^{-3}$.}
   \label{fig:AE_and_CW_selection_fraction}
\end{figure}

The AE similarly identifies the high mass signal point, but fails to identify the low mass one. This can be most easily understood by observing the selection efficiency as a function of the two jet masses for the trained AE, shown in Fig.~\ref{fig:AE_and_CW_selection_fraction}. In the left plot, we show the total number of events on the $(m_{j_{1}}, m_{j_{2}})$ plane. In the middle and right plots, we show the selection efficiencies for the $1 \, \%$ and $0.1 \, \%$ cuts. These results illustrate that the AE has learnt to treat high mass jets as anomalous (since these are rare in the training sample), and so the $(m_{j_{1}}, m_{j_{2}}) = (300, 300) \; \GeV$ signal is more easily reconstructed than high mass QCD events. In other words, high mass QCD events are regarded as more anomalous than signal events, and a sufficiently high selection cut on the AE reconstruction error will eliminate the signal. We remark again that this is one of the main limitations of the AE. Therefore, it is crucial to find the cut that maximizes the fraction of signal within the most anomalous events. As shown in Fig.~\ref{fig:AE_selection} in Appendix~\ref{sec:optimal_cuts}, that cut corresponds to the anomaly score that maximizes the SIC curve in the signal region. In contrast, the bottom row of Fig.~\ref{fig:AE_and_CW_selection_fraction} shows that CWoLa is able to learn the signal features.

\section{Conclusions}
\label{sec:conc}

In this article, we have compared weakly-supervised and unsupervised anomaly detection methods, using Classification without Labels (CWoLa) Hunting and deep autoencoders (AE) as representative of the two classes. The key difference between these two methods is that the weak labels of CWoLa Hunting allow it to utilize the specific features of the signal overdensity, making it ideal in the limit of large signal rate, while the unsupervised AE does not rely on any information about the signal and is therefore robust to small signal rates.

We have quantitatively explored this complimentarity in a concrete case study of a search for anomalous events in fully hadronic dijet resonance searches, using as the target a physics model of a heavy resonance decaying into a pair of three-prong jets. CWoLa Hunting was able to dramatically raise the significance of the signal in our benchmark points in order to breach $5\sigma$ discovery, but only if a sizeable fraction of signal is present ($S/B \gtrsim 4 \times 10^{-3}$). The AE maintained classification performance at low signal rates and had the potential to raise the significance of of one of our benchmark signals to the level of $3\sigma$ in a region where CWoLa Hunting lacked sensitivity.

Crucially, our results demonstrate that CWoLa Hunting is effective at finding diverse and moderately rare signals and the AE can provide sensitivity to rare signals, but only with certain topologies. Therefore, both techniques are complementary and can be used together for anomaly detection.  A variety of unsupervised, weakly supervised, and semi-supervised anomaly detection approaches have been recently proposed (see e.g. Ref.~\cite{Kasieczka:2021xcg}), including variations of the methods we have studied.  It will be important to explore the universality of our conclusions across a range of models for anomaly detection at the LHC and beyond.

\section*{\label{sec::acknowledgments}Acknowledgments}

BN and JC were supported by the U.S.~Department of Energy, Office of Science under contracts DE-AC02-05CH11231 and DE-AC02-76SF00515, respectively. DS is supported by DOE grant DOE-SC0010008. PMR acknowledges Berkeley LBNL, where part of this work has been developed. PMR further acknowledges support from the Spanish Research Agency (Agencia Estatal de Investigaci\'on) through the contract FPA2016-78022-P and IFT Centro de Excelencia Severo Ochoa under grant SEV-2016-0597. This project has received funding/support from the European Union’s Horizon 2020 research and innovation programme under the Marie Skłodowska-Curie grant agreement No 690575 (RISE InvisiblesPlus).

\begin{appendices}

\newpage
\clearpage
\section{\label{sec:fit}Background fit}

In this appendix, we briefly describe the details about the fit procedure and discuss results from the fit to the background events. In order to evaluate the significance of any potential excess in the signal region, the total number of predicted signal region events is calculated by summing the individual predictions from each signal region bin. The systematic uncertainty of the fit in the signal region prediction is estimated by propagating the uncertainties in the fit parameters. We test the validity of the fit using a Kolmogorov–Smirnov test.

\begin{figure}[h!]
   \begin{center}
        \includegraphics[scale=0.80]{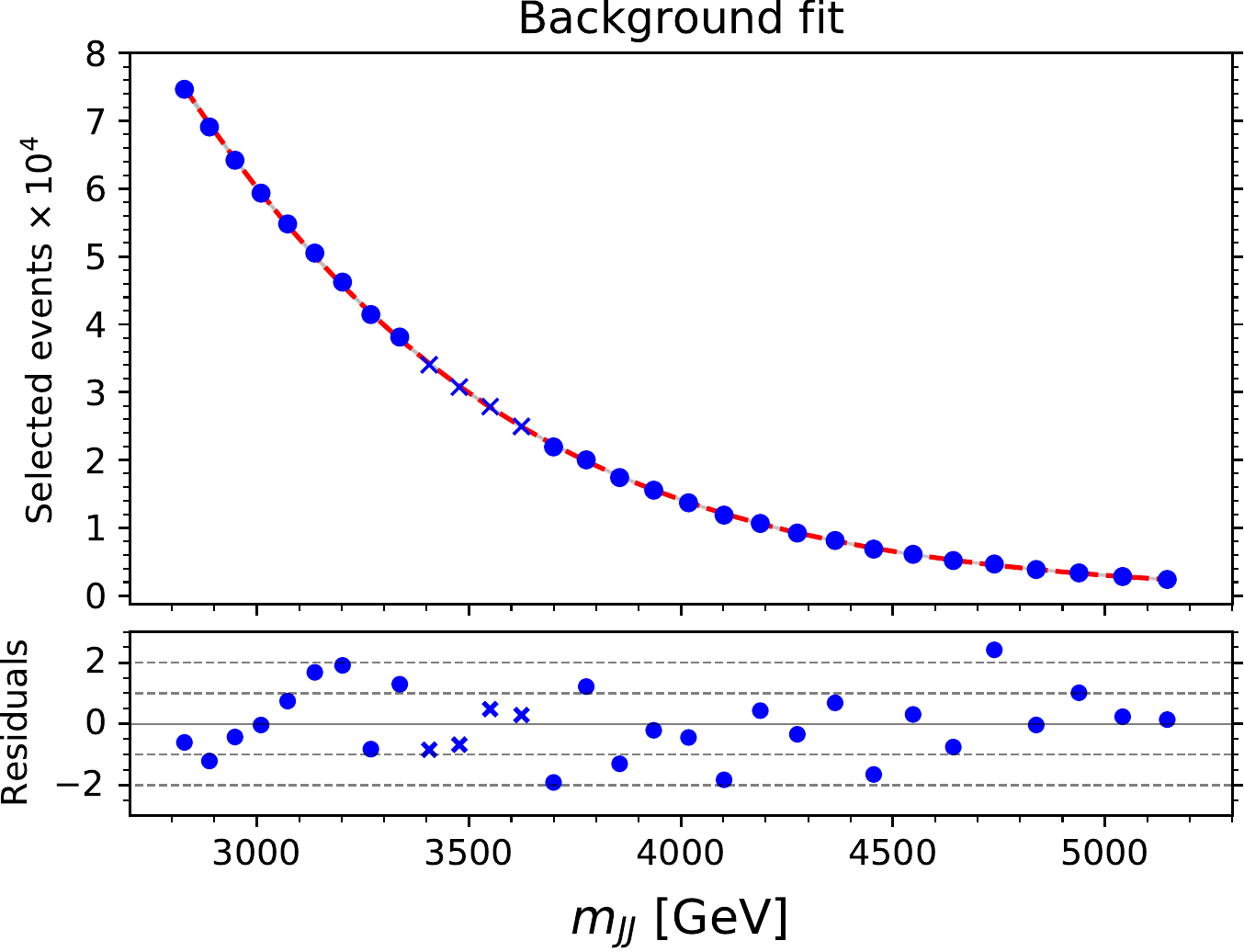}
   \end{center}
   \caption{Fit to the background distribution of dijet events and residuals from the fit. The signal region events are indicated by blue crosses. }
   \label{fig:background_fluctuation}
\end{figure}

In Fig.~\ref{fig:background_fluctuation} we show the fit to the background distribution using the 4-parameter function presented in Eq.~(\ref{eq:fit}). First, the Kolmogorov–Smirnov test yields a $p$-value of $0.99$, which means that the fit describes the background distribution well outside of the signal region. In addition, the fit result produces a $p$-value of $0.5$. However, the residuals indicate that the number of predicted events in the signal region is overestimated due to a local negative fluctuation of size $n = 123$ events\footnote{This has been validated as a fluctuation with an independent sample.}. As a result, the fit will always underestimate the excess significance when a signal is injected in the signal region. For example, if we introduce a number $n$ of signal events in the signal region, the fit prediction will match the number of observed events and therefore the excess significance will be exactly zero, even when a signal has been injected.

\newpage
\clearpage
\section{\label{sec:optimal_cuts} Density of events for the optimal cut }

\begin{figure}[h!]
   \begin{center}
        \includegraphics[scale=0.335]{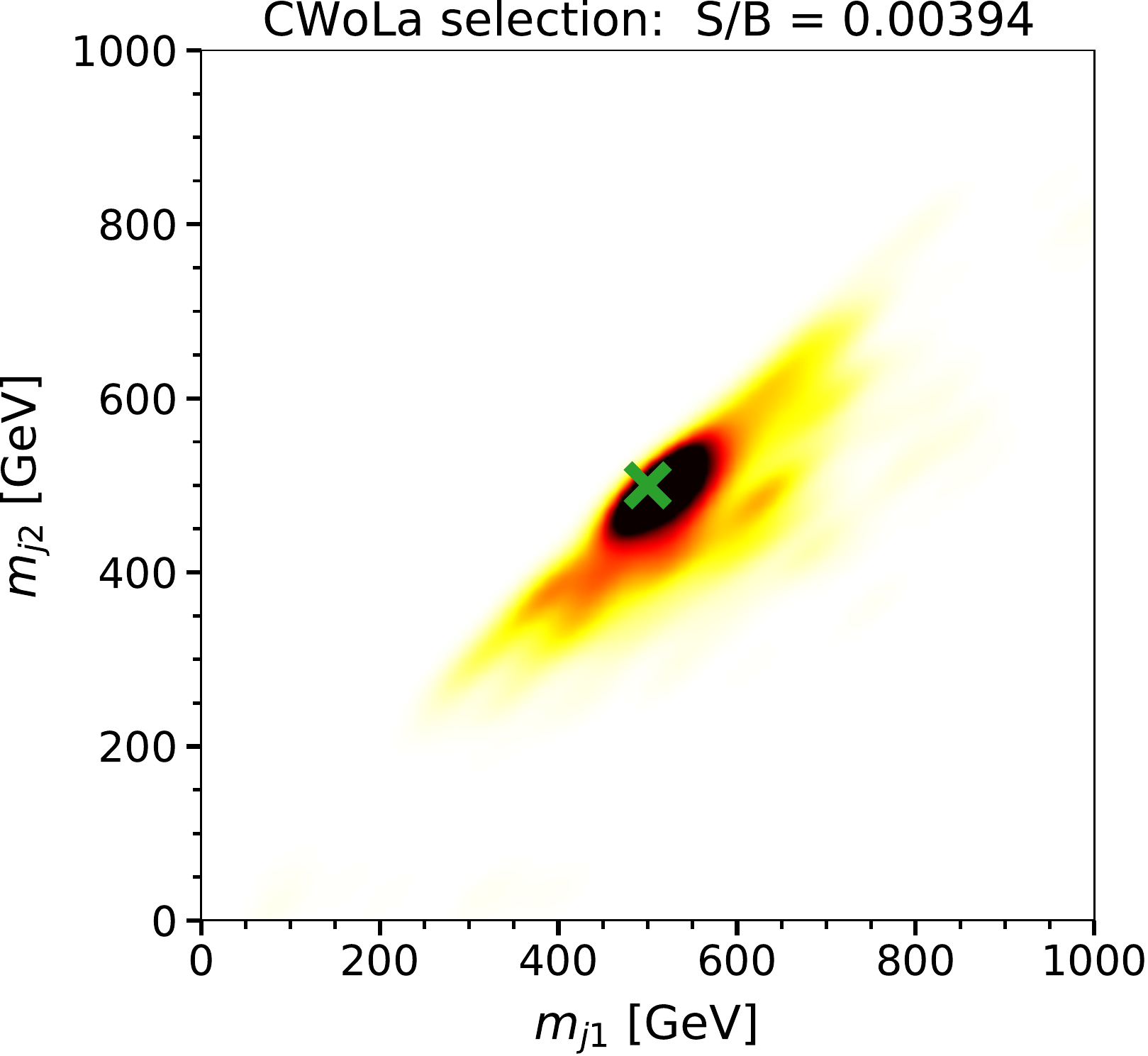}
        \includegraphics[scale=0.335]{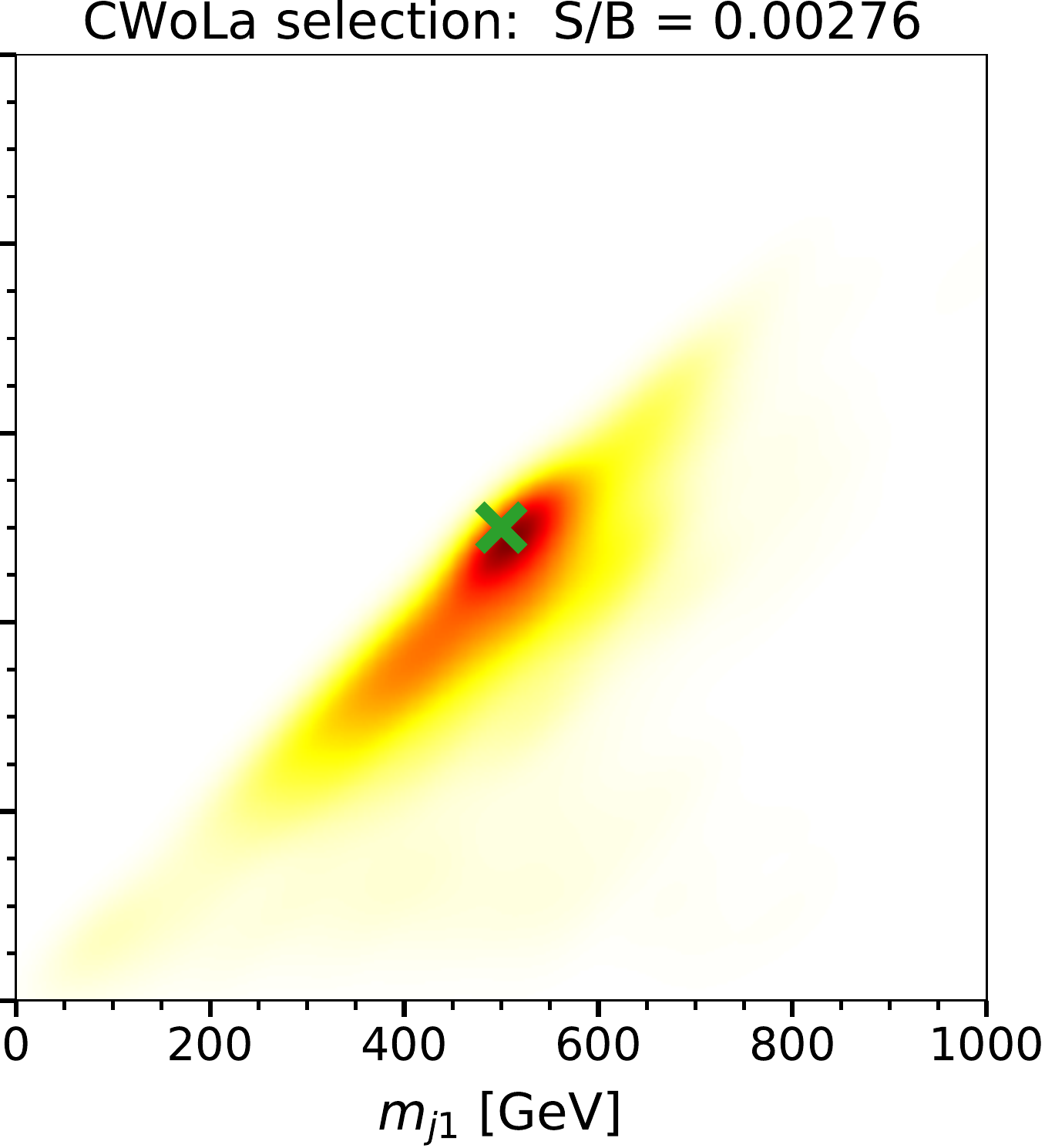}
        \includegraphics[scale=0.335]{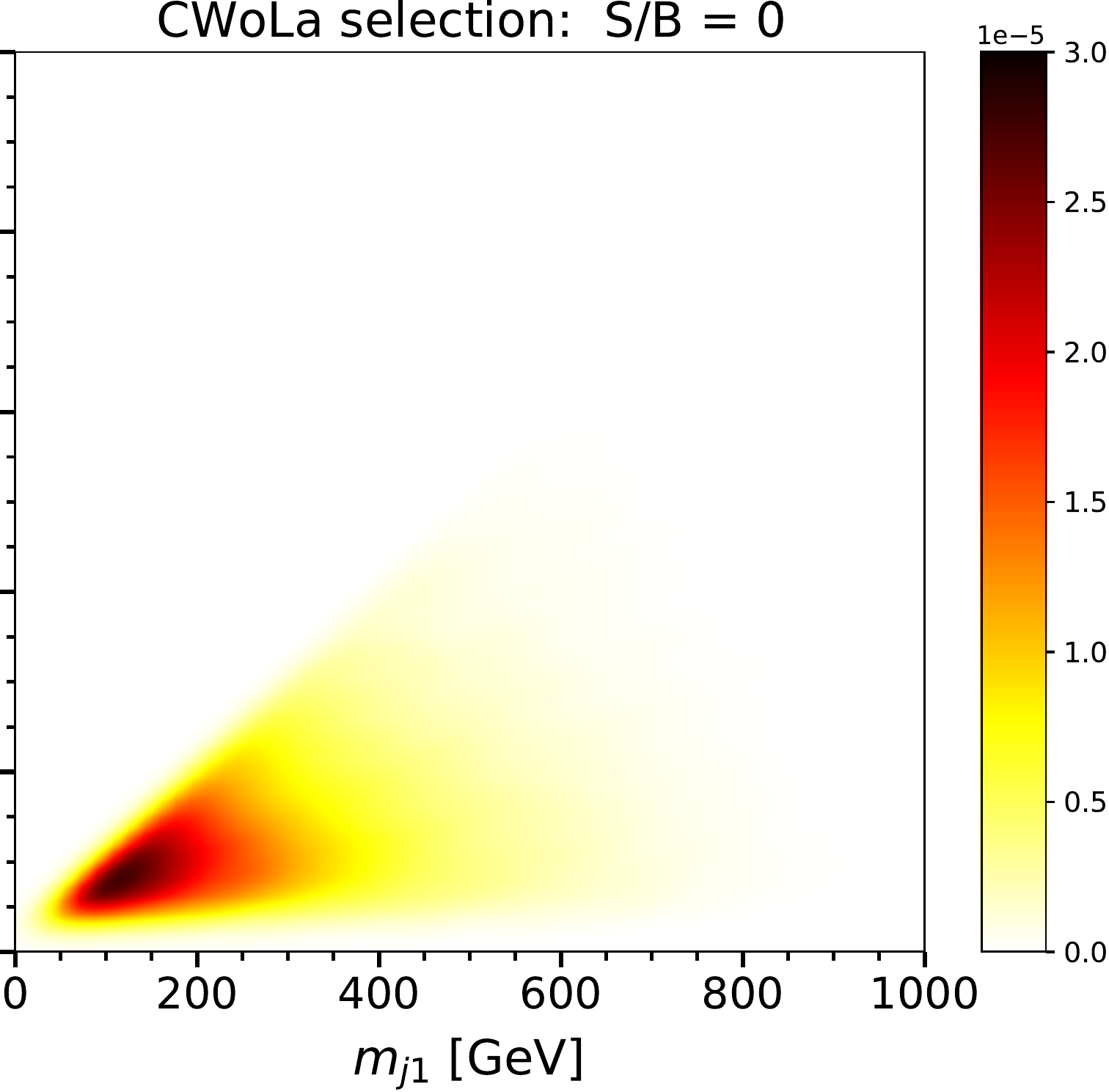} \\
        \vspace{10pt}
        \includegraphics[scale=0.335]{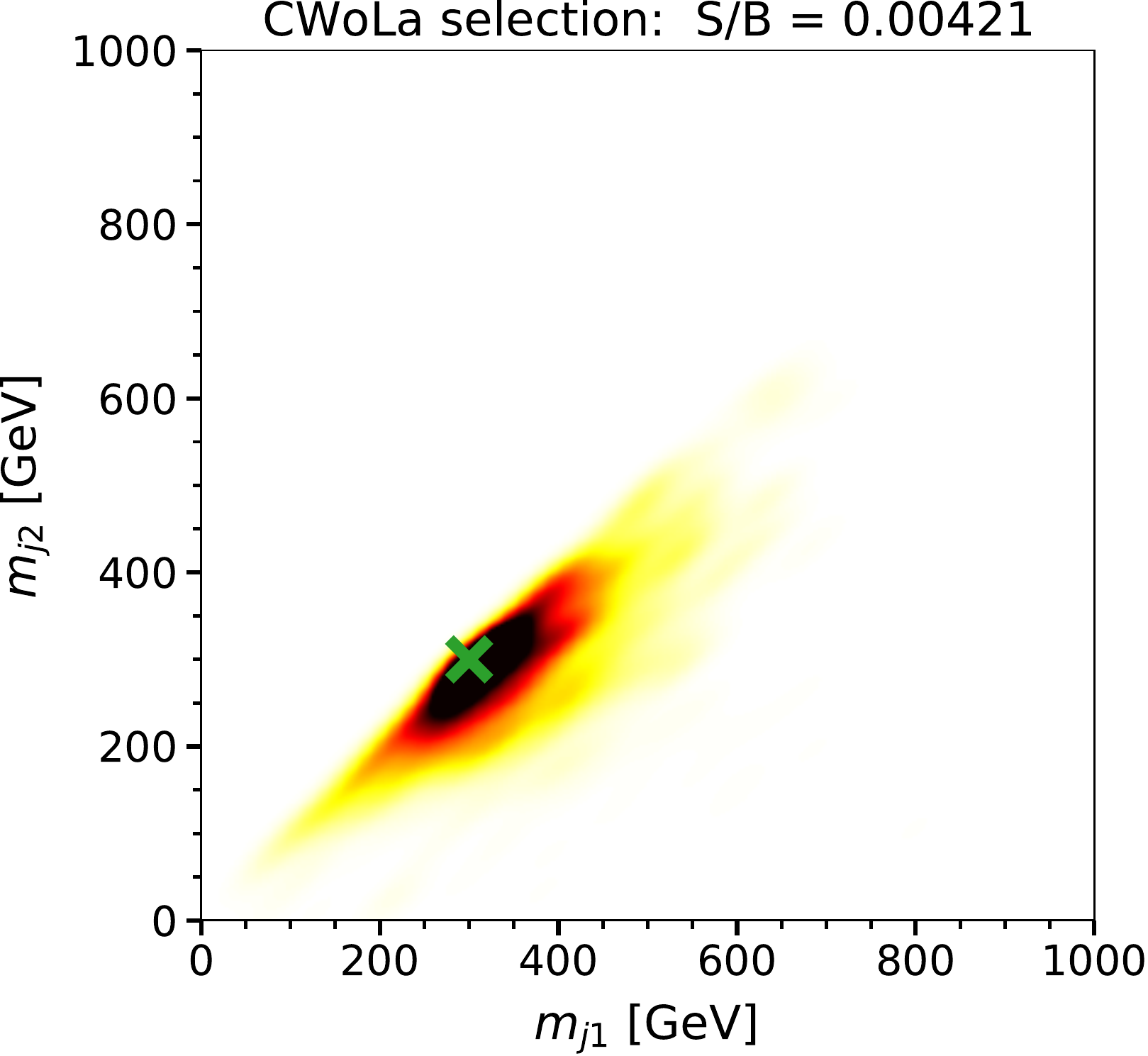}
        \includegraphics[scale=0.335]{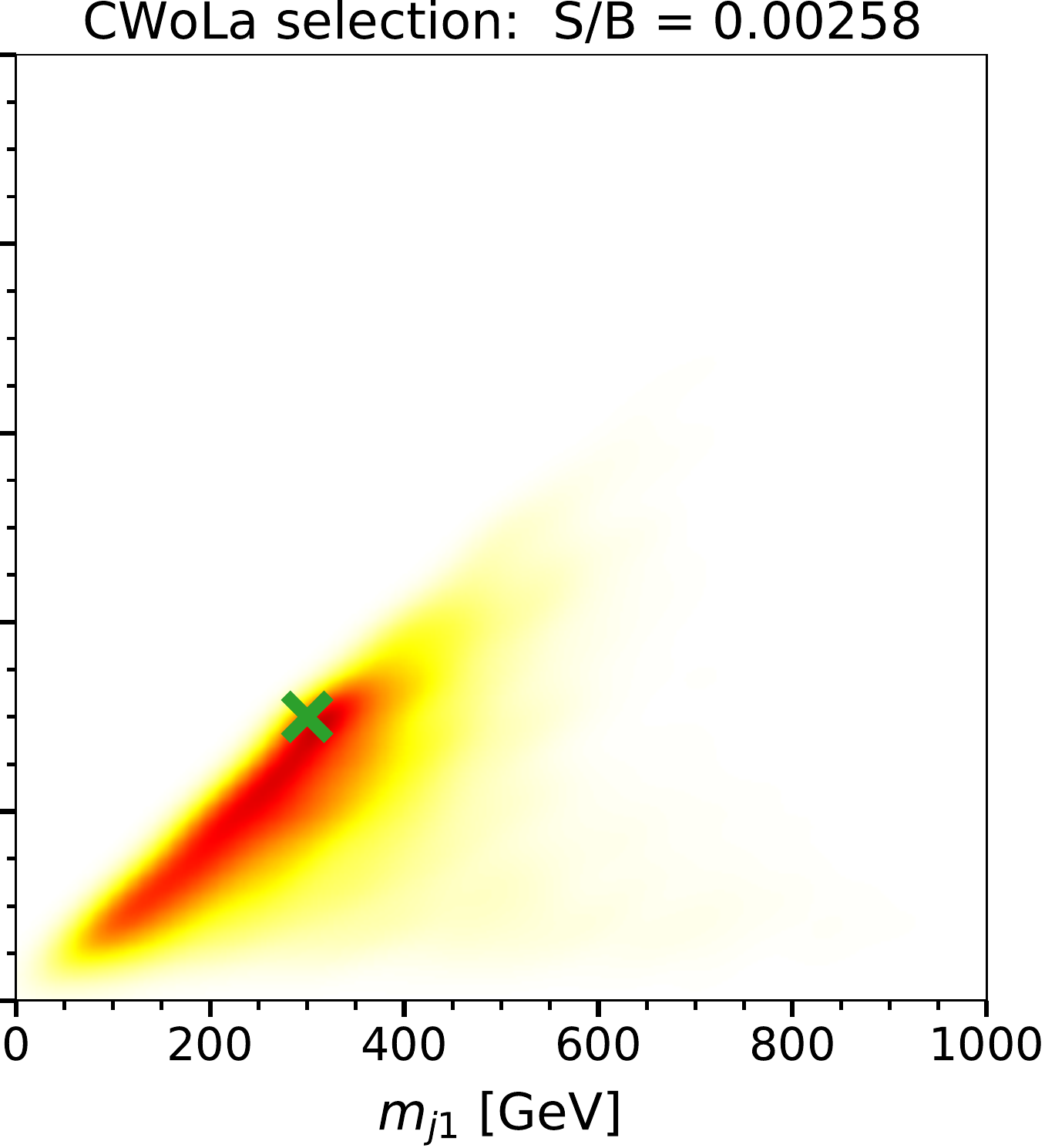}
        \includegraphics[scale=0.335]{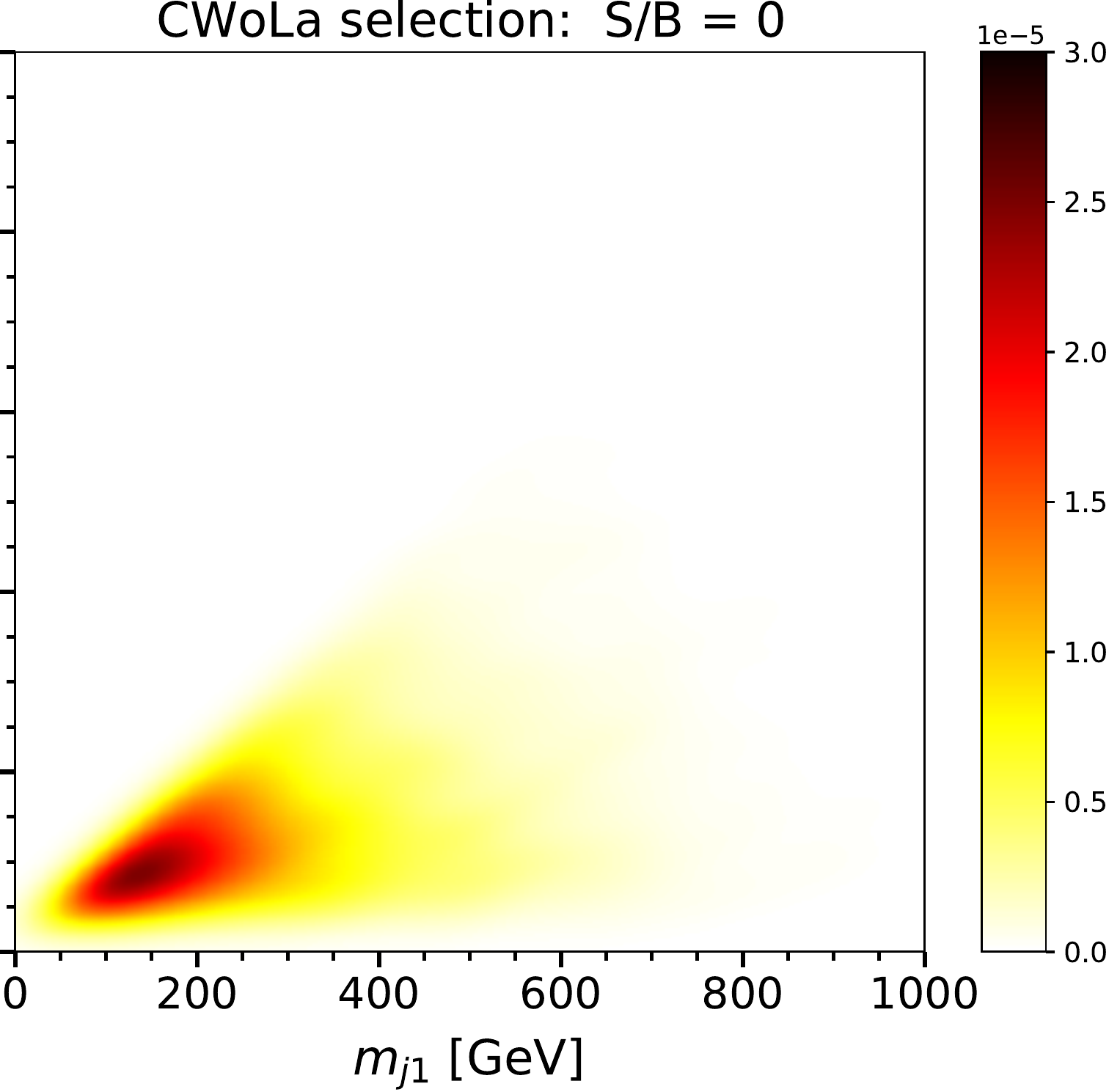}
   \end{center}
   \caption{Density of events on the $(m_{j_{1}}, m_{j_{2}})$ plane for the most signal-like events selected by CWoLa for the signal hypothesis $(m_{j_{1}}, m_{j_{2}}) = (500, 500) \; \GeV$ (top row) and  $(m_{j_{1}}, m_{j_{2}}) = (300, 300) \; \GeV$ (bottom row). The optimal cut is derived from the signal efficiency that maximizes the SIC curve. From left to right, we show results for three benchmarks with $S/B \simeq 4 \cdot 10^{-3}, 2.8 \cdot 10^{-3}, 0$. The location of the injected signal is indicated by a green cross. Note that the upper right plot shows a small statistical fluctuation that disappears when averaging over a larger number of simulations.}
   \label{fig:CWoLa_selection}
\end{figure}

\begin{figure}[h!]
   \begin{center}
        \includegraphics[scale=0.335]{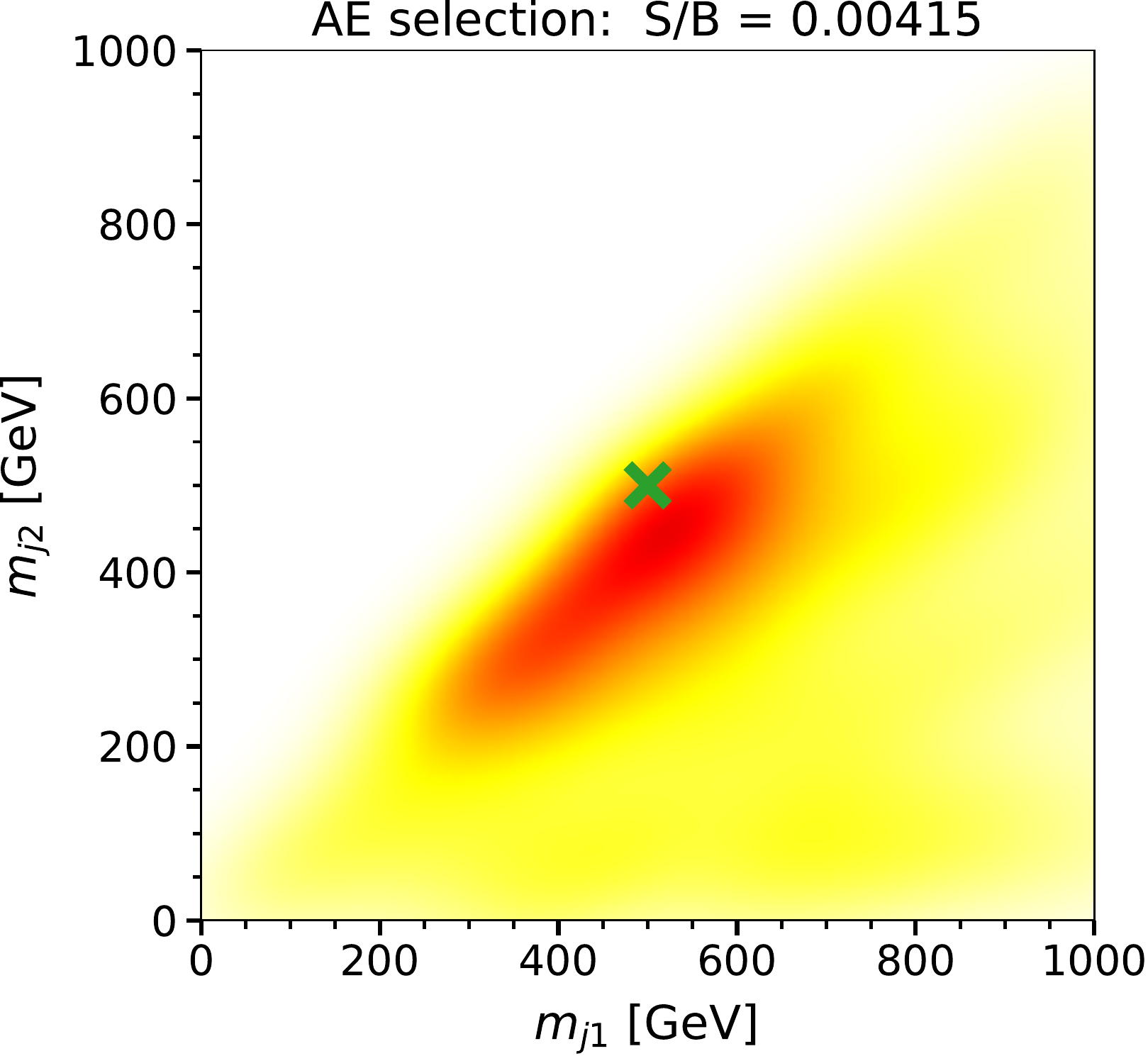}
        \includegraphics[scale=0.335]{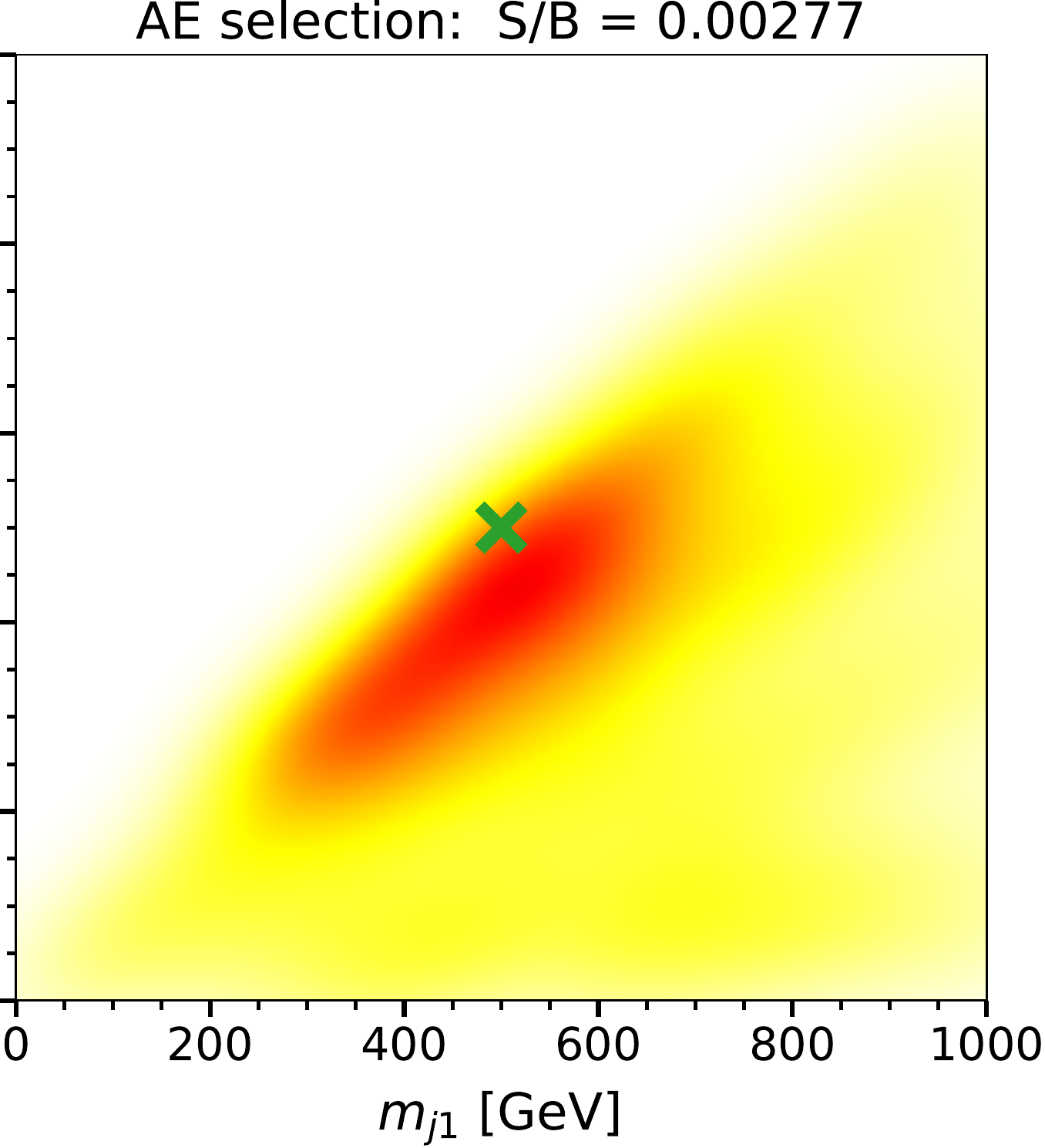}
        \includegraphics[scale=0.335]{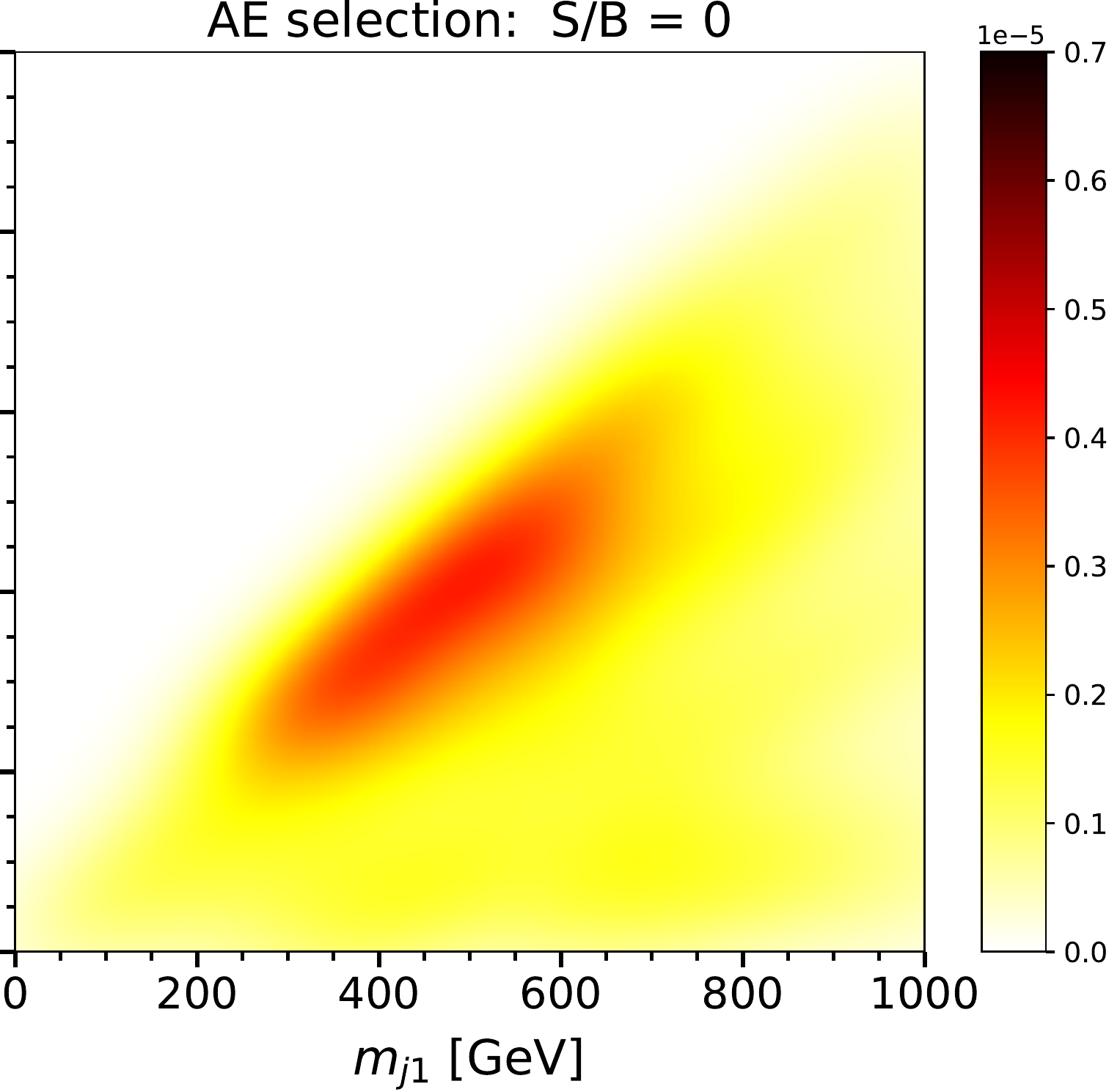} \\
        \vspace{10pt}
        \includegraphics[scale=0.335]{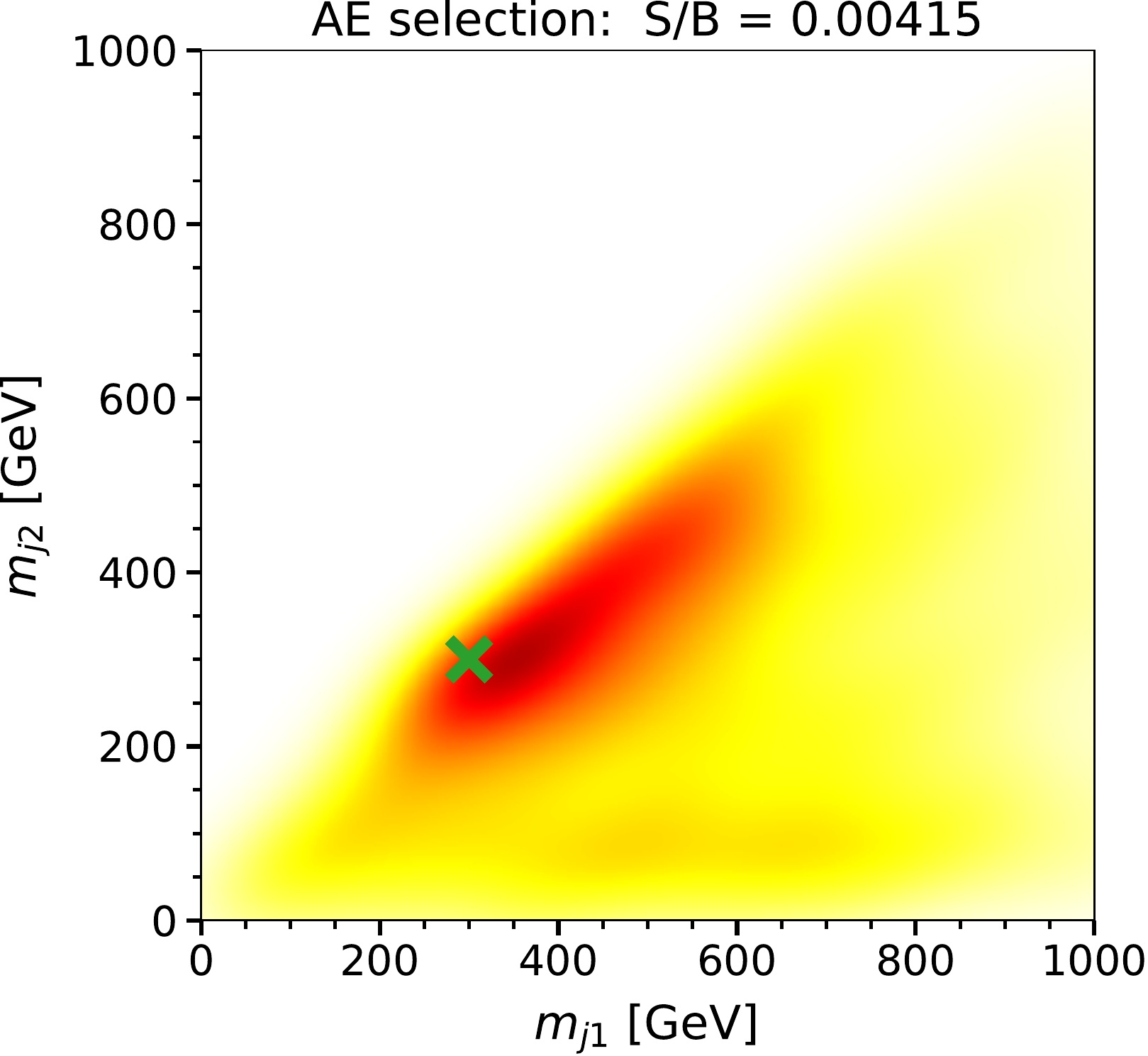}
        \includegraphics[scale=0.335]{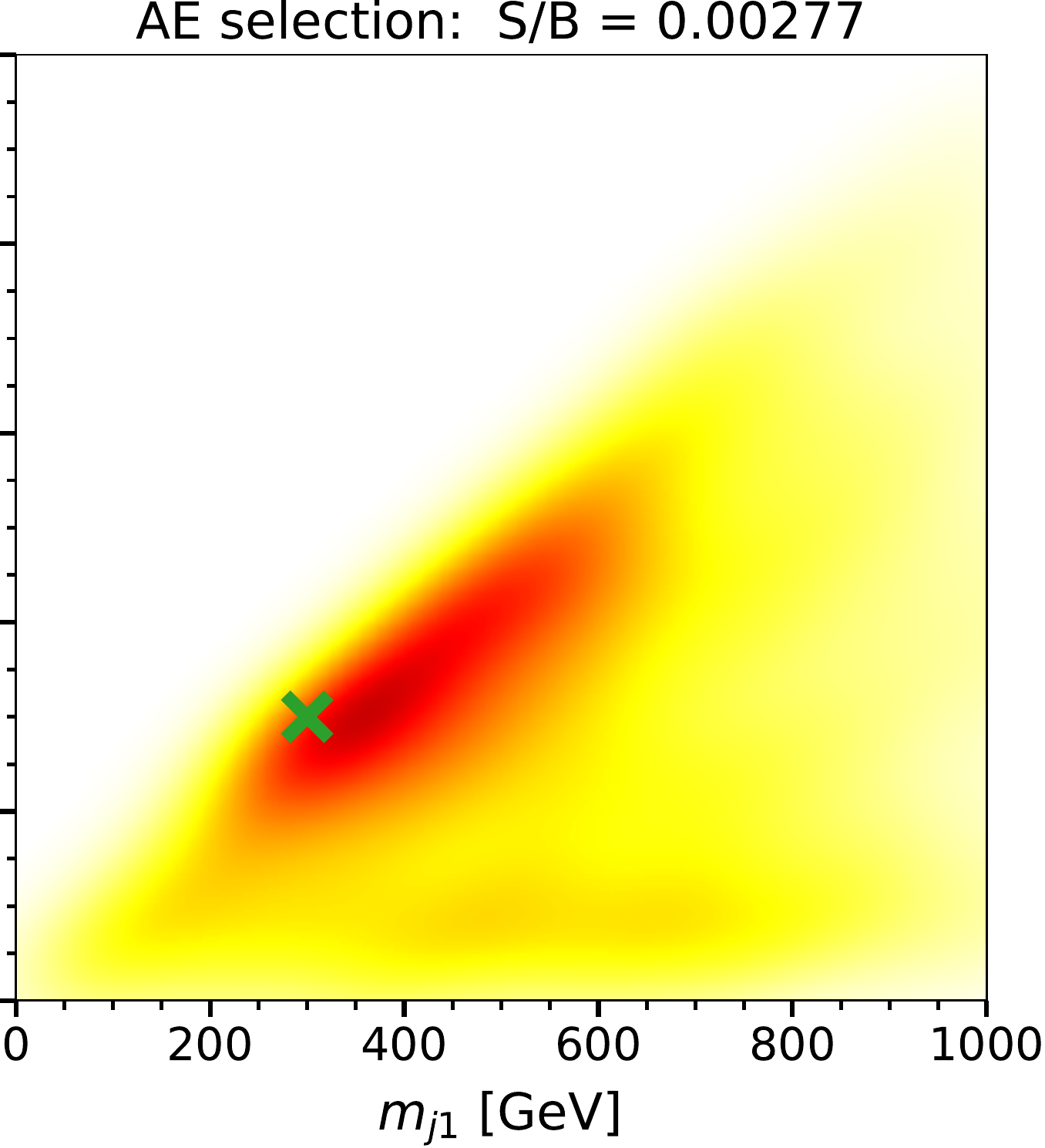}
        \includegraphics[scale=0.335]{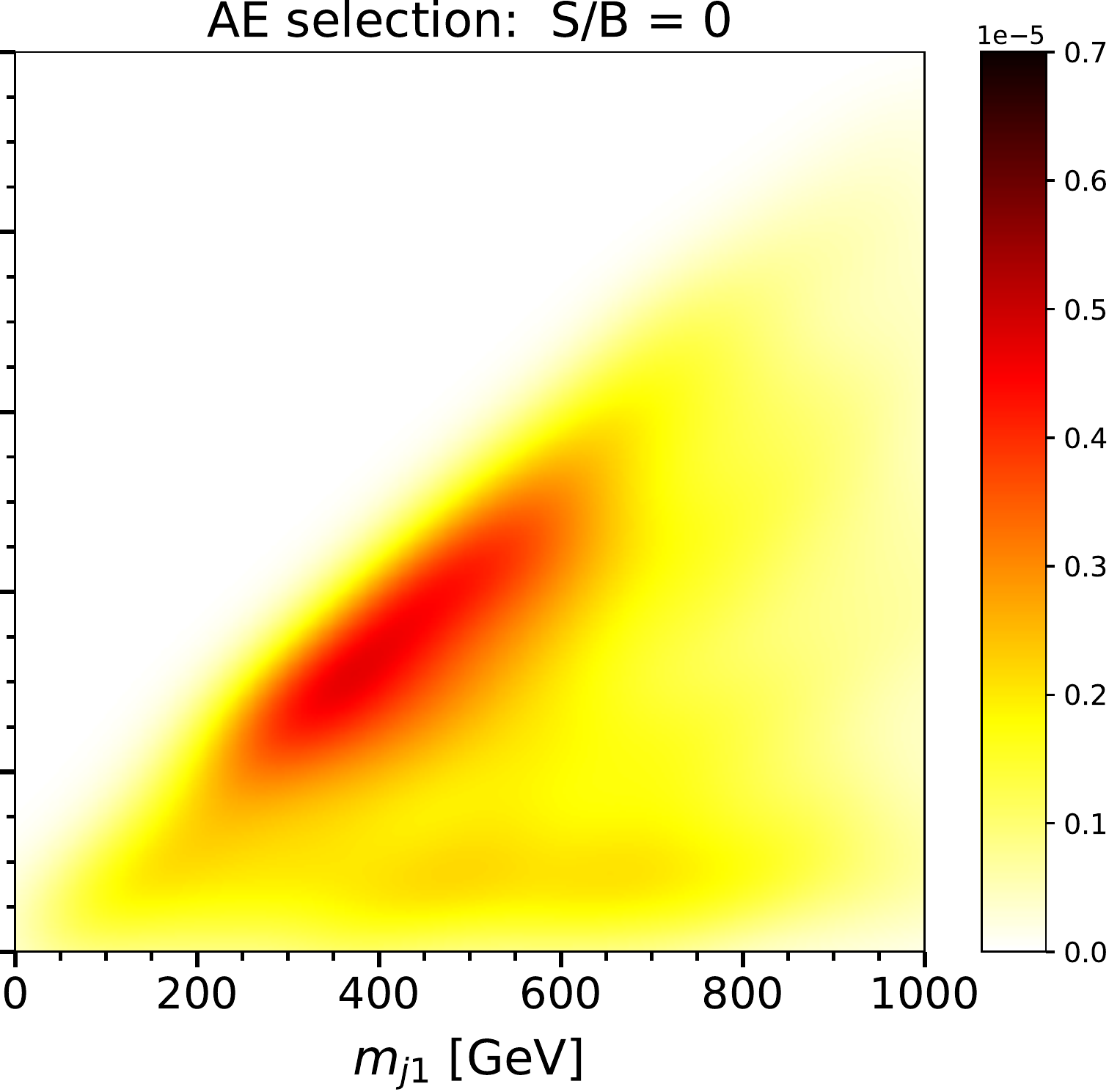}
   \end{center}
   \caption{Density of events on the $(m_{j_{1}}, m_{j_{2}})$ plane for the most signal-like events selected by the AE for the signal hypothesis $(m_{j_{1}}, m_{j_{2}}) = (500, 500) \; \GeV$ (top row) and  $(m_{j_{1}}, m_{j_{2}}) = (300, 300) \; \GeV$ (bottom row). The optimal cut is derived from the signal efficiency that maximizes the SIC curve. From left to right, we show results for three benchmarks with $S/B \simeq 4 \cdot 10^{-3}, 2.8 \cdot 10^{-3}, 0$. The location of the injected signal is indicated by a green cross.}
   \label{fig:AE_selection}
\end{figure}

\newpage
\clearpage

\section{\label{sec:AEmodel} Autoencoder model selection}

Here we will motivate the selection of the AE model used in the main body of the paper. In general, the challenge or central tension of AE model selection for anomaly detection is to choose a model that strikes a good balance between compression and expressivity, between describing the bulk of the data just well enough (i.e. with the right size latent space) without swallowing up all the anomalies as well. Here we will put forth some guidelines that could be used to find this balance in an unsupervised way. While a complete study is well beyond the scope of this work, the two signals provide some evidence for the usefulness of these guidelines.

To begin, it is useful to consider the AE as consisting of three components: 
\begin{enumerate}

\item Choice of input features.

\item Latent space dimension.

\item Rest of the architecture.

\end{enumerate}
Our philosophy is that item 1 defines the type of anomaly we are interested in, and so cannot be chosen in a fully unsupervised way. In this paper, we chose the input features to be $(m_J,p_T,\tau_{21},\tau_{32},n_{trk})$ because we observed they did well in finding the 3-prong $qqq$ signals. In contrast, item 2 and item 3 can be optimized to some extent independent of the anomaly (i.e. just from considerations of the background). 

Our main handle for model selection will be the concept of FVU: {\it fraction of variance unexplained}. This is a commonly used statistical measure of how well a regression task is performing at describing the data. Let the input data be $\vec x_i$, $i=1,\dots,N$ and the (vector-valued) regression function being
\beq
\vec y_i = f(x_i) \, .
\eeq
Let the data to be described be $\vec Y_i$. (So, for an AE, $\vec x_i=\vec Y_i$.) Then the FVU $F$ is
\beq
F = {{1\over N}\sum_{i=1}^N (\vec Y_i-\vec y_i)^2\over {1\over N}\sum_{i=1}^N(\vec Y_i-\langle\vec Y\rangle)^2}\,,
\eeq
i.e. it is the MSE of the regression divided by the sample variance of the data. In the following, we will be working with features standardized to zero mean and unit variance, in which case the denominator (the sample variance) is just $n$, the number of input features, and $F$ becomes
\beq
F = {1\over N}\sum_{i=1}^N {1\over n}\sum_{a=1}^n (Y_{ia}-y_{ia})^2 \, ,
\eeq
i.e. it is the MSE of the regression normalized to the number of input features.

\begin{figure}[!t]
    \begin{center}
        \includegraphics[scale=0.5]{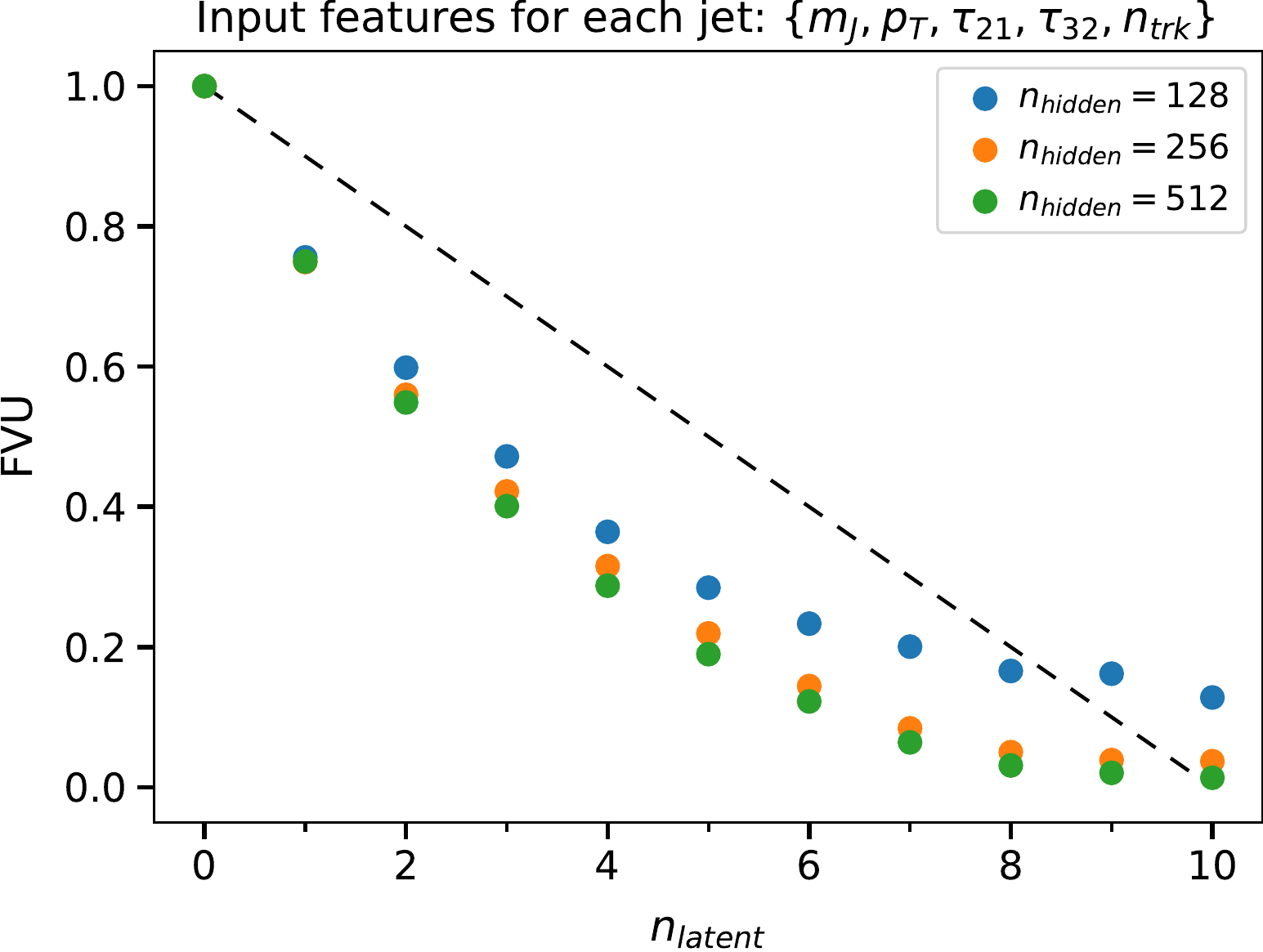}
    \end{center}
    \caption{FVU vs number of latent dimensions, for 10 input features and different AE architectures. The diagonal line is $1-n_{latent}/n=1-n_{latent}/10$ indicating the nominal case of a latent dimension just memorizing one of the input features.}
    \label{fig:FVU_vs_nlat}
\end{figure}

Our criteria for whether it is worth adding another latent space dimension to the AE is whether it substantially reduces the FVU. Here the measure of ``substantially reduces" is whether it decreases the FVU by significantly more than $1/n$. A decrease of $1/n$ (or less) suggests that the AE is merely memorizing one of the input features via the extra latent space dimension. In that case, adding the latent space dimension should not help with anomaly detection. Meanwhile, a decrease in FVU of significantly more than $1/n$ suggests that the latent space dimension is learning something nontrivial about the inner workings of the data, capturing one of the underlying correlations. In this case adding the latent space dimension may help with anomaly detection performance. 

We will demonstrate the effectiveness of this model selection criteria using the two signals considered in this paper, $Z'(3500)\rightarrow X(m)X(m)$, $X\to qqq$ events with $m=500 \; \GeV$ and $m=300 \; \GeV$. 

We scan over the size of the latent space and hidden layers, $n_{latent} = 1,2,3,4,\dots$ and $n_{hidden}=128, 256, 512$, respectively. For each architecture and choice of input features we train 10 models with random initializations on a random subset of $50000$ background jets. 

For evaluation, we feed all 1M QCD events and all the signal events to the trained models. We compute the following metrics for each model: $\langle \text{MSE}\rangle_{bg}$, $\sigma(\text{MSE})_{bg}$, $\max$(SIC) where the SIC is computed by cutting on the MSE distribution. For all three metrics, we only compute them using the MSE distribution in a window $(3300, 3700) \; \GeV$ in $m_{JJ}$.

\begin{figure}[!t]
    \begin{center}
        \includegraphics[scale=0.5]{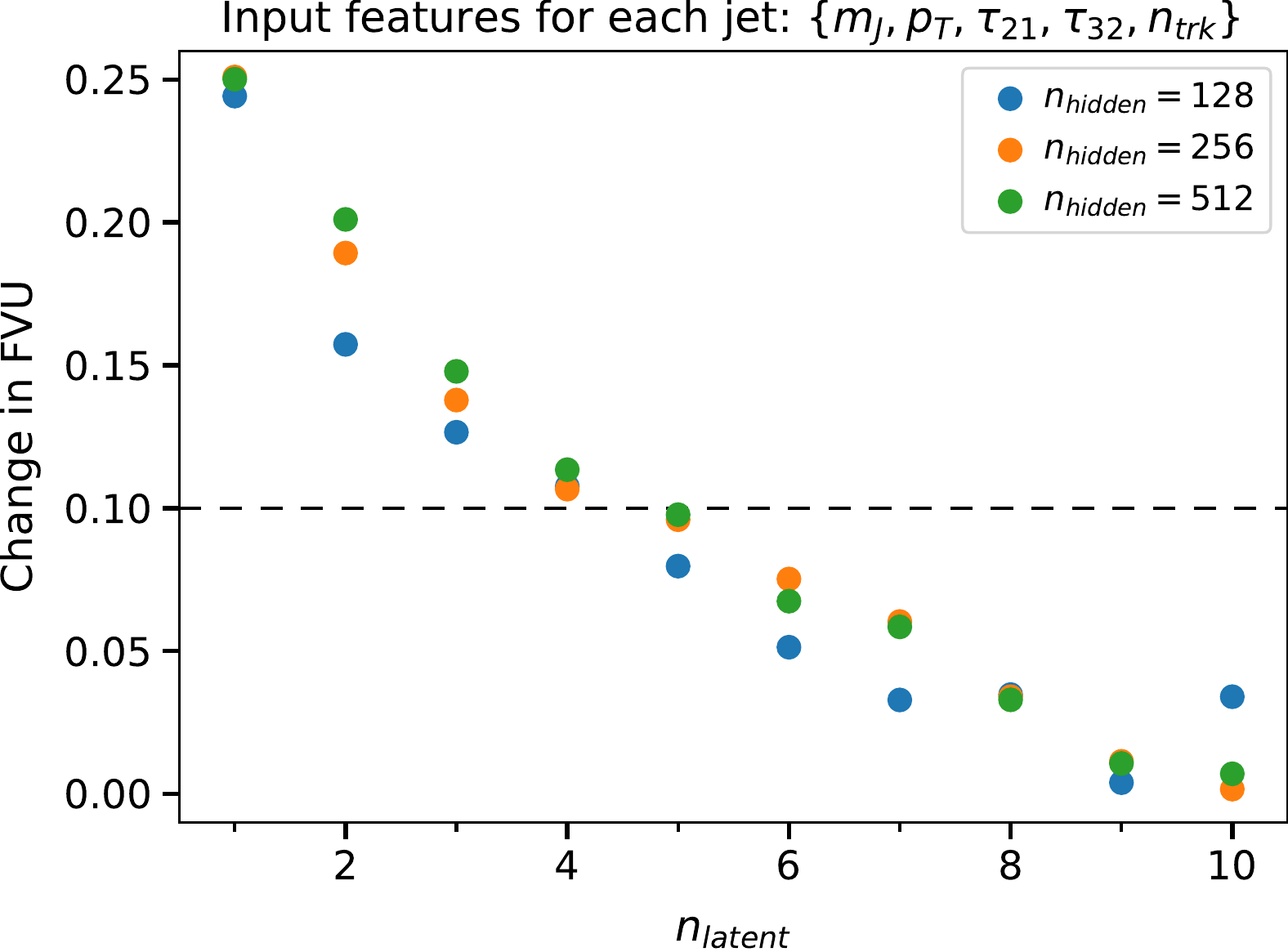}
    \end{center}
    \caption{Decrease in FVU from adding one more latent dimension  vs number of latent dimensions, for 10  input features and different AE architectures. The horizontal line is $1/n=1/10$ indicating the nominal case of a latent dimension just memorizing one of the input features.}
    \label{fig:DeltaFVU_vs_nlat}
\end{figure}

Shown in Fig.~\ref{fig:FVU_vs_nlat} is the FVU versus the number of latent dimensions, for 5 input features and different AE architectures. Each point represents the average MSE obtained from 10 independent trainings. We see that the FVU versus $n_{latent}$ plot has a characteristic shape, with faster-than-nominal decrease for small $n_{latent}$ (the AE is learning nontrivial correlations in the data) and then leveling out for larger $n_{latent}$ (the AE is not learning as much and is just starting to memorize input features). 

In Fig.~\ref{fig:DeltaFVU_vs_nlat} we show the decrease in FVU with each added latent dimension, versus the number of latent dimensions. From this we see that $n_{latent}=1,2,3$ add useful information to the AE but beyond that the AE may not be learning anything useful. 

We also see from these plots that the FVU decreases with more $n_{hidden}$ as expected, although it seems to be levelling off by the time we get to $n_{hidden}=512$. This makes sense -- for fixed $n_{latent}$ the bottleneck is fixed, so increasing $n_{hidden}$ just increases the complexity of the correlations that the AE can learn, with no danger of becoming the identity map. This suggests that the best AE anomaly detector will be the largest $n_{hidden}$ that we can take for fixed $n_{latent}$, although the gains may level off for $n_{hidden}$ sufficiently large.

\begin{figure}[!t]
    \begin{center}
        \includegraphics[scale=0.45]{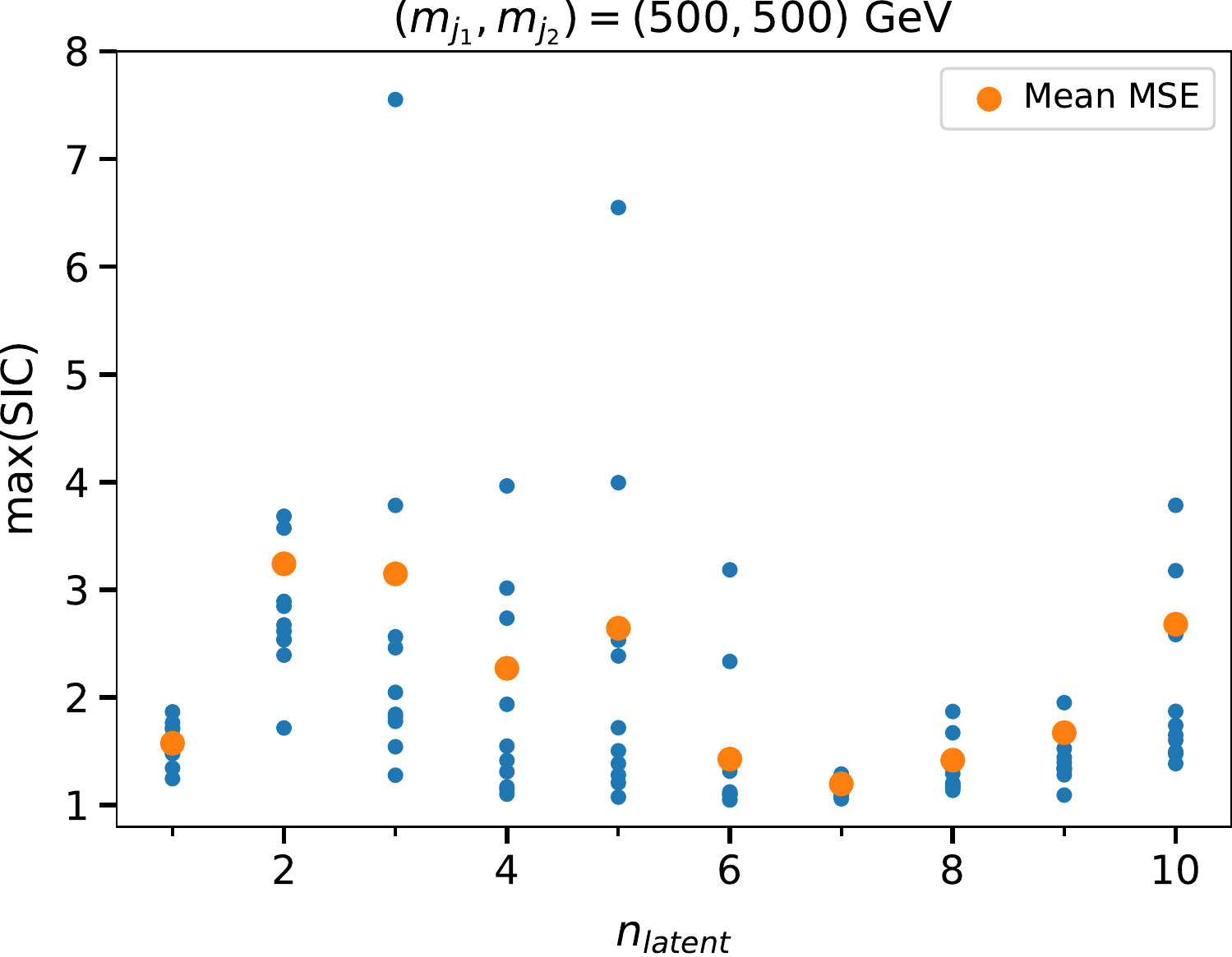}
        \hspace{10pt}
        \includegraphics[scale=0.45]{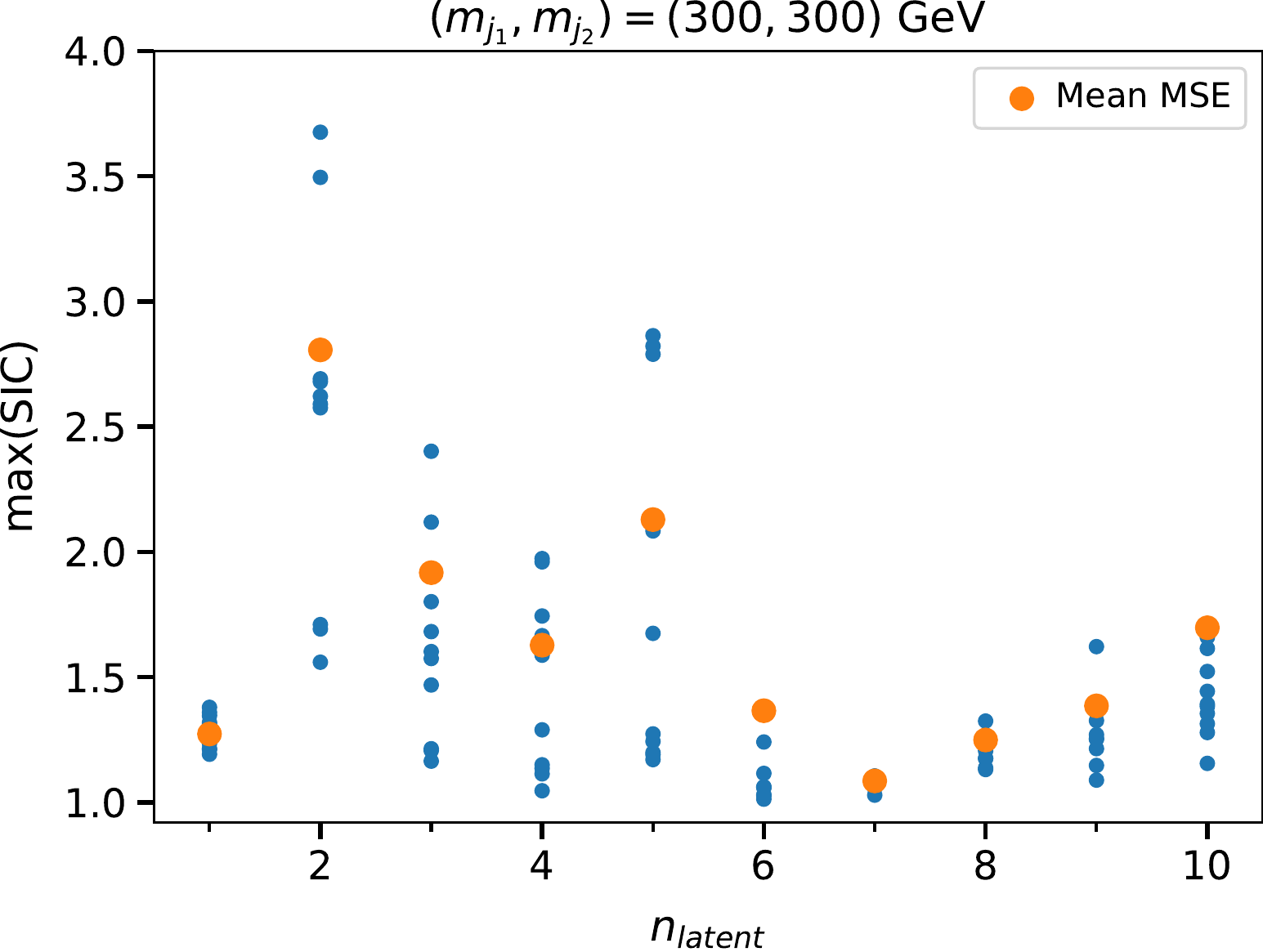}
    \end{center}
    \caption{$\max$(SIC) vs.\ $n_{latent}$ for the $500 \; \GeV$ signal (left) and $300 \; \GeV$ signal (right), 5 input features, and $n_{hidden}=512$. The blue dots are the maxSICs for each of the 10 independent trainings, while the orange dot is the max(SIC) obtained from the average of the 10 MSE distributions.}
    \label{fig:maxSIC_vs_nlat}
\end{figure}

Now we examine the performance of the various AE models on anomaly detection of the $300 \; \GeV$ and $500 \; \GeV$ 3-prong signals. The $\max$(SIC) versus $n_{latent}$ is shown in Fig.~\ref{fig:maxSIC_vs_nlat}.  We see that there is decent performance on both signals for $n_{latent}=2,3,4,5$ with $n_{latent}=2$ being especially good for both\footnote{We also see a rise in performance for very large $n_{latent}$ which is puzzling and mysterious.}. This is roughly in line with the expectations from the FVU plots. Importantly, if we restricted to $n_{latent}=2,3$ which have the larger decreases in FVU, we would not miss out on a better anomaly detector. 

\begin{figure}[!h]
    \begin{center}
        \includegraphics[scale=0.45]{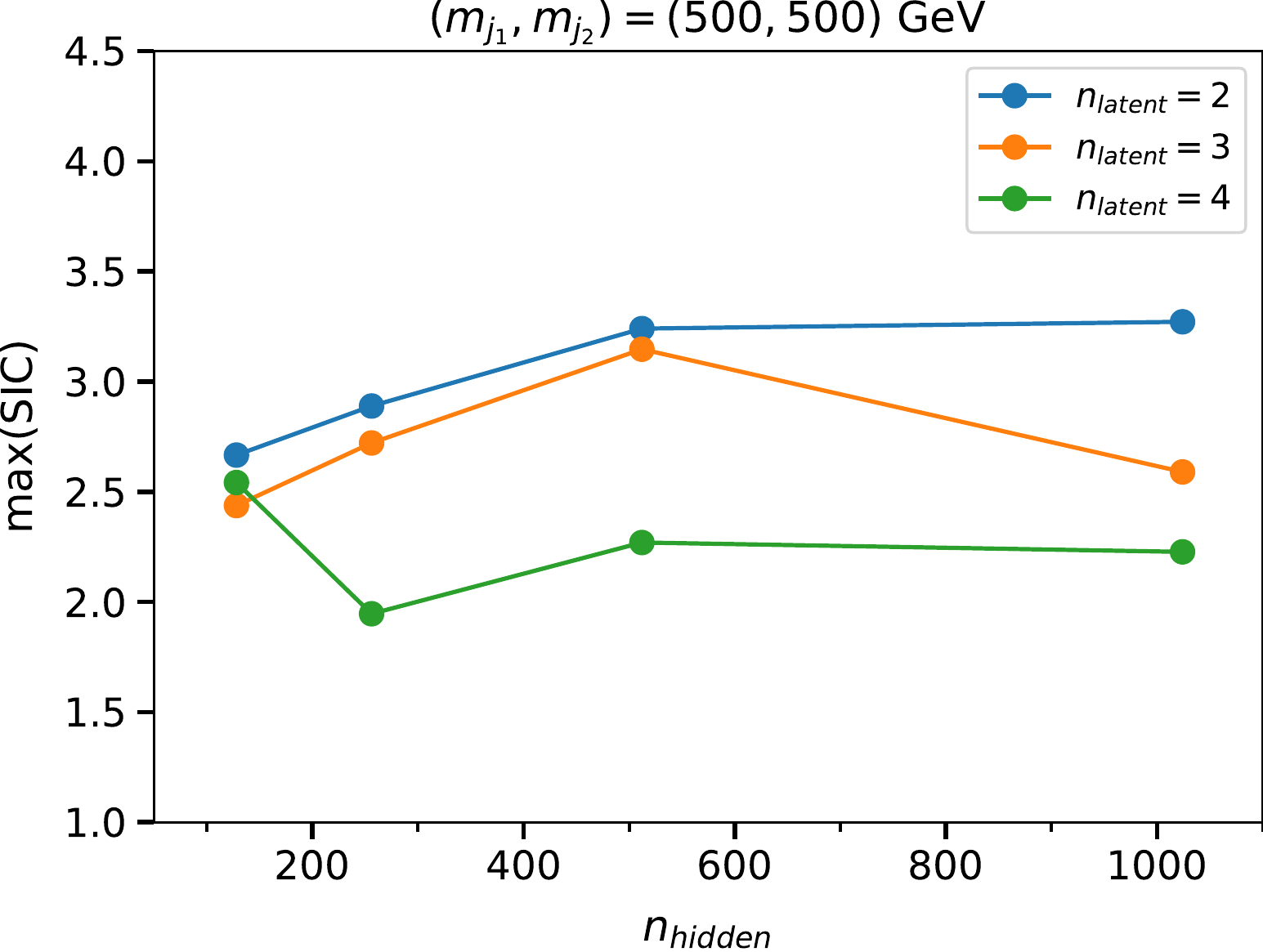}
        \hspace{10pt}
        \includegraphics[scale=0.45]{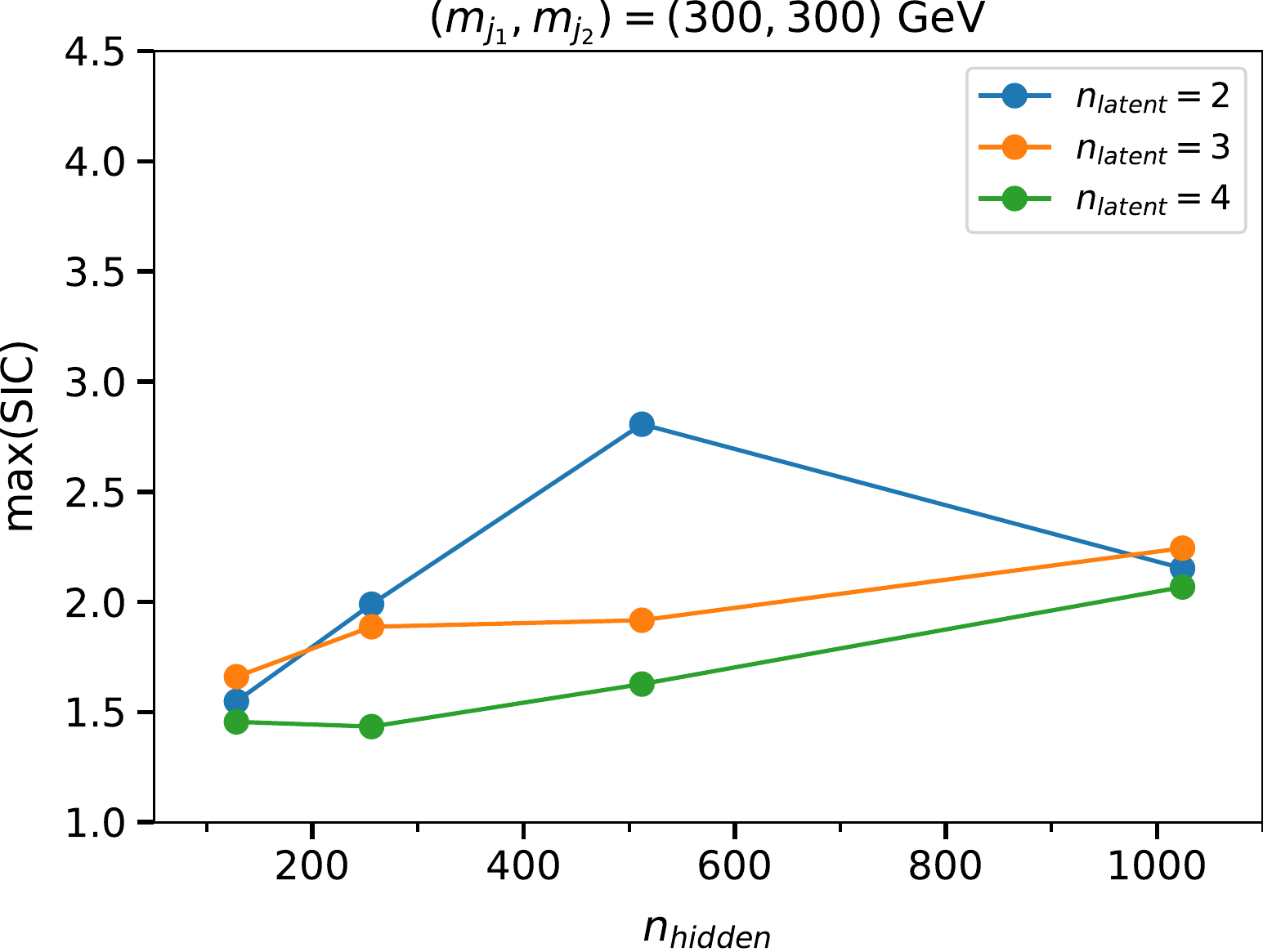}
    \end{center}
    \caption{$\max$(SIC) of the averaged MSE distributions vs.\ $n_{hidden}$ for the $500 \; \GeV$ signal (left) and $300 \; \GeV$ signal (right), 5 input features, and $n_{latent}=2,3,4$.}
    \label{fig:maxSIC_vs_nhid}
\end{figure}

Finally in Fig.~\ref{fig:maxSIC_vs_nhid}, we show the $\max$(SIC) for the MSE distributions averaged over 10 trainings vs $n_{hidden}$, for $n_{latent}=2,3,4$. We see that generally the trend is rising or flat with increasing $n_{hidden}$, which is more or less consistent with expectations.

To summarize, we believe we have a fairly model-independent set of criteria for AE model selection, based on the FVU, which empirically works well on our two signals. Admittedly this is too small of a sample size to conclude that this method really works; it would be interesting to continue to study this in future work. Based on these criteria, we fix the AE model in this paper to have $n_{latent}=2$ and $n_{hidden}=512$.

\clearpage
\end{appendices}
\clearpage

\bibliography{refs,HEPML}
\bibliographystyle{utphys}

\end{document}